\documentclass[onecolumn,a4paper,accepted=2026-11-17]{quantumarticle}
\pdfoutput=1
\usepackage{fontenc}
\usepackage{amsmath,amssymb,amsfonts,amsthm}
\usepackage{thmtools,thm-restate}
\usepackage{mathtools}
\usepackage{graphicx}
\usepackage{physics}
\usepackage{cancel}
\usepackage[dvipsnames]{xcolor}
\usepackage{dsfont}
\usepackage{geometry}
\usepackage{setspace}
\usepackage{url}
\usepackage{algorithm}
\usepackage{algorithmic}
\usepackage{nicefrac}
\usepackage{bm}
\usepackage{tcolorbox}
\usepackage{adjustbox}
\usepackage{fancybox}
\usepackage{booktabs}
\usepackage{placeins}
\usepackage{accents}
\usepackage{quantikz}
\usepackage{svg}
\usepackage{subcaption}

\usepackage{circuitikz}

\usepackage[normalem]{ulem}

\usepackage[numbers,sort&compress]{natbib}

\usepackage[colorlinks]{hyperref}
\hypersetup{
	pdfstartview={FitH},
    pdfnewwindow=true,
	colorlinks=true,
	linkcolor=blue,
	citecolor=blue,
	filecolor=blue,
	urlcolor=blue}
\usepackage{multirow}
\usepackage{physics}
\usepackage[capitalise]{cleveref}

\newtheorem{remark}{Remark}
\newtheorem{example}{Example}

\newcommand{\thm}[1]{\hyperref[thm:#1]{Theorem~\ref*{thm:#1}}}
\newcommand{\defn}[1]{\hyperref[defn:#1]{Definition~\ref*{defn:#1}}}
\newcommand{\lem}[1]{\hyperref[lem:#1]{Lemma~\ref*{lem:#1}}}
\newcommand{\prop}[1]{\hyperref[prop:#1]{Proposition~\ref*{prop:#1}}}
\newcommand{\fig}[1]{\hyperref[fig:#1]{Figure~\ref*{fig:#1}}}
\newcommand{\tab}[1]{\hyperref[tab:#1]{Table~\ref*{tab:#1}}}
\renewcommand{\sec}[1]{\hyperref[sec:#1]{Section~\ref*{sec:#1}}}
\newcommand{\app}[1]{\hyperref[app:#1]{Appendix~\ref*{app:#1}}}
\newcommand{\cor}[1]{\hyperref[cor:#1]{Corollary~\ref*{cor:#1}}}
\newcommand{\obs}[1]{\hyperref[obs:#1]{Observation~\ref*{obs:#1}}}
\newcommand{\rem}[1]{\hyperref[rem:#1]{Remark~\ref*{obs:#1}}}
\newcommand{\nn}{\nonumber \\}
\newcommand{\append}[1]{\hyperref[append:#1]{Appendix~\ref*{append:#1}}}
\renewcommand{\swap}{\mathrm{SWAP}}

\newtheorem{theorem}{Theorem}
\newtheorem{definition}[theorem]{Definition}
\newtheorem{lemma}[theorem]{Lemma}
\newtheorem*{lemma*}{Lemma}
\newtheorem{proposition}[theorem]{Proposition}
\newtheorem{corollary}[theorem]{Corollary}
\newtheorem{prob}[theorem]{Problem}

\newtheorem{result}[theorem]{Result}

\newcommand{\<}{\ensuremath{\langle}}
\renewcommand{\>}{\ensuremath{\rangle}}

\newcommand{\imag}{\mathrm{i}}

\newcommand{\e}{\mathrm{e}}
\newcommand{\dom}{S}
\newcommand{\domcomp}{S_c}
\newcommand{\domcompD}{S_c^D}
\newcommand{\domcompN}{S_c^N}

\newcommand{\orac}{\ensuremath{\mathrm{O}}}

\renewcommand{\l}{\left}
\renewcommand{\r}{\right}
\newcommand{\mc}{\mathcal}

\newcommand{\D}{\mathrm{d}}
\renewcommand{\dd}[1]{\frac{\mathrm{d}}{\mathrm{d}#1}}
\newcommand{\ddt}{\dd{t}}
\renewcommand{\proj}{{\mathsf{P}_c^\perp}}
\newcommand{\projc}{{\mathsf{P}_c}}
\newcommand{\projcD}{{\mathsf{P}_c^D}}
\newcommand{\projcN}{{\mathsf{P}_c^N}}
\newcommand{\identity}{\mathds{I}}

\newcommand{\hamt}{{\text{HAM-T}}}

\newcommand{\TE}[3]{\mathcal T \e^{\int_0^{#3} #1(#2) \mathrm \D #2}}

\renewcommand{\tr}{\mathrm{tr}}
\newcommand{\diff}{\ensuremath{\mathsf{D}}}

\newcommand{\reals}{\mathds{R}}
\newcommand{\complex}{\mathds{C}}
\newcommand{\ipic}{_I}

\usepackage{hyperref}

\newcommand{\UofT}{\affiliation{Department of Computer Science, University of Toronto, Canada.}}
\newcommand{\vectorInst}{\affiliation{Vector Institute for Artificial Intelligence, Toronto, Canada.}}
\newcommand{\lanl}{\affiliation{Theoretical Division, Los Alamos National Laboratory, Los Alamos, USA.}}

\newcommand{\YMSC}{\affiliation{Yau Mathematical Sciences Center and Department of Mathematics, Tsinghua University, Beijing, China}}
\newcommand{\BIMSA}{\affiliation{Yanqi Lake Beijing Institute of Mathematical Sciences and Applications, Beijing, China}}

\newcommand{\berkeleyone}{\affiliation{Department of Chemistry, University of California, Berkeley.}}
\newcommand{\berkeleytwo}{\affiliation{Berkeley Quantum Information and Computation Center, University of California, Berkeley.}}
\newcommand{\pnnl}{\affiliation{Pacific Northwest National Laboratory, Richland, USA.}}
\newcommand{\cias}{\affiliation{Canadian Institute for Advanced Studies, Toronto, Canada.}}
\newcommand{\UofTchem}{\affiliation{Department of Chemistry, University of Toronto, Canada.}}
\newcommand{\cifar}{\affiliation{Canadian Institute for Advanced Research, Toronto, Canada.}}
\newcommand{\ac}{\affiliation{Acceleration Consortium, Toronto, Canada.}}
\newcommand{\chemeng}{\affiliation{Department of Chemical Engineering \& Applied Chemistry, University of Toronto, Toronto, Canada.}}
\newcommand{\materials}{\affiliation{Department of Materials Science \& Engineering, University of Toronto, Toronto, Canada.}}
\newcommand{\nvidia}{\affiliation{NVIDIA, Toronto, Canada.}}
\newcommand{\lebovic}{\affiliation{Lebovic Fellow, Canadian Institute for Advanced Research, Toronto, Canada.}}

\graphicspath{{./figs/}}
\begin{document}

\title{Arbitrary  Boundary Conditions  and Constraints in Quantum Algorithms for Differential Equations via Penalty Projections}
\author{Philipp Schleich}\email{philipps@lanl.gov, philipps@cs.toronto.edu}\UofT\vectorInst\lanl
\author{Tyler Kharazi}\berkeleyone \berkeleytwo
\author{Xiangyu Li}\pnnl
\author{Jin-Peng Liu}\YMSC\BIMSA
\author{Al\'an Aspuru-Guzik}\UofT \vectorInst \UofTchem  \cifar \ac \chemeng \materials \lebovic  \nvidia
\author{Nathan Wiebe}\email{nawiebe@cs.toronto.edu}\UofT\pnnl\cias

\begin{abstract}
Complicated boundary conditions are  essential to accurately describe phenomena arising in nature and engineering.
Recently, the investigation of a potential speedup through quantum algorithms in simulating the governing ordinary and partial differential equations of such phenomena has gained increasing attention.   
We propose an approach to enforce arbitrary boundary conditions and constraints through adding a penalty projection to the governing equations, relying on known quantum differential equation solvers, leading to an overall efficient quantum algorithm for solving differential equations under boundary conditions and constraints. 
Assuming a fast-forwardable representation of the projection to ensure an efficient interaction picture simulation, the cost to enforce the constraints is at most $O(\log\lambda)$ in the strength of the penalty $\lambda$ in the gate complexity within the scenarios we considered, for precision $\varepsilon$ and a dynamical system generated by $A_0(t)$ of size $n^d\times n^d$ with forcing $b(t)$. E.g., for the heat equation, this leads roughly to a gate complexity overhead of $\widetilde O(d\log n + \log t)$.
We show constraint error bounds for the penalty approach, both constraint and solution error, and provide validating numerical experiments, and estimate the circuit complexity using the Linear Combination of Hamiltonian Simulation.  
\end{abstract}

\maketitle
\newpage
\tableofcontents


\section{Introduction}
Ordinary and partial differential equations (ODEs, PDEs) describe many processes and phenomena in science and engineering. The modelling of these phenomena typically involves describing interactions of the system studied with its environment through `boundaries', which imposes constraints on the solution of these equations. Hence, for a meaningful numerical solution it is crucial to represent such constraints and boundary conditions. 
In addition, solving DEs on classical computers is cursed by dimensionality. Quantum algorithms for differential equations have been developed in order to mitigate exponential cost in the dimension. 
Such algorithms are based on, e.g., encoding the solution in a linear system~\cite{berry2014high,berry2017quantum,berry2022quantum,krovi2023improved,childs2020quantum,childs2021highprecision}, performing explicit `time-marching'~\cite{fang2023time,bharadwaj2024compact}, or making use of integral identities~\cite{an2022theory,an2023fractal,an2023linear,an2023quantum,jin2024quantum,lu2025infinite}. Recently, algorithms that map the ODE to a Lindblad evolution have been explored in \cite{shang2024design}, drawing from quantum algorithms that are able to simulate open quantum systems.
Particular systems studied are mostly focused around popular models from classical differential equations courses such as the Poisson equation~\cite{cao2013quantum} or more general elliptic equations~\cite{childs2021highprecision,bagherimehrab2023fast}, the advection equation~\cite{novikau2024koopmanvneumann} or the wave equation~\cite{costa2019quantum} and  also the Maxwell equations~\cite{jin2024maxwell} or nonlinear equations such as reaction diffusion equations~\cite{liu2021efficient,liu2023efficient,krovi2023improved,costa2023further} or fluid dynamics~\cite{li2023potential,bharadwaj2025towards,penuel2024feasibility}.
Most studies so far focus on input models that assume a structured geometry with no (or, periodic) boundary conditions which limits applicability. 
Imposing the boundaries in the encoding of the generator of the dynamics is possible for several techniques~\cite{childs2021highprecision,novikau2023simulation,jin2024quantumboundary,liu2023efficient}, based on other classical techniques such as complex absorbing potentials also also done in \cite{mangin2024efficient}, perfect matched layers or Dirichlet-to-Neumann maps~\cite{jin2024artificialboundaries} or analytic continuation~\cite{kharazi2024explicit}. However,  there are some practical limitations to this in the way it can affect the respective access model of the system matrix, e.g. the complexity of a circuit construction of a specific block-encoding.  Additionally, it is not immediately obvious how to approach boundary conditions in the recently popular Linear Combination of Hamiltonian Simulations (LCHS) algorithm~\cite{an2023linear,an2023quantum}, which is promising due to (near-)optimal scaling both in the number of state-preparation queries as well as the system  matrix queries.  
Beyond the specific concept of boundary conditions, we may be interested in more general constraints. 

Penalty methods have long been used to impose boundary conditions and constraints in classical numerical methods~\cite{luenberger2016penalty}. The constraint is enforced approximately through an additional term in the equations or variational functional. Increasing the penalty strength improves the approximation, but can also introduce stiffness or ill-conditioning. For example, \cite{babuska1973penalty} considers a finite-element formulation of the Poisson equation in which the trial functions need not satisfy the Dirichlet boundary conditions. By choosing the boundary penalty in relation to the mesh width, the convergence rate can approach that of finite elements satisfying the boundary conditions directly; thereby, their constraint errors are static, mesh-dependent and do not give a similar penalty-strength based error bound coming from finite-time propagation. Reference~\cite{angot1999penalization} studies volume penalization for incompressible viscous flows around obstacles.
For a velocity penalty of strength $\lambda$, their analysis shows an $O(\lambda^{-1/4})$ error in the global energy norms and an $O(\lambda^{-3/4})$ error in the space-time $L^2$ norm inside the obstacle.
A formulation that also penalizes the time derivative and viscous terms gives $O(1/\lambda)$ convergence under stronger regularity and compatibility assumptions.

In this work, we implement boundary values or constraints on evolution equations by separating the dynamics into unconstrained (feasible) and constrained (infeasible) dynamics using perturbation theory. 
We build on the idea of approximating constraints through a large penalty, using an oscillatory projection to separate the feasible and infeasible dynamics. We derive bounds on both the constraint violation and the error relative to the constrained solution, uniformly over a finite time interval. These bounds give an explicit choice of the penalty strength. 
Our method is also capable of modeling interface conditions, which were also explored in the context of quantum algorithms in \cite{jin2024quantumboundary}, as well as Robin boundary conditions.
Primarily, there are ties to `complex absorbing potentials' for quantum dynamics~\cite{muga2004complex}, as previously used in quantum algorithms for non-unitary dynamics~\cite{an2023linear,ManginBrinet2024efficientsolutionof}.
Another interesting study in the simulation of acoustic waves uses counter-acting waves to annihilate the solution~\cite{tsynkov2003artificial} -- imagine numerical noise-cancelling headphones.
Specifically, our approach adds a projection scaled by $\imag \lambda$ to the generator of a dynamical system. A physical interpretation of this may be seen as measurement with (very high) frequency $\lambda$ on the domain of the projector, similar to the quantum Zeno effect. Thereby, our bounds also closely resemble respective bounds related to the Zeno effect~\cite{burgarth2019generalized,burgarth2022one}. 
General results from perturbation theory suggest what we are looking for is a small signal-to-noise ratio~\cite{van1975perturbations}. For an ODE $\ddt{v(t)} = A_0 v(t)$, this can be envisioned as the systems `energy' divided by the strength of the perturbation,  $\frac{|\<v(0), A_0 v(0)\>|}{\lambda}$; this gives us intuition about what necessary frequency in the perturbation to expect. Our bounds in \cref{subsec:constraint-error-bounds} confirm this intuition. 
Assuming a fast-forwardable representation of the projection, we can then implement the penalty through interaction-picture simulation within LCHS, without adding the penalty strength to the block-encoding normalization of the original generator.
The dependence on $\lambda$ in the remaining oracle and gate costs is discussed in \cref{sec:algorithm-and-complexity}, with overhead in gate complexity coming from enforcing the constraint roughly $O(\log(\lambda))$.

\begin{figure}[ht]
    \centering
    \includegraphics[width=0.7\linewidth]{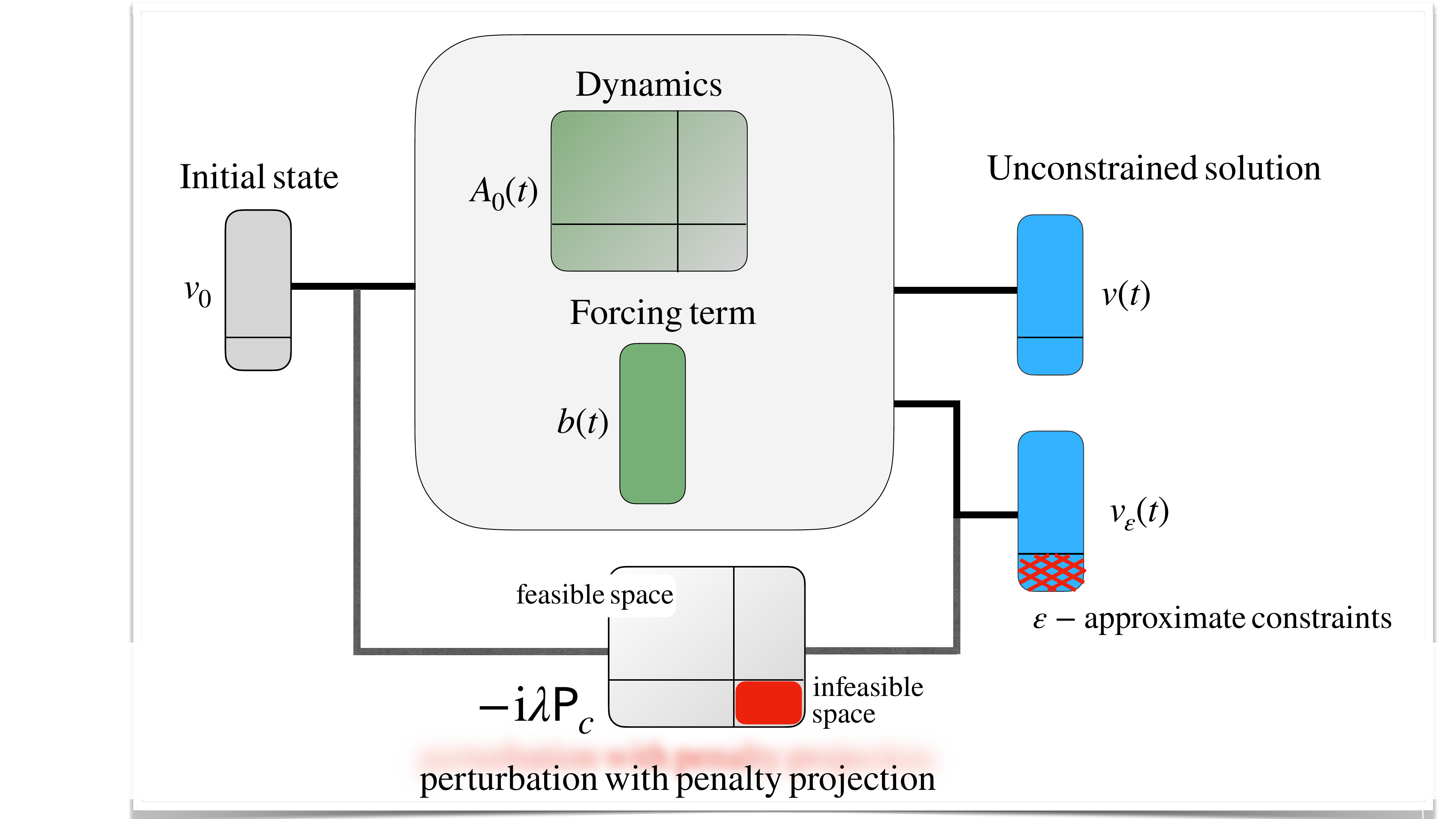}
    \caption{Enforcing constraints in solution $v(t)$ to an ordinary differential equation defined by $A_0(t)$ and source term $b(t)$, by adding a penalty function defined by a projection operation $\projc$ that projects onto the infeasible, constraint space. By appropriately choosing the penalty strength $\lambda$, this approach can enforce constraints that are efficiently representable by a projection up to arbitrarily small accuracy $\varepsilon>0$.}
    \label{fig:overview-fig}
\end{figure}

Within this work, we focus on ordinary differential equations under constraints. This is equivalent to readily discretized partial differential equations under boundary conditions. Hereby, our analysis separates the penalty error from any spatial-discretization error.  We focus on evolution equations of the kind $\ddt v(t) = A(t)v(t)+b(t)$ and a time parameter $t\ge 0$ and will consider stationary boundary value problems in future work. \cref{fig:overview-fig} depicts the general approach: When evolving an ODE, the addition of the penalty projection allows us to produce a final state that satisfies a set of constraints, as represented by the projection onto an infeasible space, up to some accuracy $\varepsilon$. 

\subsection{Outline and Main Results}
The main contribution of our work is a penalty-projection construction for constrained evolution equations, together with error guarantees and an  implementation using existing quantum ODE simulation methods. 
In particular, we provide bounds on the penalty strength to achieve a given solution error, for the case of Dirichlet, Neumann and Robin conditions under certain assumptions on the projection as outlined below.
The focal point of our approach is a perturbative term $-\imag\lambda\projc$ in the DE system
\begin{equation*}
    \ddt v(t) = A_0(t)v(t) + b(t) \colorbox{blue!10}{$-\imag\lambda\projc v(t)$}.
\end{equation*}
Here, $A_0(t)$ describes the dynamics of interest, $\lambda>0$ is the strength of the penalty term and $\projc$ is a Hermitian projection onto the constraint space. 
We say that a constraint $\projc$ is satisfied by a state $v(t)$ if $\projc v = 0$.
Simulating for time $t'>0$ gives $v(t')$ that satisfies 
\begin{equation}\label{eq:mainres-constraint}
    \|\projc v(t')\|_{\ell_2}\le \varepsilon \quad\text{ and }\quad \|\proj (v(t')-u(t'))\|_{\ell_2}\le \varepsilon,
\end{equation}
where $u(t)$ is obtained from evolving unconstrained dynamics restricted to the feasible space. Thereby, we derive error bounds to pick $\lambda$ so that the constraint is satisfied to error $\varepsilon$ and the solution in the unconstrained part of the domain is not compromised beyond a the same tolerance $\varepsilon$.
An example we focus on in this work are constraints through boundary conditions for discretized partial differential equations.

To execute this approach, we need to know,
\begin{itemize}
    \item how to choose the penalty strength $\lambda$ so that \cref{eq:mainres-constraint} holds ? \hfill\ovalbox{\cref{subsec:constraint-error-bounds}}
\end{itemize}
As we are primarily interested in constraints that come from boundary conditions in PDEs, we look at 
\begin{itemize}
   \item how to express Dirichlet, Neumann and  Robin boundary conditions via a Hermitian projection, including necessary assumptions to ensure efficient representation and simulation \hfill \ovalbox{\cref{subsec:discretized-pdes-discussion}}
\end{itemize}
Furthermore, our goal is to provide a quantum algorithm; this requires defining an input model for the projection and its simulation as well a full quantum algorithm: 
\begin{itemize}
    \item we provide necessary circuit elements to implement the projection for the types of boundary conditions mentioned above \hfill\ovalbox{\cref{subsec:input-model-for-projections}}
    \item embed the penalty into the simulation algorithm via interaction picture simulation, discuss the complexity and algorithmic overhead when adding constraints via our penalty method  \quad\quad\textcolor{white}{I just need this box to go over to the right} \hfill\ovalbox{\cref{sec:algorithm-and-complexity}}
\end{itemize}
We summarize this in the following two main results of this work:
\begin{itemize}
    \item \cref{result:error-new-kubo}: Error bound that ensures \eqref{eq:mainres-constraint} in the solution error
    \item \cref{result:complexity}: Final complexity when implementing via LCHS, accounting for the evolved-state error in the quantum algorithm
\end{itemize}

\begin{result}[Constraint and solution error]\label{result:error-new-kubo}
Let $\projc$ be an orthogonal projector, $\proj=\identity-\projc$, and $A_0(t)\in\complex^{n^d\times n^d}$ satisfy $\tfrac12(A_0(t)+A_0(t)^\dagger)\preceq\gamma$ for $0\le t\le T$, with $\gamma\ge0$.
Assume the initial condition satisfies the constraint, $\projc v_0=0$.
Consider the dynamics $\ddt v(t)=(A_0(t)-\imag\lambda\projc)v(t)+b(t)$ and the dynamics for $u(t)$ restricted to the feasible region, $\ddt u(t)=\proj A_0(t)\proj u(t)+\proj b(t)$, with $v(0)=u(0)=v_0$.

Then, we have the following bounds on the constraint and solution error, 
\begin{equation}
\begin{alignedat}{3}
  &\text{\rm \scriptsize constraint error:}\quad
  &&\max_{0\le t\le T}\norm{v(t)}_{\ell_2,\domcomp}
  &&\le \frac{E_T}{\lambda},\\
  &\text{\rm \scriptsize total solution error:}\quad
  &&\max_{0\le t\le T}\norm{v(t)-u(t)}_{\ell_2}
  &&\le \frac{(1+Ta_T)E_T}{\lambda}.
\end{alignedat}
\end{equation}
The parameters $a_T, E_T$ are defined in \cref{tab:error-parameters-summary}. 
Therefore, choosing $\lambda\ge (1+Ta_T)E_T/\varepsilon$ suffices for solution error at most $\varepsilon$.
For stable dynamics, we take $\gamma=0$.
The proof follows from \cref{app:proof-error-infeasible-inhomo-dissipative-time,app:proof-error-feasible-time}.
\end{result}
We collect error parameters in \cref{tab:error-parameters-summary}, as we will use these definitions throughout.
\begin{table}[htb]
    \centering
    \caption{Error parameters used for constraint-error bounds; assume sufficient regularity, $A_0, b \in C^1([0,T])$}
    \begin{tabular}{l|l}
    \toprule
    \textbf{Parameter} & \textbf{Definition} \\
    \midrule
        $a_T$ & $\max_{0\le t\le T}\|A_0(t)\|$  \\
        $V_T$ & $\int_0^T\D s\,\|\dd{s}A_0(s)\|$ \\
        $C_T$ & $\|v_0\|+\int_0^T\D s\,\|b(s)\|$ \\
        $K_T$ & $2a_T+V_T+2Ta_T^2$ \\
        $E_T$ & $\e^{\gamma T}\l[ C_TK_T +2\max_{0\le s\le T}\|\projc b(s)\|
                 +\int_0^T\D s\,\left(\left\|\projc\frac{\D b(s)}{\D s}\right\|+a_T\|\projc b(s)\|\right)\r]$ \\     
    \bottomrule
    \end{tabular}
    \label{tab:error-parameters-summary}
\end{table}

The final algorithm we propose follows the Linear Combination of Hamiltonian Simulation \cite{an2023quantum,an2023linear}, carried out in the interaction picture with respect to the constraint projection $\projc$. The necessary circuit primitives beyond LCHS then are Hamiltonian simulations of the projection, $\e^{\imag\lambda t\projc}$. These are provided in \cref{subsec:input-model-for-projections}, whilst the full algorithm can be found in \cref{sec:algorithm-and-complexity}.

\begin{result}[Quantum implementation of constraints (simplified version of~\cref{thm:complexity-linear-constrained-ode})]
\label{result:complexity}
Under the same assumptions as \cref{result:error-new-kubo} with time-independent $A_0, b$, for $0<\varepsilon<1$, and with $\tfrac12(A_0+A_0^\dagger)\preceq 0$, choose
\[
    \lambda=
    \frac{8(1+Ta)^2}{\varepsilon v_{\min}}
    \left(aC_T+\|\projc b\|\right).
\]
Here, $a=\norm{A_0}$, $C_T=\norm{v_0}+T\norm{b}$ and $\norm{u(T)}\ge v_{\min}>0$.

Using the same access model as
\cref{thm:complexity-linear-constrained-ode}, interaction-picture
LCHS solves the penalized equation to normalized-state error
$\varepsilon/2$, giving an $\varepsilon$-approximation of the solution within the feasible region
$u(T)/\|u(T)\|$ with constant success probability, while the constraint error on the infeasible region is at most $\varepsilon$.
This uses
$\widetilde O\!\left(C_T(1+\alpha_{A_0}T)/v_{\min}\right)$
queries to the block-encoding of $A_0$, with block-encoding normalization $\alpha_{A_0}$.

For nonzero forcing, a sufficient number of quadrature points in the time discretization is
\[
\begin{aligned}
    M'
    &\in\widetilde O\!\left(1+T(a+\|b\|+\lambda)\right)\\
    &=\widetilde O\!\left(
        1+T(a+\|b\|)
        +\frac{T(1+Ta)^2
        \left(aC_T+\|\projc b\|\right)}
        {\varepsilon v_{\min}}
    \right).
\end{aligned}
\]
The notation $\widetilde O$ suppresses logarithmic factors;
the corresponding quadrature register uses $O(\log M')$ qubits.
Full implementation bounds, including internal time discretization and time-dependent generators and forcing, are given in \cref{thm:complexity-linear-constrained-ode}.
\end{result}
Thus, the penalty affects the time discretization within each Hamiltonian simulation and, for an inhomogeneous equation, the discretization of the quadrature.
The query complexity depends on $\lambda$ through the solution norm, while implementing the time registers and projection phases also contributes to the gate complexity.
We account for these costs in \cref{sec:algorithm-and-complexity}.

Numerical experiments for the heat and wave equations are given in \cref{sec:numerical-experiments}, where we compare the analytical bounds with constraint errors obtained from numerics.
In particular, we consider the heat equation, which is dissipative, and the wave equation, which is not dissipative and may experience norm-growth, under Dirichlet and Neumann boundary conditions.

\subsection{Notation}
Throughout this work we assume operators as finite-dimensional representations, typically as $n^d$-dimensional complex matrices, where $n$ is the spatial grid number in each coordinate and $d$ is the spatial dimension. Results reported pertaining to partial differential equations thus assume suitable numerical discretization. We often use bracket notation for inner products with a comma for non-normalized (non-quantum states) vectors, $\langle a, b \rangle \equiv a^\dagger b$. 
Whenever we do not specify the type of norm, we use the spectral or operator norm for matrices and $\norm{A}$ and the $\ell_2$-norm for vectors $\norm{v}= \norm{v}_{\ell_2}$.
Bold notation $\bm{j}$ denotes multi-indices so that $\bm{j}\in\mathbb{N}^d$, or quantities after spatial discretization. 
We define stability by a non-positive logarithmic norm of a matrix, i.e., we assume that $A$ satisfies that $\tfrac12(A+A^\dagger)\preceq 0$, or $\mu_2(A)=\lambda_{\max}[(A+A^\dagger)/2]\le 0$. For normal matrices, this is equivalent to spectral stability ($\max_j \Re(\lambda_j)\le 0$). 
More generally, if $\mu_2(A(t))\le\gamma$ with $\gamma\ge0$ for $0\le t\le T$, then $\|\mc T\exp(\int_s^t\D r\,A(r))\|\le\e^{\gamma(t-s)}$ for $0\le s\le t\le T$.
These bounds apply to non-normal matrices as well.
We use square-bracket notation for ordered sets of integers:  $[n,m] = \{n, n+1, \ldots, m\}$ for any $n<m\in \mathds{Z}$. Most commonly, for $n>0$,   $[n] \equiv [1,n] $, $[n]_0 \equiv [0,n]$.
With $\identity$ we denote the identity matrix; if dimensionality is not clear from context, $\identity_N$ acts on $N$ qubits, i.e., on $\complex^{2^N}$.
Some operators/matrices are denoted by sans-serif font, e.g., $\mathsf{P}$ for projections.
The relations $\eqsim, \gtrsim, \lesssim$ are used to denote the relations up to a constant or asymptotically similar/greater/smaller.

\subsection{Problem Setting}
Within this work, we consider the solution of ODEs given constraints, using a quantum algorithm. 
To motivate the setting, let $v(t):[0,T]\to\complex^{n^d}$, and let $\projc$ be the orthogonal projection onto the constraint-inadmissible subspace.
The constraint-admissible subspace is its kernel, with projection $\proj=\identity-\projc$.
We seek a solution $v(t)$ to the constrained ODE such that
\begin{align}\label{eq:abstract-constrained-de}
 \proj\l(\ddt v(t)-A_0(t)v(t)-b(t)\r)&=0,\\
 \projc v(t)&=0.\nonumber
\end{align}
Our nomenclature follows penalty methods in constrained optimization~\cite{luenberger2016penalty}. The scalar penalty $\<v,\,\projc v\>=\|\projc v\|^2$ is non-negative and vanishes precisely when $v$ satisfies the constraint.
We use the associated projection in the evolution equation and analyze convergence to the constrained dynamics as $\lambda$ increases.
Moreover, there is an equivalence between stationary solutions of some dynamical systems and optimization.
There are quantum algorithms that approach optimizatin using dynamics, as in \cite{catli2025exponentially}, and could be extended by the penalty projections in our work to satisfy constraints.

We are particularly interested in settings where the constraint arises from  boundary conditions on spatially discretized PDEs. 
In this case, $\projc$ projects onto the space of violated value or derivative constraints on a subset of grid points. 
Dirichlet boundary conditions are conditions that constrain the solution to follow specific values, whereas Neumann conditions constrain the derivative (within domains in 3D space, typically the surface normal derivative). Linear combination of  Dirichlet and Neumann condition leads to what is typically called Robin boundary conditions. 
\begin{definition}[Computational domain]\label{defn:geometry-index-sets}
    We consider a problem over a $n^d$-dimensional finite space $\complex^{n^d}$. 
    This can be expressed in terms of basis elements $\{\bm{j}\}$ where each label $\bm{j}=(j_1,\ldots,j_d)\in([n-1]_0)^{d}$ is a $d$-dimensional tuple indexing a basis element; there are $n\in\mathbb{N}$ elements per dimension, such as representing each spatial dimension of the problem domain. The following definitions are used throughout the paper to describe the relevant subsets of the problem domain that appear in our work.
    \begin{enumerate}
    \item We first define the index set $\mathcal{I}_\Omega$ making up the unconstrained space, or ``inside the domain'', consisting of all basis elements 
    \begin{equation*}
    \mathcal{I}_\Omega=\{\bm{j}\in[n-1]_0^d \mid \bm{j} \textnormal{ is unconstrained}\}.\end{equation*}
    This corresponds to points inside the domain.
    \item Further, there is the set of points with constrained values
    \[\mathcal{I}_{\Gamma_D} = \{\bm{j}\in[n-1]_0^d \mid \bm{j} \text{ has a constraint on the value}\}\]
    as is the case for a Dirichlet boundary conditions and the set of points with derivative constraints
    \[\mathcal{I}_{\Gamma_N} = \{\bm{j}\in[n-1]_0^d \mid \bm{j}  \text{ has a derivative constraint}   \} \]
    such as on a Neumann boundary. 
    \item Associated to every $ \bm{j}\in\mathcal{I}_{\Gamma_N}$, we have the neighbour set
    \[  \zeta_{\bm{j}} \subseteq \{ \bm{k} \in [n-1]_0^d \mid \norm{\bm{k} - \bm{j}}_1 =1 
    \}.  \]

    \item We call the constrained set, or boundary set, $\mathcal{I}_\Gamma = \mathcal{I}_{\Gamma_D}\sqcup\mathcal{I}_{\Gamma_N}$, 
    \item To satisfy value and derivative constraints, index sets must be disjoint, $\mathcal{I}_{\Gamma_D}\cap\mathcal{I}_{\Gamma_N} = \emptyset, \mathcal{I}_{\Omega}\cap\mathcal{I}_{\Gamma}=\emptyset$. As derivative constraints also involve neighbouring points, we specify compatibility when constructing projections.
    \end{enumerate}
\end{definition}

\begin{figure}
    \centering
    \includegraphics[width=\textwidth]{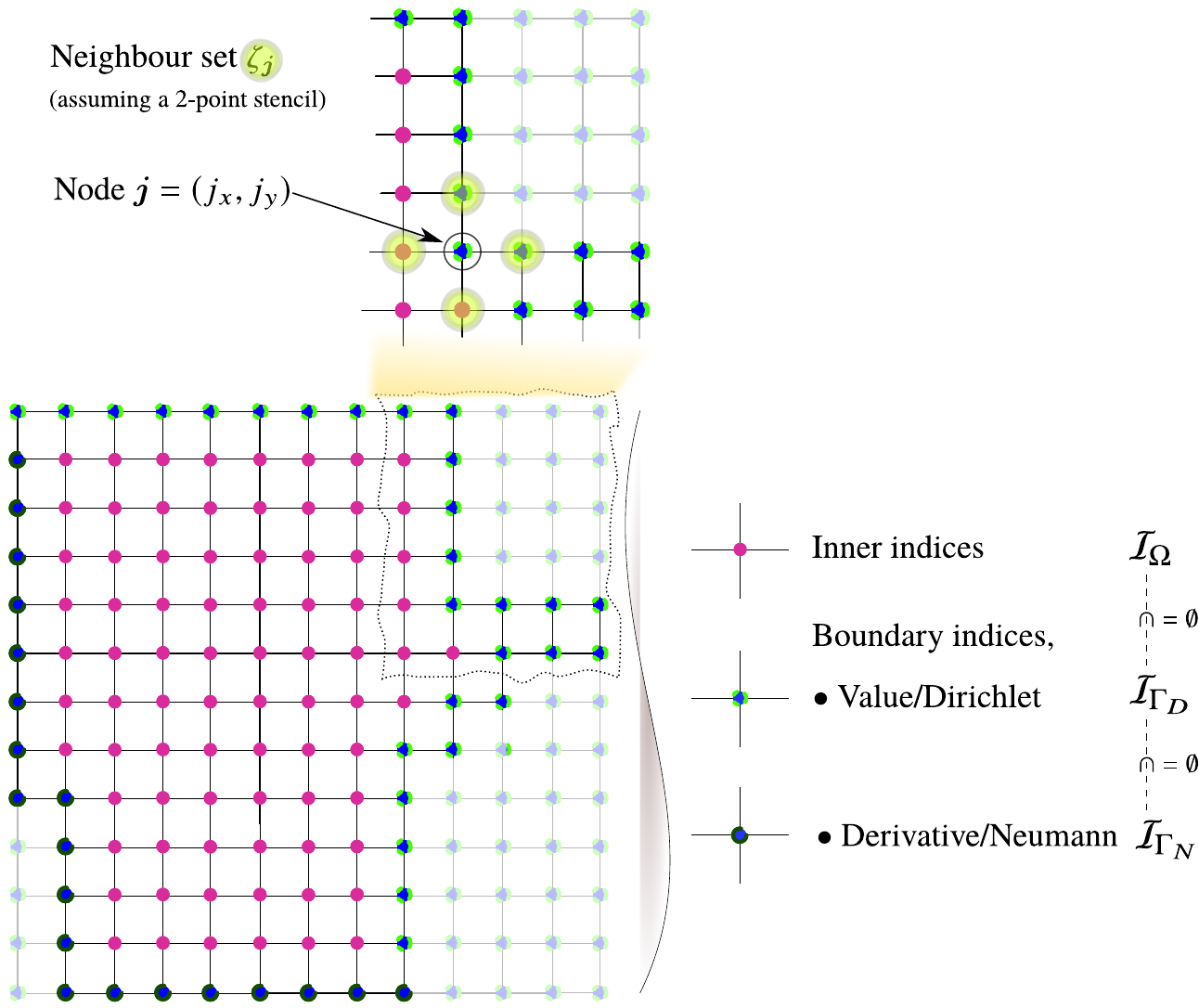}
    \caption{Visualization of relevant index sets as a graph depicted as a uniform grid. Recall that ``inner'' points span the unconstrained space $\dom$. The boundary, i.e.,  the constrained space, is composed by $\domcomp = \domcompD\oplus \domcompN$.
    Note that neighbour sets are excluded from the set of inner indices.}
    \label{fig:index-sets-etc}
\end{figure}

\begin{definition}[Linear space associated to computational domain]\label{defn:geometry-spaces}
     Given constraints following \cref{defn:geometry-index-sets}, call the space of vectors satisfying the homogeneous constraints the feasible or unconstrained space $\dom$.
     Its orthogonal complement is the constrained space $\domcomp$.
     For Dirichlet conditions, $\domcompD=\mathrm{span}\{\ket{\bm j}\mid\bm j\in\mathcal I_{\Gamma_D}\}$.
     For derivative conditions, $\domcompN$ is spanned by the vectors defining the finite differences; for example, a two-point difference has constraint vector $(\ket{\bm j}-\ket{\bm k})/\sqrt2$.
     When $\domcompD\perp\domcompN$, we have $\domcomp=\domcompD\oplus\domcompN$.
     The solutions live in $\overline S=\dom\oplus\domcomp=\complex^{n^d}$.
\end{definition}
We further define the following projections. 
\begin{definition}[Projection on constrained space]\label{defn:infeasible-projector}
Consider a feasible (unconstrained) space $\dom$ and infeasible (constrained) spaces $\domcomp, \domcompD, \domcompN$ as defined in \cref{defn:geometry-index-sets}.
Then, for $v\in \overline{S}$, we have the following projections, 
\begin{align}
   \quad \projc v &=\begin{cases} v, & v \in \domcomp \\ 0, & v \in \dom \end{cases},   \\
  \proj &= \identity - \projc,
\end{align}
so that $\projc$ projects onto the constrained space $S_c$ and $\proj$ to its complement $S$. Here, $\identity$ is the identity on $\overline{S}$.
We can further decompose $\projc = \projcD + \projcN$ so that $ (\domcompD)^\perp = S\oplus\domcompN = \ker(\projcD)$ and $(\domcompN)^\perp = S\oplus\domcompD = \ker(\projcN)$. 
\end{definition}
{\begin{definition}[$\ell_2$-norms over computational spaces]\label{defn:norms-for-infeasible-spaces}
    Let $a\in \overline{\dom}$, then  
    \begin{align}\label{eq:projc-error-defn-1}
        \lVert a \rVert_{\ell_2,\domcomp} = \lVert \projc a \rVert _{\ell_2}
        = \sqrt{\left< a, \projc a\right>}  
    \end{align}
    Similarly, $ \lVert a \rVert_{\ell_2,\dom} =   \lVert \proj a \rVert_{\ell_2} $.
\end{definition}}
Next, we define the problem setting we generally consider in this work. 
{
\begin{prob}[Constrained Discrete Initial Value Problem]\label{prob:ivp-bvp-defn}
We consider a finite-dimensional initial value problem with solution vector  $v(t):  [0,T]  \to \overline{S}$ and a matrix $A(t):[0,T]\to \overline{S}\times\overline{S}$ as dynamical generator.
We seek approximate solutions to the constrained dynamics
\begin{align}\label{eq:ivp-bvp-defn}
 \proj\l(\ddt v(t)-A(t)v(t)-b(t)\r)&=0,\quad t\ge0,\\
 \projcD v(t)&=g(t),\quad t\ge0,\nn
 \projcN v(t)&=h(t),\quad t\ge0,\label{eq:dynamics-defn-neumann}\\
 v(0)&=v_0,\quad\projcD v_0=g(0),\quad\projcN v_0=h(0).\nonumber
\end{align}
We require $\projcD\projcN=0$, with $g(t)\in\domcompD$ and $h(t)\in\domcompN$.
The initial data specify the full vector $v_0$ and agree with the boundary values at $t=0$.
Projected representation of Dirichlet and Neumann boundary values are given by $g(t)$ and $h(t)$, respectively.
\end{prob}
}
\begin{remark}[Projection matrices as constraint functions are penalty functions]
    Following \cite{luenberger2016penalty}, the penalty function associated with $\projc$ is $\<v,\,\projc v\>=\|\projc v\|^2$.
    It is non-negative and vanishes exactly on the feasible space in \cref{defn:infeasible-projector}.
\end{remark}
\begin{remark}
    For the derivative constraint in~\cref{eq:dynamics-defn-neumann}, $\projcN$ already embodies the notion of a finite difference approximation of a derivative constraint.
    Given a proper construction of $\projcD$ and $\projcN$, we can treat Dirichlet and Neumann conditions in a similar manner.
    Later on, we use orthogonal projections satisfying $\projcD\projcN=0$ for efficient simulation.
    A sufficient condition is that the value constraints do not involve any point used by a derivative constraint,
    $\mathcal I_{\Gamma_D}\cap(\mathcal I_{\Gamma_N}\cup\bigcup_{\bm j\in\mathcal I_{\Gamma_N}}\zeta_{\bm j})=\emptyset$.
    Then, the projections have disjoint support.
\end{remark}

\section{Approximating Boundary Conditions with a Projection}
\begin{figure}[ht]
    \centering
    \includegraphics[width=\textwidth]{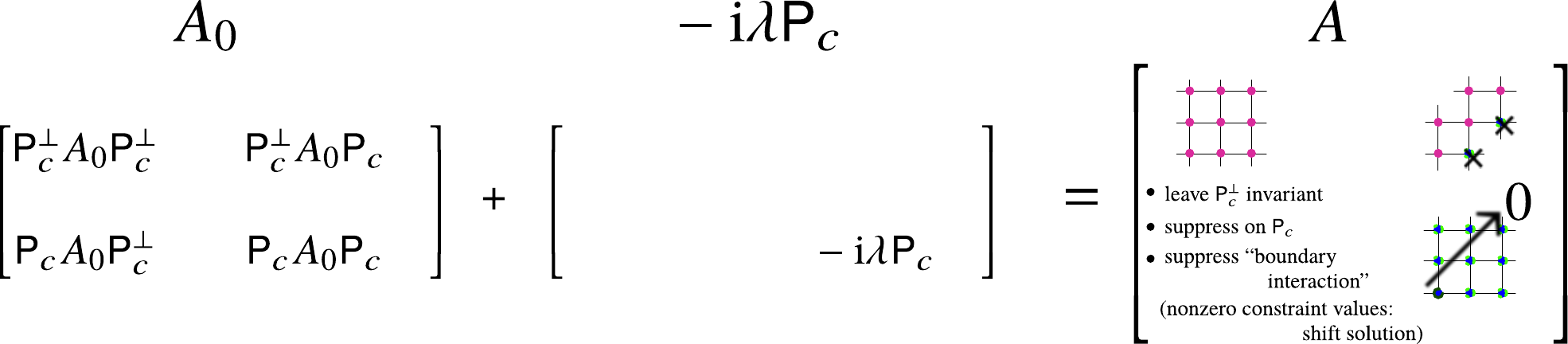}
    \caption{Inducing a constraint subspace via a penalty projection $-\imag\lambda\projc$ with $\lambda>0$ and $\projc$ the projection onto the constraint subspace. 
    For initial values supported on the $\proj$-subspace, evolution under $A=A_0-\imag\lambda\projc$ approaches the dynamics constrained to the feasible $\proj$-subspace for increasing $\lambda$.
    }
    \label{fig:perturbed-generator}
\end{figure}

\subsection{Motivation}
Now, we discuss a method to efficiently implement the solution of such boundary problems on a quantum computer.
To that end, we take inspiration from the so-called complex absorbing potentials in quantum physics and chemistry~\cite{muga2004complex}. 
Note that for the case when \cref{eq:ivp-bvp-defn} has $A = -\imag H$  anti-Hermitian and operator and a normalized initial state, the dynamics of $\psi(t,x)$ follows the Schr\"odinger equation, i.e., 
\begin{equation}\label{eq:tdep-schreq}
    \imag \pdv{t} \psi(t,x)  = H {\psi(t,x)}. 
\end{equation}
Then, a modification of the Hamiltonian with $\lambda>0 $ 
\begin{equation}\label{eq:modified-hamiltonian}
   H \to H - \imag \lambda \projc  
\end{equation}
leads to enforcing that $\psi(t,x')\to 0$ for all $x'\in\Gamma$ on the boundary in the limit of $\lambda\to\infty$ if $
\projc = \int_{\Gamma} \D\ketbra{x'}$. In particular, we want $\lambda \gg \norm{H}$.
This has the effect that in the solution
\begin{equation}
    \psi(t,x) = \e^{-\imag (H - \imag\lambda \projc )t   }   \psi(0, x),
\end{equation}
components on $\Gamma$ are damped.

In what follows, we will show that the modification in \cref{eq:modified-hamiltonian} for boundary conditions over more general, i.e. not necessarily quantum, dynamics can work in the same way. 
In this more general setting, we consider the following modified dynamics governed by a (potentially time-dependent) system matrix $A_0(t)$ and a forcing term $b(t)$, 
\begin{equation}\label{eq:modified-dynamics}
    \ddt  v(t) = (A_0(t) - \imag \lambda \projc)  v(t) + b(t).
\end{equation}
Compared to the case of complex absorbing potentials in \cref{eq:modified-hamiltonian}, $A_0(t)$ need not be Hermitian. As we will be working with orthogonal projections $\projc,\proj$ exclusively, this implies their Hermiticity; a more careful analysis could relax the assumptions to non-Hermitian penalty functions in future work. Additionally, a time-dependent constraint region can be thought of via a $\projc(t)$. 

We make the following observation here. 
In the case of quantum dynamics, \cref{eq:tdep-schreq,eq:modified-hamiltonian}, a perturbation of the form $\imag\lambda$ as in complex absorbing potentials ``dissipates'' the wavefunction on the boundaries. 
Our approach in \cref{eq:modified-dynamics} with positive $\lambda$ does not add dissipation per se.
Rather, we can understand this e.g. under the realm of the rotating-wave approximation, where the evolution under a Hamiltonian $H=H_0+H_1$ shows a separation of time-scales with respect to dynamics generated by $H_0$ and $H_1$, respectively. Then, one can say that the solution coming from the perturbative generator is negligible with respect to the initial dynamics in the sense that the on the unperturbed system's time-scale, the highly oscillatory perturbation averages to zero. In other words, the solution on the infeasible region then is perceived negligible compared to the solution on the feasible region.

Our analysis follows perturbation theoretic arguments, with the penalty strength $\lambda$ controlling transitions between the feasible and infeasible spaces.
A natural way to estimate the error is to consider `transition elements' with respect to the projections $\proj,\projc$, as expectations $\<v(t'),\,P v(t)\>$.
For $t=t'$, the Kubo formula~\cite{kubo1957statistical} provides a pathway to estimate this error.
We are interested in a non-Hermitian formulation, which was developed in \cite{sticlet2022kubo} and further discussed in \cite{geier2022non}.
Building on this, we provide a generalization that also covers time-dependent generators for the unperturbed dynamics, and a forcing term.

Before stating the result, we introduce some context and notation.
We consider a dynamical system $\ddt{v_\zeta(t)}=A(t)v_\zeta(t)+b(t)$, where $A(t)=H(t)+\zeta V(t)$ and $\zeta\in\reals$ is a perturbation parameter.
The matrices $H(t),V(t)$, the initial condition $v_0$, and the forcing $b(t)$ are independent of $\zeta$.
The Kubo formula describes the difference between the expectation of an observable $P$ with respect to a perturbed and an unperturbed solution vector.
We use the following quantities:
\begin{itemize}
    \item We denote the propagator of $H(t)$ by $G_H(t,s)=\mc T\exp(\int_s^t\D r\,H(r))$.
    The unperturbed solution is $v_H(t)=G_H(t,0)v_0+\int_0^t\D s\,G_H(t,s)b(s)$.
    \item We use $\sigma(t',t)$ for the outer product of the unperturbed solutions, $\sigma(t',t)=v_H(t')v_H(t)^\dagger$.
    This notion is similar to the density matrix of pure quantum states, however with non-normalized complex vectors $v_H(t)$.
\end{itemize}

\begin{proposition}[A non-Hermitian, time-dependent, inhomogeneous Kubo formula]\label{prop:generalized-kubo}
Let $A(t)=H(t)+\zeta V(t)$, and let $v_\zeta(t)$ solve $\ddt{v_\zeta(t)}=A(t)v_\zeta(t)+b(t)$ with initial condition $v_\zeta(0)=v_0$.
For a matrix $P$, write
\[
 \<P(t)\>_\zeta=\frac{\<v_\zeta(t),\,P v_\zeta(t)\>}{\<v_\zeta(t),\,v_\zeta(t)\>}.
\]
Suppose $v_{\min}=\min_{0\le t\le T}\|v_H(t)\|>0$.
For sufficiently small $|\zeta|$, the response to the perturbation is
\begin{equation}
    \<P(t)\>_\zeta-\<P(t)\>_0
    =\zeta\int_0^t\D s\,
    \frac{\tr[\{P-\<P(t)\>_0\identity,\,G_H(t,s)V(s)\sigma(s,t)\}_\sim]}
         {\tr[\sigma(t,t)]}
    +O(\zeta^2),
\end{equation}
with the error uniform for $0\le t\le T$.
Here, $\{X,Y\}_\sim=XY+Y^\dagger X$ is the modified anticommutator, and $\sigma(s,t)=v_H(s)v_H(t)^\dagger$ includes the forcing through $v_H$.
The exact response is
\begin{equation}\label{eq:kubo-exact-response-main}
 \<P(t)\>_\zeta-\<P(t)\>_0
 =\int_0^\zeta\D\eta\,\int_0^t\D s\,\chi_\eta(t,s),
\end{equation}
where $\chi_\eta$ is the same kernel evaluated using the solution and propagator of $H+\eta V$.
\end{proposition}
For completeness, we briefly restate the original Kubo formula, as derived in \cite{kubo1957statistical}, with adapted notation.
\begin{remark}[Original Kubo formula]
We are given Hermitian $H,V(t)$ and an observable $P=P^\dagger$, with dynamics $\imag\ddt{\psi_\zeta(t)}=(H+\zeta V(t))\psi_\zeta(t)$ and a normalized initial state.
    Writing $P_I(t)=\e^{\imag Ht}P\e^{-\imag Ht}$ and $V_I(s)=\e^{\imag Hs}V(s)\e^{-\imag Hs}$, the first-order response is
    \begin{equation}
        \<P\>_{t,\zeta}-\<P\>_{t,0}
        =-\imag\zeta\int_0^t\D s\,\<[P_I(t),V_I(s)]\>_0+O(\zeta^2),
    \end{equation}
    where $\<\cdot\>_0$ denotes expectation in the initial state.
    Non-Hermitian dynamics require the normalization correction discussed in \cite{sticlet2022kubo,geier2022non} and in \cref{prop:generalized-kubo}.
    We account for time dependence through $G_H(t,s)$ and include the forcing in $\sigma(s,t)$.
\end{remark}
The proof of \cref{prop:generalized-kubo} is given in \cref{app:proof-of-gen-kubo}.
In the next section, we apply the exact Kubo formulation to the constraint error and consider the resulting bounds for time-independent and time-dependent dynamics, with and without forcing.

\subsection{Approximation Guarantees}\label{subsec:constraint-error-bounds}
In the upcoming section, we provide error bounds on the approximation of constraints such as given in \cref{prob:ivp-bvp-defn}. We will structure this argumentation as follows. First, we consider the case of only one type of constraint, i.e., we do not split $\domcomp$ into $\domcompD$ and $\domcompN$ using only the projective properties of $\projc$. 
This is sufficient, as we will show how to construct the projections  so that the constraints on the different boundary strips will not affect each other, and $ [\projcD, \projcN]=0$, aligning with our assumption in \cref{defn:geometry-spaces} that $\domcompD\perp\domcompN$. Once we have derived guarantees for general constraints, we then proceed to connect this to the case of Dirichlet and Neumann boundary conditions in discretized PDEs in \cref{subsec:discretized-pdes-discussion}.

\begin{prob}[Projection-Constrained Discrete Initial Value Problem]\label{prob:proj-ivp-bvp-defn}
Under the same assumptions as in \cref{prob:ivp-bvp-defn}, we define
\begin{equation}\label{eq:proj-ivp-bvp-defn-proj}
   \ddt v(t) = (A_0(t) -\imag \lambda \projc) v(t)  + b(t) , \quad t\ge 0,
\end{equation}
where $\projc$ is a Hermitian projection and $\lambda>0$ is a penalty term that needs to be chosen.
Then, given $\varepsilon>0$, we seek $\lambda$ such that
\begin{equation}\label{eq:constraint-error-general}
   \lVert v(t)  \rVert_{\ell_2,\domcomp} \le\varepsilon,
\end{equation}
where $v(t)$ is understood to satisfy a constraint with respect to a projection $\projc$ if $\projc v(t) = 0$ (\cref{defn:infeasible-projector}).
\end{prob}
To verify that \cref{prob:proj-ivp-bvp-defn} using the modified dynamics shows the sought-after behaviour, we need to show the following:
\begin{enumerate}
    \item\label{item:conditions-for-soln} There is $\lambda>0$ so that the approximation error on the boundary quantified by \cref{eq:projc-error-defn-1} is small.
    \item\label{item:one-requirements-for-conditions} The error in the solution within the feasible (unconstrained) space $S$ due to adding the constraint projection is small.
\end{enumerate}
The relevant lemmas are summarized in \cref{tab:lemma-overview-error-bounds}. They generally  assume that $A_0$ is stable, i.e., its real part is negative semi-definite, $\tfrac12(A_0+A_0^\dagger)\preceq 0$. 
\begin{table}[t!]
    \centering
    \caption{Overview over lemmas for approximation guarantees }
    \label{tab:lemma-overview-error-bounds}
    \begin{tabular}{l l l}
    \toprule
        & error bound on feasible space $\dom$  & error bound on infeasible space $\domcomp$ \\
    \midrule
      $A$ constant, $b=0$  &  \cref{lem:error-feasible}& \cref{lem:error-infeasible-dissipative}  \\
          $A$ constant, $b(t)\neq 0$  & & \cref{lem:error-infeasible-dissipative-inhomo}  \\
      $A(t)$, $b=0$   & & \cref{lem:error-infeasible-dissipative-time} \\ 
      $A(t)$, $b(t)\neq0$   & \cref{lem:error-feasible-time} & \cref{lem:error-infeasible-inhomo-dissipative-time}\\
      \bottomrule
    \end{tabular}
\end{table}
Additionally, we consider generators satisfying $\tfrac12(A_0(t)+A_0(t)^\dagger)\preceq \gamma$ for $\gamma\ge 0$ and any $0\le t \le T$. In this case, error bounds incur an additional factor $\e^{\gamma T}$. See \cref{lem:error-infeasible-nondissipative}.

We assume that the initial data satisfies the constraints. For the quantum implementation, this can be guaranteed by the approach outlined in \cref{lem:ensuring-satisfaction}. We use the parameters $a_T, V_T, C_T, K_T, E_T$ as defined in \cref{result:error-new-kubo}, with $\gamma=0$ in the dissipative case.

We begin by showing that the following holds true under the problem setup in \cref{prob:proj-ivp-bvp-defn} for stable dynamics with a time-independent generator and then move on to the time-dependent case.

\subsubsection{Constant dynamics}
The following lemma establishes that the error in the constrained region is bounded.
\begin{lemma}[Error in the infeasible space under stable  dynamics]\label{lem:error-infeasible-dissipative}
We consider a system as in \cref{prob:ivp-bvp-defn} with $b=0$ and dynamics generated by $A=A_0-\imag\lambda\projc$, where $\tfrac12(A_0+A_0^\dagger)\preceq0$, $\lambda>0$, and $\projc$ is the projector on the infeasible space as defined in \cref{defn:infeasible-projector}.
Further, we require that the initial data satisfies $\projc v(0)=0$.
Then, for every $t\ge0$,
\begin{equation}
 \|v(t)\|_{\ell_2,\domcomp}\le\|v(0)\|\sqrt{\frac{2\|A_0\|}{\lambda}}.
\end{equation}
In particular, $\lambda\ge2\|A_0\|\|v(0)\|^2/\varepsilon^2$ suffices for constraint error at most $\varepsilon$.
\end{lemma}
\noindent
The proof for \cref{lem:error-infeasible-dissipative} can be found in \cref{app:proof-error-infeasible-dissipative}.

Additionally, we extend the result to the case when there is an inhomogeneity $b(t)$.
\begin{lemma}[Error in the infeasible space under dissipative dynamics with an inhomogeneity]\label{lem:error-infeasible-dissipative-inhomo}
We consider a system as in \cref{prob:ivp-bvp-defn} with dynamics generated by $A=A_0-\imag\lambda\projc$, where $\tfrac12(A_0+A_0^\dagger)\preceq0$, $\lambda>0$, and $\projc v(0)=0$.
We allow a time-dependent inhomogeneity $b(t)$ on $[0,T]$.
Then,
\begin{equation}
 \max_{0\le t\le T}\|v(t)\|_{\ell_2,\domcomp}\le\frac{E_T}{\lambda},
\end{equation}
where $a_T=\|A_0\|$ and $V_T=0$ in the definitions of $K_T,E_T$ in \cref{result:error-new-kubo}.
Thus, $\lambda\ge E_T/\varepsilon$ suffices for constraint error at most $\varepsilon$.
\end{lemma}
The proof for \cref{lem:error-infeasible-dissipative-inhomo} can be found in \cref{app:proof-error-infeasible-dissipative-inhomo}.

Furthermore, we want to show that the perturbation does not impact the solution in the feasible region to a large extent. Here, the metric of error is the difference between the constrained solution, propagated by $\exp((A_0-\imag\lambda\projc)t)$, vs. the solution where the generator is restricted to the feasible space, $\exp(\proj A_0 \proj t)$.
\begin{lemma}[Error within the feasible space]\label{lem:error-feasible}
Consider the system from \cref{prob:ivp-bvp-defn}, with $b=0$, $A=A_0-\imag\lambda\projc$, and $\tfrac12(A_0+A_0^\dagger)\preceq0$.
For any initial condition $v_0$ with $\proj v_0=v_0$ and any $0\le t\le T$,
\begin{equation}
 \|\proj(\e^{\proj A_0\proj t}-\e^{(A_0-\imag\lambda\projc)t})v_0\|
 \le T\|A_0\|\|v_0\|\sqrt{\frac{2\|A_0\|}{\lambda}}.
\end{equation}
In particular, $\lambda\ge2T^2\|A_0\|^3\|v_0\|^2/\varepsilon^2$ suffices for error at most $\varepsilon$.
\end{lemma}
\noindent
The proof for \cref{lem:error-feasible} can be found in \cref{app:proof-error-feasible}.

We additionally study the case of $A_0$ that are not stable.
In particular, consider $A_0$ with $\tfrac12(A_0+A_0^\dagger)\preceq \gamma$ for positive $\gamma\ge0$. This assumptions bounds the growth of the solution as follows.
\begin{lemma}[Error in the infeasible space under non-stable dynamics]\label{lem:error-infeasible-nondissipative}
Under the same assumptions as \cref{lem:error-infeasible-dissipative}, except that $\tfrac12(A_0+A_0^\dagger)\preceq\gamma$ for some $\gamma\ge0$, we have
\begin{equation}
 \max_{0\le t\le T}\|v(t)\|_{\ell_2,\domcomp}\le\e^{\gamma T}\|v(0)\|\sqrt{\frac{2\|A_0\|}{\lambda}}.
\end{equation}
Thus, $\lambda\ge2\e^{2\gamma T}\|A_0\|\|v(0)\|^2/\varepsilon^2$ suffices for constraint error at most $\varepsilon$.
\end{lemma}
\begin{proof}
We have $\|\e^{At}\|\le\e^{\gamma t}$.
Since $Av(0)=A_0v(0)$ and $A$ commutes with $\e^{At}$,
\[
 |\<v(t),\,Av(t)\>|\le\|\e^{At}v(0)\|\,\|\e^{At}A_0v(0)\|
 \le\e^{2\gamma t}\|A_0\|\|v(0)\|^2.
\]
The same upper bound holds for $|\<v(t),\,A_0v(t)\>|$.
The reverse triangle inequality in \cref{eq:bound-on-first-error-lemma} then implies the claim.
\end{proof}

\subsubsection{Time-dependent dynamics}
Next, we look into enforcing constraints under time-dependent dynamics. 
Here, we use the exact response formula in \cref{prop:generalized-kubo} to compare dynamicswith different couplings between the feasible and the infeasible space. We first consider $b=0$ in \cref{lem:error-infeasible-dissipative-time} and include a time-dependent inhomogeneity in \cref{app:proof-error-infeasible-dissipative-inhomo}.
\begin{lemma}[Error in the infeasible space under stable dynamics with a time-dependent generator]\label{lem:error-infeasible-dissipative-time}
We use the same assumptions as in \cref{lem:error-infeasible-dissipative}, with a time-dependent generator $A(t)=A_0(t)-\imag\lambda\projc$ and $\tfrac12(A_0(t)+A_0(t)^\dagger)\preceq0$ on $0\le t\le T$.
The solution is
\begin{equation}
 v(t)=\mathcal T\exp\!\left(\int_0^t\D s\,A(s)\right)v(0).
\end{equation}
Then,
\begin{equation}
 \max_{0\le t\le T}\|v(t)\|_{\ell_2,\domcomp}\le\frac{\|v(0)\|K_T}{\lambda}.
\end{equation}
Thus, $\lambda\ge\|v(0)\|K_T/\varepsilon$ suffices for constraint error at most $\varepsilon$.
\end{lemma}
\noindent
The proof to \cref{lem:error-infeasible-dissipative-time} can be found in \cref{app:proof-error-infeasible-dissipative-time}. 

We point out that the framework of Kubo's formula would also allow us to look at time-dependent strengths of the projection, which may allow to make the overall choice more optimal, and also time-dependent forms of the constraint projection $\projc$. This is left up to future research.

\begin{lemma}[Error in the infeasible space under stable dynamics with a time-dependent generator and inhomogeneity with a generalized non-Hermitian, inhomogeneous Kubo formula]\label{lem:error-infeasible-inhomo-dissipative-time}
Under the same assumptions as in \cref{lem:error-infeasible-dissipative-time}, and allow a time-dependent inhomogeneity $b(t)$ in $\ddt{v(t)}=A(t)v(t)+b(t)$.
The general solution is
\begin{equation}\label{eq:solution-general-inhomo-for-lemma}
 v(t)=\mathcal T\exp\!\left(\int_0^t\D s\,A(s)\right)v(0)
 +\int_0^t\D s\,\mathcal T\exp\!\left(\int_s^t\D r\,A(r)\right)b(s).
\end{equation}
Then,
\begin{equation}
 \max_{0\le t\le T}\|v(t)\|_{\ell_2,\domcomp}\le\frac{E_T}{\lambda}.
\end{equation}
Thus, $\lambda\ge E_T/\varepsilon$ suffices for constraint error at most $\varepsilon$.
\end{lemma}
\noindent
The proof to \cref{lem:error-infeasible-inhomo-dissipative-time} can be found in \cref{app:proof-error-infeasible-inhomo-dissipative-time}.

We also bound the error within the feasible region for time-dependent dynamics with an inhomogeneity.
\begin{lemma}[Error within the feasible space with time-dependent generator]\label{lem:error-feasible-time}
Using the same assumptions as \cref{lem:error-infeasible-inhomo-dissipative-time}, let $v_{\perp,\rm h}$ and $v_{\perp,\rm p}$ be the homogeneous and particular solutions restricted to the feasible space,
\begin{align*}
 v_{\perp,\rm h}(t)&=\mathcal T\exp\!\left(\int_0^t\D s\,\proj A_0(s)\proj\right)v_0,\\
 v_{\perp,\rm p}(t)&=\int_0^t\D s\,\mathcal T\exp\!\left(\int_s^t\D r\,\proj A_0(r)\proj\right)\proj b(s).
\end{align*}
Then,
\begin{equation}
 \max_{0\le t\le T}\|\proj v(t)-v_{\perp,\rm h}(t)-v_{\perp,\rm p}(t)\|
 \le\frac{Ta_TE_T}{\lambda}.
\end{equation}
In particular, $\lambda\ge Ta_TE_T/\varepsilon$ suffices for error in the feasible region at most $\varepsilon$.
\end{lemma}
\noindent
The proof for \cref{lem:error-feasible-time} can be found in \cref{app:proof-error-feasible-time}.

\subsection{Application to Discretized Partial Differential Equations}\label{subsec:discretized-pdes-discussion}
In this work, we are mainly interested in constraints coming from boundary conditions in the discretization of linear, evolutionary partial differential equations. 
Namely, for $\mathcal{L}$ a linear differential operator, then for $t\in\reals_+$ and $ r\in\Omega\subseteq\reals^{d}$, and $v(r,t)$ comes from a function space so that the PDE with respect to $\mathcal{L}$ is well-posed,
\begin{align}
       \partial_t {v}(r,t) - \mathcal{L}v(r,t)&= f(r,t), &r\in \Omega \nn
        v(r,t) &= g(r,t),  &r\in \Gamma_D \label{eq:application-to-pdes-setup}\\
        \<\partial_n, \,v(r,t)\> &= h(r,t) ,   &r\in \Gamma_N\nn
        v(r,0) &= v_0(r).  & 
\end{align}
The solution lives on a domain $\Omega$ and constrained to a specific value on the Dirichlet boundary $\Gamma_D$ or a normal derivative on the Neumann boundary $\Gamma_N$. 
Robin boundary conditions, which specify a linear combination of derivative and value at the boundary can be obtained as a corollary, see~\cref{subsec:robin-conditions}.

In the context of \cref{prob:proj-ivp-bvp-defn}, we have that on a discretized version of $v(r,t)$,  $\projcD$ projects onto the domain $\Gamma_D$, $\projcN$ projects onto differences along the surface normal on $\Gamma_N$ (for details see \cref{subsubsec:neumann-construction}) and the differential operator $\mc L$ becomes $A_0(t)$. There are no further assumptions beyond linearity on $\mc L$ so far, though most of our error bounds in the previous section require that, for the discretized operator, $\tfrac12(A_0+A_0^\dagger)\preceq 0$.

The methodology we developed so far is able to enforce $g_c(t) = 0$ and $h_c(t)=0$, assuming access to boundary projections $\projcD, \projcN$. 
Thus, we next consider how to deal with non-zero values as this is an essential aspect of the utility of our method.

\subsubsection{Non-zero Dirichlet boundaries}\label{subsec:nonzero-dirichlet}
We start by considering the problem in \cref{eq:application-to-pdes-setup}, assuming the condition on $\Gamma_D$.
Let us further assume for simplicity that $\overline{\Omega} = [0,L]^d$. We will construct a function $w(r,t)$ that satisfies the constraint on $\Gamma_D$ in \eqref{eq:application-to-pdes-setup}. For the one-dimensional case, $d=1$, this can be found quite easily:
\begin{equation}
    w(r, t) = \frac{r}{L}g(L,t) + \left(1-\frac{r}{L} \right)g(0,t).
\end{equation}
For $d\ge1$, we interpolate between opposite faces and subtract the values counted more than once on their intersections. In the case of $d=2$, this would apply to the corner points.
Let $\bar w_0(s)=1-s/L$ and $\bar w_1(s)=s/L$, and let $r^{\mc I,\sigma}$ denote $r$ with $r_i$ replaced by $\sigma_i L$ for each $i\in \mc I$. Then
\begin{equation}
w(r,t)= \sum_{\emptyset\neq\mc I\subseteq[d]} (-1)^{|\mc I|+1} \sum_{\sigma\in\{0,1\}^{\mc I}} \left(\prod_{i\in \mc I}\bar w_{\sigma_i}(r_i)\right) g(r^{\mc I,\sigma},t).
\end{equation}
Constraint values on intersections of different faces are required to be compatible.
If $\Gamma_D$ contains only a part of the boundary, we first extend $g$ to the remaining faces with compatible values on their intersections.
Now, let us consider a new function 
\begin{equation}
    v'(r,t) = v(r,t) - w(r,t),
\end{equation}
and compute the following,
\begin{align}
        \partial_t v'(r,t) -\mathcal{L}v'(r,t) &= \partial_t v(r,t) - \mathcal{L}v(r,t)  - \partial_t w(r,t) +\mathcal{L}w(r,t) 
        \nn
        &= f(r,t) +  \mathcal{L}w(r,t) - \partial_t w(r,t)\\
        &=: \widetilde{f}(r,t)
\end{align}
We obtain $\partial_t w$ using the same construction as above by replacing $g$ with $\partial_t g$. 
The initial condition becomes
\begin{equation}
    v'(r,0) = v(r,0) - w(r,0)
    = v_0(r) - w(r,0)
    =: v_0'(r)
\end{equation}
and the boundary condition
\begin{equation}
    v'(r,t)\big|_{r\in\Gamma_D} = g(r,t) - \underbrace{w(r,t)\big|_{x \in \Gamma_D}}_{=g(r,t)}
    = 0.
\end{equation}
Therefore, the differential equation for $v'$ is given by
\begin{align}
        \partial_t v' (r,t) - \mathcal{L}(t)v'(r,t) &= \widetilde{f}(r,t), \quad r \in \Omega,  t > 0\nn
        v'(r,t) &= 0 \quad r \in \Gamma_D,  t\ge 0,\nn
        v'(r,0) &= v'_0(r) \quad r  \in \overline{\Omega},  t =0.
    \label{eq:homDBVP}
\end{align}
That means that the PDE in $v'(r,t)$ now has Dirichlet boundary conditions of value zero.
The solution $v(r,t)$ to the original problem can then be obtained by
\begin{equation}
    v(r,t) = v'(r,t) + w(r,t).
\end{equation}
Therefore we obtain the following corollary to \cref{result:error-new-kubo} for a suitable discretization with error at most $O(\varepsilon)$, where the differential operator $\mathcal{L}$ has the discrete representation $A_0$.

\begin{corollary}[Error in the infeasible space under stable dynamics with non-zero Dirichlet Boundary Conditions.]\label{cor:error-infeasible-nonzero-dirichlet}
We have the same assumptions as in \cref{result:error-new-kubo} and consider non-zero value constraints
\begin{align}\label{eq:nonzero-dirichlet-conditions}
\begin{split}
    \proj\ddt{v(t)} &= \proj\big(A_0(t)v(t)+b(t)\big), \\
    \projc v(t) &= g_c(t)\in\domcomp,\quad t\geq0.
\end{split}
\end{align}
Let $v'(t)=v(t)-g_c(t)$ and $v'_0=v_0-g_c(0)$, where $\projc v_0=g_c(0)$.
Then $\projc v'(t)=0$ and, with $b'(t)=A_0(t)g_c(t)-\dot g_c(t)+b(t)$, we obtain the equivalent system
\begin{align}\label{eq:shifted-dirichlet-problem}
\begin{split}
    \ddt{v'(t)} &= \proj A_0(t)\proj v'(t)+\proj b'(t), \\
    \projc v'(t) &= 0,\quad v'(0)=v'_0.
\end{split}
\end{align}
We approximate this system with the penalized equation
\begin{equation}
    \ddt{v'(t)}=\big(A_0(t)-\imag\lambda\projc\big)v'(t)+b'(t),\quad v'(0)=v'_0.
\end{equation}
We choose $\lambda$ according to \cref{result:error-new-kubo} with initial value $v'_0$ and source $b'$.
Then the reconstructed solution $v(t)=v'(t)+g_c(t)$ has boundary error
\begin{equation}
    \norm{\projc v(t)-g_c(t)}=\norm{\projc v'(t)}\leq\varepsilon,
\end{equation}
and the solution error follows \cref{result:error-new-kubo}.
\end{corollary}

Assuming a suitable discretization technique~\cite{ames2014classicalpdesbook1,evans2012classicalpdesbook2} using a finite-dimensional basis of $n$ basis functions per dimension, 
we obtain a semi-continuous solution vector $v(t) \in \reals^{n^d}$. In alignment with  \cref{defn:geometry-index-sets,defn:geometry-spaces} we call the discrete domain $\dom$, the value boundary $\domcompD$ and the derivative boundary $\domcompN$. 
Then, we retrieve the formulation readily posed in \cref{prob:proj-ivp-bvp-defn}.

\subsubsection{Neumann boundary conditions}\label{subsubsec:neumann-construction}
Within this work, we will not worry about the approximation error with respect to the continuum limit and assume that we are provided with an ODE that follows from a sensible discretization, achieving a target discretization error $\varepsilon>0$.
Thus, we will continue only with bounding the error  in representing a difference formula, denoted by $\mathsf{D}$, using our projection method.
The difference formula uses the two-point neighbour pairs defined by $\zeta_{\bm j}$ in \cref{defn:geometry-index-sets}. We implement the corresponding projection through SWAP operations.

Errors from enforcing the contraint via the penalty projection follow similar as in the case for Dirichlet conditions. 
This is because fundamentally what we do is suppress the overlap on the subspace given by the penalty projection. Then, if the penalty projection approximates a derivative reasonably well, as in \cref{eq:projector-for-neumann}, the approximation of a derivative follows.
We give an example using zero Neumann boundary conditions as the simplest case later in \cref{subsubsec:proj-neumann-section}.


Moreover, the same approach of `shifting the data' can be applied to the case of Neumann boundary conditions, i.e., constraints on the derivative such as $\< \partial_{{n}(r)}, \,v(r,t)\> = h(r,t)$ for $r\in\Gamma_N$ in \cref{eq:application-to-pdes-setup}.
Here, $n(r)$ is the outward facing normal vector on the Neumann boundary of the domain, $\Gamma_N$. 
When $\Gamma_N=\partial\Omega$, the divergence theorem implies
\begin{equation}\label{eq:NeumConstCondn}
    \int_\Omega\D V\, \Delta v(r,t)=\int_{\partial\Omega} \D S\, \partial_{n(r)}v(r,t) = \int_{\partial\Omega} \D S\, h(r,t).
\end{equation}
Volume and surface elements are described by $\D V$ and $\D S$, respectively.

Similar to the Dirichlet case, we  seek a function $w(r,t)$ that complies with the constraint,
\begin{equation}
   \partial_{n(r)}  w(r,t),  = h(r,t)  \quad \text{for any } r \in \Gamma_N.
\end{equation}
Now, let 
\begin{equation}
    v'(r,t) = v(r,t)-w(r,t).
\end{equation}
The respective PDE in $v'$ then is
\begin{align}
    \partial_t v' (r,t) - \mathcal{L} v' (r,t)
    &=\partial_t v (r,t) - \mathcal{L} v (r,t) - \partial_t w (r,t) + \mathcal{L} w (r,t)
    \nn
    &= f(r,t)  + \mathcal{L} w (r,t) - \partial_t w (r,t)
    =: \widetilde{f}(r,t),
\end{align}
with initial condition
        $v'(r,0) = v_0(r) - w(r,0)
        =: v_0'(r)$,
and boundary condition
        $ \partial_{n(r)} v_0'(r) = 0$.
Therefore, in total, $v'$ will satisfy the homogeneous Neumann problem
\begin{align}
        \partial_t v' (r,t) - \mathcal{L} v' (r,t) &= \widetilde{f}(r,t), \quad & r \in \Omega,\;  t > 0
        \nn
        v'(r,0) &= v'_0(r),  \quad & r  \in \overline{\Omega}, t =0 
        \nn
         \partial_{n(r)}v'(r,t) &= 0,\quad & r \in \Gamma_N,  t\geq 0.
    \label{eq:homNBVP}
\end{align}
Analogous to the Dirichlet case, this gives rise to a means to solve  \eqref{eq:homNBVP} for $v'$ and adding to it the function $w(r,t)$. Then, $v'+w$ satisfies the Neumann problem as in \cref{eq:application-to-pdes-setup}. 

When $\Gamma_N=\partial\Omega$, homogeneous Neumann conditions imply
\begin{equation}
    \int_\Omega\D V\,\Delta v'(r,t)=0.\label{eq:vol-int-vanishes}
\end{equation}

We enforce the normal derivative at each boundary point through $\projcN$.
We quantify the error by applying \cref{result:error-new-kubo} to the shifted problem with $\projcN$, and then converting the projection error to an error in the derivative approximation as in \cref{lem:error-neumann-dissipative}.

\paragraph*{Construction of shifting function for Neumann boundary conditions}
Following the approach in \cref{subsec:nonzero-dirichlet}, given $\overline{\Omega}= [0,L]^d$, we now aim to find a function $w(r,t)$ so that it satisfies the Neumann condition in \cref{eq:application-to-pdes-setup}, namely that $\<\partial_{n(r)}, w(r,t)\> = h(r,t)$ for $r\in\Gamma_N$. 

For $d=1$, one may choose
\begin{align}
    w(r,t)&=\int\D r\,\left[\frac{r}{L}h(L,t)-\left(1-\frac{r}{L}\right)h(0,t)\right]\nn
    &=\frac{r^2}{2L}h(L,t)-\left(r-\frac{r^2}{2L}\right)h(0,t).
\end{align}
Then $-\partial_r w(0,t)=h(0,t)$ and $\partial_r w(L,t)=h(L,t)$, as required by the outward normals at the two endpoints.
For $d>1$, start with $w_0(r,t)=0$ and define for $i=1,\ldots,d$,
\begin{align}
 w_i(r,t)={}&w_{i-1}(r,t)
 +\frac{r_i^2}{2L}
 \left.\left(h-\partial_{r_i}w_{i-1}\right)\right|_{r_i=L}
 \nonumber\\
 &-\left(r_i-\frac{r_i^2}{2L}\right)
 \left.\left(h+\partial_{r_i}w_{i-1}\right)\right|_{r_i=0}.
\end{align}
The correction terms impose the outward derivatives on the two
faces $r_i=0,L$. We assume sufficient smoothness and compatibility on intersecting faces $\Gamma_i,\Gamma_j$, with outward normals $n_i,n_j$, 
$\partial_{n_j}\left(h|_{\Gamma_i}\right)
 =\partial_{n_i}\left(h|_{\Gamma_j}\right)$.
On a previously considered face $\Gamma_j$, this gives $\partial_{n_j}(h|_{\Gamma_i}-\partial_{n_i}w_{i-1})=0$.
The correction therefore has zero normal derivative on that face, preserving the conditions that have been already imposed previously, and $w=w_d$ satisfies all prescribed Neumann conditions.
If Neumann conditions are defined only on a part of the boundary, we first extend $h$ to the other faces with these compatibility conditions.


\subsection{Input Model  of Constraint Projections for Quantum  Implementation}\label{subsec:input-model-for-projections}

We continue by discussing constructing the relevant projections and their quantum implementations.
The quantity we want to encode is the solution vector $v(t)$, which is expressed with $n_l$ points per dimension $l\in[d]$; recall the definitions of $\mathcal{I}_\Omega, \mathcal{I}_{\Gamma_D}, \mathcal{I}_{\Gamma_N}$ from \cref{defn:geometry-index-sets}. 
This vector is represented in terms of an amplitude encoding:
\begin{equation}\label{eq:amplitude-encoding-at-t}
    \ket{v(t)} =\frac{1}{\lVert v(t)\rVert}  \sum_{\bm{j}\in\mathcal{I}_{\Omega\cup\Gamma}} v_{\bm{j}}(t) \ket{\bm{j}} .
\end{equation}
Here, $\bm{j} = (j_1, j_2, \dots,j_d) $ is a multi-index over $d$ dimensions so that $j_l\in[n_l-1]_0$ for all $l\in[d]$.

In order to specify boundaries, we assume access to an oracle $\orac_{\text{bdry}}$. For Neumann and Robin constraints, the oracle marks both points of each pair with the same boundary type. 
\begin{equation}\label{eq:def-bdry-oracles}
    \orac_{\text{bdry}}: \ket{\bm{j}} \ket{0}_\text{bdry} \to \begin{cases}
        \ket{\bm{j}}\ket{0}_\text{bdry}, &  \bm{j}\in \mathcal{I}_\Omega: \text{ internal point}\\
        \ket{\bm{j}}\ket{1}_\text{bdry}, & \bm{j}\in\mathcal{I}_{\Gamma_D}: \text{ Dirichlet condition}\\
        \ket{\bm{j}}\ket{2}_\text{bdry}, & \bm{j}\in\mathcal{I}_{\Gamma_N} \cup\bigcup_{\bm k \in \mc I_{\Gamma_N}} \zeta_{\bm k}: \text{ Neumann condition}\\
        \ket{\bm{j}}\ket{3}_\text{bdry}, & \bm{j}\in\mathcal{I}_{\Gamma_R} \cup\bigcup_{\bm k \in \mc I_{\Gamma_R}} \zeta_{\bm k} : \text{ Robin condition}\\
    \end{cases}
\end{equation}
We use disjoint sets for the three boundary types in this construction to ensure compatibility. 
In particular, neighbours with respect to a derivative condition are not marked as internal points. 
Changing between two points of a pairs preserves the boundary flag and does not impact uncomputation.
The algorithm we present in \cref{sec:algorithm-and-complexity} will consider the dynamics we study exclusively in an interaction picture with respect to $\projc$. Therefore, the algorithm makes use of on Hamiltonian simulation of the projection rather than direct applications of the projection operator.   
An important point then so that the interaction picture simulation algorithm in \cite{low2018hamiltonian} is efficient is that the projection's Hamiltonian simulation can be fast-forwarded~\cite[Definition~1]{atia2017fast}, so that the complexity of implementation depends at most logarithmically on $\lambda$. 

\paragraph*{Fast-forwarding constraint projections}
\begin{theorem}[Theorem 3 in \cite{atia2017fast}]
If $H$ is an $n$-qubit normalized commuting $k$-local Hamiltonian, with $k\in O(\log n)$, then it can be $(T,\alpha)$-fast forwarded for $T\in 2^{O(n)}$ and arbitrary exponentially small $\alpha$.
\end{theorem}
\noindent Note that $(T,\alpha)$ fast-forwarding means that there exists a circuit with at most poly($n$) gates and poly($n$) ancillas which approximates evolution of $H$ by for time $T$ with error at most $\alpha$ in the operator norm.

Therefore, our goal is to devise the constraint projections for Dirichlet, Neumann and Robin conditions which are commuting. 
Furthermore, we provide constructions for such circuits, which will make it obvious that their circuit cost does not explicitly depend on the time parameter. The precision required to represent the phase is logarithmic in time. 

\subsubsection{Projections for Dirichlet conditions and value constraints}
Dirichlet boundary conditions in PDEs are point-wise value constraints. On a grid, this naturally induces a projection matrix by projecting directly onto basis elements. 
We can define $\projcD$ to be 
\begin{equation}\label{eq:def-dirichlet-projection}
    \projcD = \sum_{\bm j\in\mathcal{I}_{\Gamma_D}} \op{\bm j},
\end{equation}
which is equivalent to conditioning on the boundary flag from the boundary oracle $\orac_{\rm bdry}$ in \cref{eq:def-bdry-oracles} being in 1-state, so $\bm j\in\mathcal{I}_{\Gamma_D}$.
Given this oracle $\orac_{\rm bdry}$ it is straightforward to construct a circuit for a value constraint projector,
\begin{equation}\label{eq:circuit-projcd}
    U_{\projcD}: \ket{\bm j}\ket{0}_{\rm bdry} \mapsto \orac_{\rm bdry}^\dagger \left(\identity \otimes \op{1}_{\rm bdry}\right) \, \orac_{\rm bdry} \ket{\bm j}\ket{0}_{\rm bdry}.
\end{equation}
In terms of LCU, we can see $\orac_{\rm bdry}$ as ${\rm PREP}$ and $\identity\otimes\op{1}_{\rm bdry}$ as (a part of) ${\rm SEL}$.

\paragraph*{Hamiltonian simulation of the value constraint projection.}
We notice that $[\op{\bm{j}}, \op{\bm k}] = 0 $ for any $\bm{j},\bm{k}$ in $\mathcal{I}_{\overline\Omega}$. 
This means for the evolution that  
\begin{equation}
    \e^{-\imag\lambda t \projcD} = \prod_{\bm j \in \mathcal{I}_{\Gamma_D}} \e^{-\imag \lambda t \op{\bm j}}. 
\end{equation}
So for time-independent constraints, this can be implemented simply by phase gates parametrized by the penalty $\lambda$, controlled on $\ket{\bm{j}}$.

\begin{equation}\label{eq:hamsim-for-dirichlet}
    {\rm HamSim}(\projcD, \lambda t) \ket{\bm j}\ket{0}_{\rm bdry} =       
    \orac_{\rm bdry} ^\dagger
    \left(\identity +  \e^{-\imag \lambda t} \otimes \op{1}_{{\rm bdry}}   \right) 
    \orac_{{\rm bdry}}  
    \ket{\bm j } \ket{0}_{{\rm bdry}} 
\end{equation}
Therefore, \cref{eq:hamsim-for-dirichlet} is fast-forwarded by this construction. It relies on two boundary-oracle calls and one phase gate for every $\lambda t$.

\subsubsection{Projectors for Neumann conditions and derivative constraints}\label{subsubsec:proj-neumann-section}
To obtain a finite-dimensional approximation to a Neumann projection, we employ a discrete finite-difference derivative ${\sf D}$. 
As observed in the statement of \cref{prob:ivp-bvp-defn}, a crucial ingredient for our method to work is that the projection $\projcN$ for the approximate derivative condition is orthogonal to, and thereby commutes with, the projection onto the feasible region and the projection onto the value constraint region. 
We discuss next how these requirements are satisfied by a simple finite difference stencil evaluated on two points assuming a grid discretization.
The motivation is that this allows us to enforce that the value on the boundary has the same value as the next-closest value within the domain in the direction of interest. 

First, we look at a one-dimensional example. Take $v = \sum_{l=1}^L v_l e_l$ with the canonical basis $\{e_l\}_{l=1}^L$. 
Consider $\partial_n\cdot v\mid_L = 0$ and $v$ representing a line split into $L$ points, then $v_L$ falls into the derivative constraint. Then, a ``inward'' finite difference stencil would be $\frac{v_L - v_{L-1}}{h}$. 
The projection enforces equality of the two values, while the derivative approximation is $\mathsf{D}v\mid_{\Gamma_N}=(v_L-v_{L-1})/h$.
Consequently, what we want to enforce is that $v_{L}=v_{L-1}$. The feasible and infeasible spaces then are  
\begin{equation}
    \dom = {\rm span}\{v: v_L = v_{L-1}\},
    \quad 
    \domcompN = {\rm span}\{e_L-e_{L-1}\},
\end{equation}
with unit vectors $e_l = (0, \ldots, 1_{l}, \ldots, 0)^T$.
We aim to enforce the approximative derivative constraint 
\begin{equation}
      \lvert v_L - v_{L-1} \rvert < h\varepsilon,
\end{equation}
for $\varepsilon>0$.     
A swap operation ${\sf S}$ so that ${\sf S}v=\sum_{l=1}^{L-2}v_le_l+v_Le_{L-1}+v_{L-1}e_L$ leaves $\dom$ invariant.
Then the projection $\projcN=\frac12(\identity-{\sf S})$ satisfies $\ker(\projcN)=\dom$ and ${\rm image}(\projcN)=\domcompN$.
Its image is spanned by the difference $e_L-e_{L-1}$.
As ${\sf S}={\sf S}^\dagger$ and ${\sf S}^2=\identity$, both $\projcN$ and its complement $\proj=\frac12(\identity+{\sf S})$ are orthogonal projections, as required in \cref{defn:infeasible-projector}.

We continue by discussing how to create the projector for $\lvert\mathcal{I}_{\Gamma_N}\rvert>1$. 
A general solution vector $v$ represents a spatial discretization of the continuous solution over a domain $\Omega\subseteq\reals^d$, with overall $n=\prod_{l=1}^d n_l$ basis elements.  
For every point $\bm j\in\mathcal I_{\Gamma_N}$ on the derivative constraint region or Neumann boundary, the set $\zeta_{\bm j}$ contains the neighbour used in the two-point finite difference formula.
As assumed in \cref{defn:geometry-index-sets}, the corresponding SWAP operations commute, so the pairs are disjoint.
The algorithm in \cref{sec:algorithm-and-complexity} also applies to more general derivative constraints when Hamiltonian simulation of their orthogonal constraint projection is provided. 
This leads to the following as a formulation for the sought after projector,
\begin{equation}\label{eq:projector-for-neumann}
    \projcN  
     =  \frac{1}{{2}}\left(\identity - \prod_{\bm j \in \mathcal{I}_{\Gamma_N}, \bm k \in \zeta_{\bm j}}\mathrm{SWAP}(\bm j, \bm k)\right) = \frac{1}{2}(\identity - {\sf S}).
\end{equation}
Since the swaps commute and each swap is a Hermitian involution, ${\sf S}^\dagger={\sf S}$ and ${\sf S}^2=\identity$.
The disjoint pairs therefore imply
\begin{equation}
    \projcN=\frac12\sum_{\bm j\in\mathcal I_{\Gamma_N},\,\bm k\in\zeta_{\bm j}}
    (\ket{\bm j}-\ket{\bm k})(\bra{\bm j}-\bra{\bm k}).
\end{equation}
Thus $\projcN$ is the orthogonal projection onto the pairwise differences, and its kernel satisfies every two-point constraint.
Direct implementation by a linear combination of unitaries yields,
\begin{equation}\label{eq:circuit-projcn}
\color{black}
    \begin{quantikz}
        \lstick{$\ket{0}$} & \gate{X} & \gate{{\rm Had}} & \ctrl{1} & \gate{{\rm Had}} & \qw\rstick{$\bra0$}\\
        \lstick{$\ket{\psi}$} & \qwbundle{} & \qw & \gate{{\sf S}} & \qw & \qw\rstick{$\frac12(\identity-{\sf S})\ket{\psi}$}
    \end{quantikz}
\normalcolor
\end{equation}
Circuit constructions for the SWAP's ${\sf S}$ are discussed in the upcoming paragraphs.

\begin{figure}
    \centering
    \includegraphics[width=0.7\linewidth]{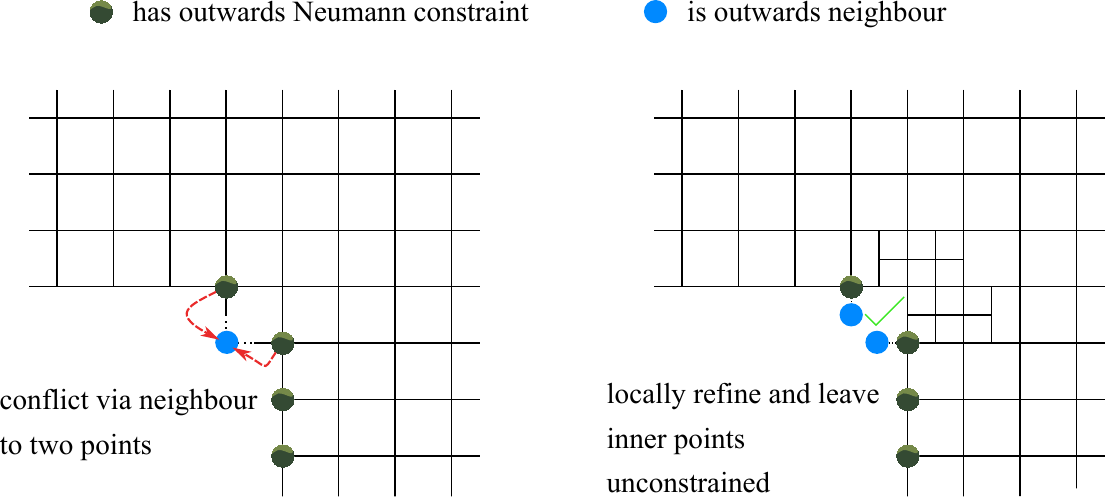}
    \caption{Conflicts arising in derivative constraints on a regular grid (left). Mitigation is possible by either only allowing one of the constraints, or a local grid refinement and ignoring the constraint on the finer level (right).}
    \label{fig:refinement-conflict}
\end{figure}

For the two-point stencils, we express the disjointness assumption in terms of the boundary points and their neighbours as follows:
\begin{itemize}
    \item  Any point that is neighbour to a constraint point for the sake of a derivative condition (i.e., part of the finite difference formula) cannot be affected by a derivative constraint itself.
    \begin{equation}\label{eq:constraint-only-one}
        \forall \bm j \in \mc I_{\Gamma_N}, \zeta_{\bm j} \cap  \mc I_{\Gamma_N}=\emptyset
    \end{equation}
    \item Any point can  be  neighbour  in a finite difference formula for a derivative constraint to at most one point on $\Gamma_N$.
    \begin{equation}\label{eq:neighbour-only-one}
        \forall \bm j,\bm k\in \mc I_{\Gamma_N}, \bm j \neq \bm k:\;  \zeta_{\bm j}\cap\zeta_{\bm k}=\emptyset
    \end{equation}
\end{itemize}
\cref{fig:refinement-conflict} illustrates how local refinement can separate two-point stencils to satisfy the commutation assumption. 
If the problem-at-hand is a more general constrained ODE, one can think of the same strategy together with introducing additional dummy variables that are trivially coupled with their neighbours instead of a global `refinement'. 
\begin{example}[Error on a derivative constraint for stable dynamics]\label{lem:error-neumann-dissipative}
We consider the dynamics and assumptions in \cref{result:error-new-kubo}, with $\projcN$ enforcing two-point derivative constraints on disjoint pairs of points separated by grid spacing $h>0$.
Then the error in the derivative approximation is
\begin{equation}
    \norm{\diff v(t)}_{\ell_2,\Gamma_N}
    =\frac{\sqrt2}{h}\norm{\projcN v(t)}.
\end{equation}
In particular, choosing $\lambda$ in \cref{result:error-new-kubo} so that $\norm{\projcN v(t)}\leq h\varepsilon/\sqrt2$ ensures $\norm{\diff v(t)}_{\ell_2,\Gamma_N}\leq\varepsilon$.
\begin{proof}
For each boundary point $\bm j$ and its neighbour $\bm k\in\zeta_{\bm j}$, the derivative approximation is $(v_{\bm j}-v_{\bm k})/h$.
The projection onto $(\ket{\bm j}-\ket{\bm k})/\sqrt2$ has squared norm $|v_{\bm j}-v_{\bm k}|^2/2$.
Summing over the disjoint pairs, we obtain
\begin{equation}
    \norm{\diff v(t)}_{\ell_2,\Gamma_N}^2
    =\frac{2}{h^2}\<v(t),\,\projcN v(t)\>.
\end{equation}
We apply \cref{result:error-new-kubo} to bound $\norm{\projcN v(t)}$.
\end{proof}
\end{example}

\paragraph{Input model for $\swap$ circuit.}
This section discusses a circuit construction for the swap circuit ${\sf S}$ in \cref{eq:projector-for-neumann}.
The first ingredient is the neighbour oracle $\orac_\zeta$, which encodes the displacement $\bm k-\bm j$ from a point $\bm j$ to its neighbour $\bm k\in\zeta_{\bm j}$.
We use the disjoint two-point pairs from \cref{defn:geometry-index-sets}; the oracle acts on both points of each pair.
Then, we define the neighbour oracle by its action on a state $v_{\bm j} \ket{\bm j}$ as
\begin{align} 
    \orac_\zeta: & \,  
    v_{j_1\cdots j_d}\ket{j_1 \cdots j_d} \ket{0} 
    \nn \label{eq:input-model-neumann-after-Ozeta}
    &\to
    v_{j_1\cdots j_d} \ket{j_1 \cdots j_d} 
    \left(\frac{1}{\sqrt{\lvert \zeta_{\bm j}\rvert}}\sum_{(k_1\cdots k_d) \in \zeta_{\bm j}} \ket{k_1-j_1}\ket{k_2-j_2}\cdots\ket{k_d-j_d}\right) 
    .
\end{align}
The action of SWAP$(\bm j, \bm k)$ for $\bm k \in \zeta_{\bm j}$ with displacement $(k_1-j_1, \ldots k_d - j_d)\in\zeta_{\bm{j}}$ is given as follows:
\begin{equation}\label{eq:swap-general-example}
    v_{j_1\cdots j_d}\ket{j_1\cdots j_d} 
    \mapsto 
    v_{j_1\cdots j_d}\ket{k_1}\ket{k_2}\cdots\ket{k_d}.
\end{equation}
Namely, it effects a shift by distances $k_w-j_w$, $w\in[d]$, where the distances are given by the neighbour oracle $\orac_\zeta$.
For a two-point stencil orthogonal to the boundary, only one coordinate ($k_w$) differes between the two points
Then, \cref{eq:swap-general-example} becomes
\begin{equation}
     v_{j_1\cdots j_d}\ket{j_1\cdots j_d} 
    \mapsto 
    v_{j_1\cdots j_d}\ket{j_1}\ket{j_2}\cdots\ket{k_w}\cdots\ket{j_d} 
\end{equation}
Therefore, these swaps can be implemented a sequence of controlled additions of $j_w$ with the distances $k_w-j_w$ and call this subroutine. The arithmetic that needs to be done here is fairly simple as we only need to add bit-strings.
Recall that CADD is defined as 
\begin{align}
    {\rm CADD} \,:\,
    \ket{\bm j}\ket{\bm k} &\to \ket{\bm j + \bm k} \ket{\bm k} 
    \\ \nonumber
    \equiv 
    \ket{j_1}\ket{j_2}
    \cdots\ket{j_d}
    \ket{k_1}\ket{k_2}
    \cdots\ket{k_d}
    &\to
    \ket{j_1+ k_1}\ket{j_2+ k_2}
    \cdots\ket{j_d+ k_d}
    \ket{k_1}\ket{k_2}
    \cdots\ket{k_d},
\end{align}
where each coordinate is stored in a bitstring of length $\lceil \log_2 n\rceil$ and addition is modulo $2^{\lceil \log_2 n\rceil}$. We represent negative displacements in the same modular arithmetic. The definition of index sets prevents overflow. For a displacement $\bm k - \bm j$ as defined by the neighbour oracle, resulting indices $\bm k$ are valid grid points.

For each pair $\bm j,\bm k$, we extend the neighbour oracle to act on both points, which yields displacement $\bm k-\bm j$ at $\bm j$ and $\bm j-\bm k$ at $\bm k$.
One application of CADD performs at most $d$ coordinate additions, each on $\lceil\log_2n\rceil$ qubits for at most $n$ basis elements per dimension.
The arithmetic therefore uses $O(d\log n)$ gates~\cite{gidney2018halving}, on top of the neighbour-oracle calls.
For a pair $\bm j,\bm k$, the circuit acts as
\begin{align}
    v_{\bm j}\ket{\bm j}\ket0
    &\overset{\orac_\zeta}{\longmapsto}
    v_{\bm j}\ket{\bm j}\ket{\bm k-\bm j}\nn
    &\overset{\mathrm{CADD}}{\longmapsto}
    v_{\bm j}\ket{\bm k}\ket{\bm k-\bm j}\nn
    &\overset{\text{negate}}{\longmapsto}
    v_{\bm j}\ket{\bm k}\ket{\bm j-\bm k}\nn
    &\overset{\orac_\zeta^{-1}}{\longmapsto}
    v_{\bm j}\ket{\bm k}\ket0.
\end{align}
Negating the displacement before applying $\orac_\zeta^{-1}$ is necessary because the state register now contains $\bm k$.
The same sequence maps $\bm k$ to $\bm j$.
This sequence conditioned on the boundary flag $2$.
Finally, we uncompute the boundary flag using $\orac_{\rm bdry}^{-1}$.
Thus,
\begin{align}\label{eq:swaps-circuit-jk}
    U_{\mathrm{SWAPS}}\big(v_{\bm j}\ket{\bm j}_v\ket0_{{\rm bdry},\zeta}\big)
    =\begin{cases}
        v_{\bm j}\ket{\bm k}_v\ket0_{{\rm bdry},\zeta},&\bm j,\bm k\text{ form a Neumann pair},\\
        v_{\bm j}\ket{\bm j}_v\ket0_{{\rm bdry},\zeta},&\text{otherwise}.
    \end{cases}
\end{align}

Overall, we have the following registers,
\begin{itemize}
    \item a state register $\ket{\bm j}_v$, consisting of $d\lceil\log_2(n)\rceil$ qubits for $n$ basis elements per dimension with binary representation; 
    \item ancillas $\ket{\cdot}_{{\rm bdry}}$ with 2 qubits to encode the type of basis element (inner, Dirichlet boundary, Neumann boundary);
    \item ancillas $\ket{\cdot}_{\zeta}$ to hold the neighbour set values coming from $\orac_\zeta$,  with $d\lceil\log_2 n\rceil$ qubits.   
\end{itemize}

\paragraph{Hamiltonian simulation of the derivative constraint projection.}
In order to time-evolve a derivative constraint, we consider \cref{lem:evolution-projection} that allow to avoid compiling $\e^{-\imag\lambda t \projcN}$ using Hamiltonian simulation techniques of the Hermitian projector $\projcN$. 
The following formula illustrates how this can be performed easily in the case of orthogonal projections. 
\begin{lemma}[Exponential of an orthogonal projection.]\label{lem:evolution-projection}
    Let $\sf P, Q$ be orthogonal projections so that ${\sf P}+{\sf Q} = \identity$. Then, for any $\xi\in\complex$, 
    \begin{equation}
        \e^{ \xi {\sf P}} = {\sf Q} + \e^{ \xi }{\sf P}.
    \end{equation}
    The case of $\xi \in\imag\reals$ recovers  Hamiltonian simulation of an orthogonal projection.
\end{lemma}
\noindent Hence, access to $\projcN$ and its orthogonal complement is sufficient for simulation in this context as only the image of the penalty projection is meant to experience a phase. 
Recall that the situation was the same for value constraints before, however, the complement there was simply the identity on the unconstrained domain and thus does not require an additional circuit implementation.
By \cref{lem:evolution-projection}, we have 
    $\e^{-\imag \lambda t \projcN} = \identity - \projcN + \e^{-\imag\lambda t}\projcN$. 
More specifically, this is 
\begin{equation}\label{eq:hamsim-neumann-projector}
    {\rm HamSim}(\projcN,\lambda t) = \frac{1}{2}(\identity + {\sf S}) + \frac{\e^{-\imag \lambda t}}{2}(\identity-{\sf S}).
\end{equation}
Indeed, this form is now equivalent to the value constraints up to modified projectors, where in the former case we can identify ${\sf Q}$ with $\sum_{\bm j \in \Omega \setminus \Gamma_D}\op{\bm j}$ and ${\sf P}$ with $\sum_{\bm j \in \Gamma_D} \op{\bm j}$.
A circuit construction for ${\sf S}$ was given above in~\cref{eq:swaps-circuit-jk}.
Then, constructing a circuit for \cref{eq:hamsim-neumann-projector} is simple.  
We compute the two eigenspaces of $\mathsf{S}$ into an ancilla, apply the phase, and uncompute the ancilla 
\begin{equation}
\color{black}
    \begin{quantikz}
        \lstick{$\ket{0}$} & \gate{{\rm Had}} & \ctrl{1} & \gate{{\rm Had}} & \gate{\mathrm{diag}(1,\e^{-\imag\lambda t})} & \gate{{\rm Had}} & \ctrl{1} & \gate{{\rm Had}} & \qw\rstick{$\bra{0}$}\\
        \lstick{$\ket{\psi}$} & \qwbundle{} & \gate{{\sf S}} & \qw & \qw & \qw & \gate{{\sf S}} & \qw & \qw\rstick{$\e^{-\imag\lambda t\projcN}\ket{\psi}$}
    \end{quantikz}
\normalcolor\label{eq:circuit-for-neumann-evolution}
\end{equation}
The final Had--controlled-${\sf S}$--Had sequence returns the ancilla to $\ket0$.
The phase gate is the same as in the value constraint evolution in \cref{eq:hamsim-for-dirichlet}.
This allows us to effectively fast-forward the evolution of $\projcN$. The circuit in \cref{eq:circuit-for-neumann-evolution} uses two controlled applications of $\mathsf S$ and one phase gate for every $\lambda t$. 
The ability to fast-forward also follows from \cite[Theorem~3.1 and Fig.~2]{gu2021fast}.

\paragraph{Interface conditions via `derivative projection'.}
The finite difference construction to approximate Neumann conditions is also useful to represent interface conditions, i.e., conditions between regimes that are governed by different PDEs.
Suppose there are $\eta$ subdomains which represent a different physical model (expressed by a different linear PDE)
\begin{equation}
    (\partial_t - \mc L_1 ) u_1(x;t) = 0, \; x \in \Omega_1 , \quad \ldots, (\partial_t - \mc L_\eta ) u_\eta(x;t) = 0 , \; x \in \Omega_\eta,
\end{equation}
then one can define interface conditions describe relationships at any intersections of the domains $k\neq l \in [\eta]$, 
\begin{equation}
    u_k(x;t) = \omega\,u_l(x;t), \quad x\in\Omega_k \cap\Omega_l,
\end{equation}
with $\omega\in\complex$; one can also think of more general relationships. 
The numerical treatment of interface conditions in classical numerical methods is described e.g. in~\cite{cen2015multiphysics}.
Our approach somewhat resembles the `penalty method' described there, in the sense that it also introduces a notion virtual work that is minimized by admissible solutions.

In order to use `Neumann projections' as in \cref{eq:projector-for-neumann} we need an index set ${\mc I}_{\rm interface}$ and neighbour sets $\bigcup_{\bm j \in{\mc I}_{\rm interface}} \zeta_{\bm j} $. 
Using a grid discretization of the computational domain, then there are no more `true' boundary points that are in $\Omega_k\cap\Omega_l$ but every point is in either $k$ or $l$; therefore, what we get are bipartite sets that give immediate rise to the necessary boundary point---neighbour relation.
For a pair of points $\bm j$ and $\bm k$ with interface condition $v_{\bm j}=\omega v_{\bm k}$, the projection onto violations of the condition is
\begin{equation}
    \frac{\big(\ket{\bm j}-\overline\omega\ket{\bm k}\big)
    \big(\bra{\bm j}-\omega\bra{\bm k}\big)}{1+|\omega|^2}.
\end{equation}
For $\omega=1$, this is the two-point Neumann projection.
For general $\omega$, we rotate the normalized constraint vector to $\ket{\bm j}$, apply the phase $\e^{-\imag\lambda t}$ to $\ket{\bm j}$, and undo the rotation.
We mark $\bm j$ as interface points by adjusting the boundary oracle accordingly by a flag for $\mathcal I_{\rm interface}$.
For disjoint pairs, these projections are orthogonal and their evolutions commute.

\subsubsection{Input Model for Robin Boundary Conditions}\label{subsec:robin-conditions}
The ability to implement value constraints and derivative constraints allows us to extend the applicability to another class of boundary conditions.
Constraints described by a superposition of Dirichlet and Neumann conditions are called  Robin boundary conditions~\cite{gustafson1998third}; they can be expressed as
\begin{equation}
    \alpha v(x;t) + \beta \,\partial_n\cdot v(x;t) = f(x;t), \quad x\in\Gamma_R,
\end{equation}
with real constants $\alpha,\beta$.
If $f(x;t)$ is nonzero, we first subtract a function satisfying the Robin constraint, following the constructions above.
For homogeneous conditions, a two-point stencil with grid spacing $h$ imposes
\begin{equation}
    \left(\alpha+\frac\beta h\right)v_{\bm j}-\frac\beta h v_{\bm k}=0,
    \quad \bm j\in\mathcal I_{\Gamma_R},\,\bm k\in\zeta_{\bm j}.
\end{equation}
The corresponding projection on disjoint pairs is
\begin{equation}
    \mathsf{P}_c^R=\sum_{\bm j\in\mathcal I_{\Gamma_R},\,\bm k\in\zeta_{\bm j}}
    \frac{\big[(\alpha+\beta/h)\ket{\bm j}-(\beta/h)\ket{\bm k}\big]
    \big[(\alpha+\beta/h)\bra{\bm j}-(\beta/h)\bra{\bm k}\big]}
    {|\alpha+\beta/h|^2+|\beta/h|^2}.
\end{equation}
Then, the kernel of this projection consists vectors that satisfy the Robin conditions.
Using \cref{lem:evolution-projection}, we obtain
\begin{equation}\label{eq:hamsim-robin-projector}
    {\rm HamSim}(\mathsf{P}_c^R,\lambda t)
    =\identity-\mathsf{P}_c^R+\e^{-\imag\lambda t}\mathsf{P}_c^R.
\end{equation}
For each pair of indices on the Robin boundary conditioned on $\ket{3}_{\rm bdry}$, we rotate to $\ket{\bm j}$ (using the coefficients $\alpha+\beta/h$ and $-\beta/h$), apply the phase $\e^{-\imag\lambda t}$ and undo the rotation. 
For more general stencils, we provide Hamiltonian simulation of the joint orthogonal constraint projection following \cref{sec:algorithm-and-complexity}.

\subsection{Hamiltonian simulation of combined projections}
Using the results from the previous two sections, we can express the Hamiltonian simulation of orthogonal value and derivative projections as follows,
\begin{align}\label{eq:overall-hamsim-equation}
    {\rm HamSim}(\projc, \lambda t) = \orac_{\rm bdry}^{-1} \; \Big(
        &\identity\;\;\;\;\;\;\;\;\;\;\;\;\;\;\;\;\;\;\;\;\;\,\;\;\;
    \;\;\otimes \op{0}_{\rm bdry}\nn +
    &{\rm HamSim}(\projcD,\lambda t)\;\;\otimes \op{1}_{\rm bdry}\nn
    +
    &{\rm HamSim}(\projcN,\lambda t)\;\;\otimes \op{2}_{\rm bdry}\nn
    +
    &{\rm HamSim}(\mathsf{P}_c^R, \lambda t) 
    \;\;\otimes\op{3}_{\rm bdry}
    \Big)
     \orac_{\rm bdry},\,\,
\end{align}
where the controlled evolutions are implemented using \cref{eq:hamsim-for-dirichlet,eq:hamsim-neumann-projector,eq:hamsim-robin-projector}.
The Dirichlet evolution requires a phase gate, the Neumann evolution additionally uses the swap ${\sf S}$, and the Robin evolution uses the weighted constraint vectors above.
Recall that $\ket{1}_{\rm bdry}$ marks Dirichlet points, $\ket{2}_{\rm bdry}$ Neumann points and $\ket{3}_{\rm bdry}$ Robin boundary points.


\section{Quantum Algorithm via Interaction-Picture simulation}\label{sec:algorithm-and-complexity}
Recall that the constrained ODEs we consider in this work have the form
\begin{equation}\label{eq:sec-iv-ode}
    \ddt v(t) = A_0(t) v(t) -\imag\lambda \projc v(t) +b(t) \equiv A(t) v(t) +b(t),
\end{equation}
with initial data $v(t=0)=v_0$ and an orthogonal projection $\projc$.
The general solution to this via the variation of parameters formula is
\begin{equation}
    v(t) = \mathcal{T} \e^{\int_0^t \D s\, A(s)} v_0 + \int_0^t \D s\, \mathcal{T} \e^{\int_{s}^t \D s'\, A(s')} b(s).
\end{equation}
There exist many quantum DE solvers that could be employed here. Their cost commonly depends on a block-encoding factor $\alpha_A\ge\max_{0\le t\le T}\|A(t)\|$.
A direct block encoding of \cref{eq:sec-iv-ode} has normalization $\alpha_{A_0}+\lambda$.
The sufficient penalties in \cref{subsec:constraint-error-bounds} can therefore make such an implementation costly.

We can overcome this dependence in the block encoding by simulating \cref{eq:sec-iv-ode} in the interaction frame of the projection.
The interaction picture transformation is unitary, so a block encoding of $A_I(t)=\e^{\imag\lambda\projc t}A_0(t)\e^{-\imag\lambda\projc t}$ has the same normalization $\alpha_{A_0}$ as the supplied block encoding of $A_0(t)$.
The time discretization still depends on $\lambda$, since
\begin{equation}
    \|\ddt A_I(t)\|\le\|\ddt A_0(t)\|+2\lambda\|A_0(t)\|.
\end{equation}
The truncated Dyson series algorithm stores this discretization in a time register~\cite{low2018hamiltonian}.
We know from \cref{subsec:input-model-for-projections} that $\projc$ is fast-forwardable by construction.
Its realization relies on a phase determined by $\lambda t$, whose implementation cost must be included at the required precision.

Therefore, let ${v\ipic}(t) = \e^{\imag\lambda\projc t}v(t)$. 
A quick calculation shows that
\begin{align}\label{eq:interaction_pic}
    \ddt \e^{\imag\lambda\projc t}v(t) &=  \e^{\imag\lambda\projc t} \imag\lambda\projc v(t) + \e^{\imag\lambda\projc t} \ddt v(t)   \\
    &= \e^{\imag\lambda\projc t} \imag\lambda\projc v(t) + \e^{\imag\lambda\projc t} (A_0(t) - \imag \lambda \projc ) \e^{-\imag\lambda\projc t} \e^{\imag\lambda\projc t} v(t)
    \nn
    &=  \underbrace{\e^{\imag\lambda\projc t} A(t)  \e^{-\imag\lambda\projc t}}_{={A\ipic}(t)}  \e^{\imag \lambda \projc t} v(t) = A\ipic(t) v\ipic(t).\label{eq:ipic-equation-two} 
\end{align}
This means that if we solve for   
\begin{equation}
    v_I(t) = \mathcal{T} \e^{\int_0^t \D s\, A_I(s)} v_0 + \int_0^t \D s\, \mathcal{T} \e^{\int_{s}^t \D s'\, A_I(s')} b_I(s) =: S_{I}(t) v_I(0),
\end{equation}
we can retrieve the sought-after solution $v(t)$ through implementing
\begin{equation}
    {\rm HamSim}(\projc, -\lambda)
    \underbrace{S_{I}(t)}_{\text{ODE solver}} 
    {\rm HamSim}(\projc, +\lambda).
\end{equation}

\subsection{Time Evolution by LCHS}
We use the near-optimal Linear Combination of Hamiltonian simulations approach~\cite{an2023linear,an2023quantum} as a quantum ODE solver. 
For our purposes, this choice is more natural to tackle ODEs or discretized PDEs compared to Schr\"odingerisation~\cite{jin2024quantum,jin2025schr}, has better complexity with respect to approximation error and time compared to time-marching approaches~\cite{fang2023time}, and has better complexity with respect to state preparation compared to the linear-systems based approaches. 
The LCHS theorem in \cite{an2023linear} allows us to write the (time-ordered) exponential of a matrix $A(t)$ as an integral of a unitary time evolution with a specific kernel function $\frac{f(k)}{1-\imag k}$, namely
\begin{equation}\label{eq:lchs-formula}
    \TE{A}{s}{t} = \int_{\reals} \frac{f(k)}{1-\imag k }\TE{\imag(kA_{\Re} + A_{\Im})}{s}{t} \D k,
\end{equation}
where the real and imaginary part of a matrix $A$ respectively  are
\begin{equation}
  A_{\Re} = \frac{A+A^\dagger}{2}, \quad A_{\Im} = \frac{A-A^\dagger}{2\imag}.
\end{equation}
Duhamel's formula gives an extension to particular solutions given a source term~\cite{an2023linear}.
In our case, the ODE system is defined by the interaction picture matrix
\begin{equation}
    A_I(t) = \e^{\imag \lambda \projc t } (A_{0}) \e^{-\imag\lambda\projc t} =  \e^{\imag \lambda \projc t } (A_{0,\Re} +\imag A_{0,\Im}) \e^{-\imag\lambda\projc t}.
\end{equation}  
We note that \cite{an2023fractal} provides an algorithm of LCHS with interaction picture, however in a  different setting. 
In our case, the operator for the interaction frame itself is a Hamiltonian simulation and does not require treatment by the LCHS formula.
The improved kernel in \cite{an2023quantum} allows a cutoff $K$ that grows as a power of the logarithm of the inverse error.
    Choosing $\lambda\in\imag\reals$ would recover the approach in \cite[Lemma 17]{an2023fractal} which could  easily be extended to the higher-order quadratures and improved kernel function of \cite{an2023quantum}. However, this would lead to higher simulation cost as we would need to simulate the linear combination $\e^{-\lvert{\lambda}\rvert \projc t \cdot k_j}$ for all $ k_j = -K, \ldots,  K$ with $[-K,K]$ the discretization of the Fourier domain in the LCHS formula~\cref{eq:lchs-formula}. 

The solution of ODEs using LCHS requires access to a $\hamt$ oracle (Hermitian block-encoding access `HAM' with a time parameter `T') for $A_I(t)$ and a fast-forwarded Hamiltonian simulation $\e^{-\imag\lambda t\projc}$.
For completeness, we briefly recall the LCHS algorithm from \cite{an2023quantum} applied to $\mathcal{T}\e^{\int_0^t A_I(s) \D s}$.
Given $k\in\reals$, consider the Hamiltonian simulation integrand in~\eqref{eq:lchs-formula},  
\begin{equation}
    U(k;t) = \mathcal{T} \e^{\imag \int_0^t  \D s\, k A_{I,\Re} + A_{I,\Im} }.
\end{equation}
This unitary satisfies
\begin{equation}
    \ddt U(k;t) = \imag (k A_{I,\Re}(t) + A_{I,\Im}(t)) U(k;t), \quad U(k;0) = \identity, 
\end{equation}
where
\begin{equation}
    A_{I,\Re}(t) = \e^{\imag \lambda \projc t } A_{0,\Re}(t)\e^{-\imag \lambda \projc t } , \quad A_{I, \Im}(t) =  \e^{\imag \lambda \projc t } A_{0,\Im}(t) \e^{-\imag\lambda\projc t}.
\end{equation}
Simulating $U(k;t)$ is achieved by a truncated Dyson series algorithm~\cite{low2018hamiltonian}.
Then, the integral over $k$ in \cref{eq:lchs-formula} is truncated to some finite interval $[-K;K]$ and evaluated with a higher-order quadrature; these steps amount to a linear combination the unitaries $U(k;t)$.  
These routines thus require \hamt access to $A_{0,\Re}(t)$, $A_{0,\Im}(t)$ and access to $\e^{-\imag \lambda t \projc}$.

\subsubsection{Oracles}\label{subsec:oracles-for-algorithm}
\paragraph*{$\hamt$ oracles}
The $\hamt$-oracles follow the convention from \cite{an2023quantum,an2023linear}. 
The truncated Dyson series is implemented across short-time integrators with length $\Delta t$ using $M_D$ time-steps. Let $q\in[M_D]$ be a short-time segment and $\ket{0}_a$ a register of ancillas, then
\begin{equation}
    \bra{0}_a \hamt_{A_I,q} \ket{0}_a =   \sum_{l=0}^{M_D-1} \op{l}\otimes \frac{A_I(q\Delta t + l\frac{\Delta t}{M_D})}{\alpha_A}.
\end{equation}
This leads to the oracle for long-time integration with Hamiltonian $kA_{I,\Re} + A_{I,\Im}$,
\begin{equation}
    \bra{0}_a \hamt_{kA_{I,\Re} + A_{I,\Im}, q} \ket{0}_a
    =
    \sum_{j=0}^{M-1}
    \sum_{l=0}^{M_D-1} \op{j} \otimes \op{l}\otimes 
    \frac{k_j A_{I,\Re}(q\Delta t + l\frac{\Delta t}{M_D})+A_{I,\Im}(q\Delta t + l\frac{\Delta t}{M_D})}{\kappa \alpha_{A_{I,\Re}} + \alpha_{A_{I,\Im}}},
\end{equation}
where $\kappa=\max_j\abs{k_j}$, $M$ is the number of quadrature nodes for the integral over $k$ and we have the short-time segments $q\in[M_D]$.
Every oracle call to this HAM-T oracle will require one call to Hamiltonian simulation as in \cref{eq:overall-hamsim-equation} and one to its inverse.

\paragraph*{Further oracles}
Additionally, we require the following oracles for the LCHS algorithm:
\begin{itemize}
    \item Circuits for the HamSim routines in \cref{eq:overall-hamsim-equation}, within the HAM-T construction and to, if desired, transform back from the interaction picture to the ``Schr\"odinger picture'' of the untransformed solution state.
    \item A state preparation oracle $\orac_v$: $\ket{0}\to\ket{v(0)}$.
    \item An oracle to prepare the forcing term $\orac_b$: $\ket{0}\ket{\tau}\to\ket{b(\tau)}\ket{\tau}$ for all time-steps $\tau\in[n_t-1]_0$. Subsequent application of $\e^{\imag\lambda\tau\projc}$ prepares $\ket{b_I(\tau)}$.
    \item State preparation for the quadrature coefficients, including phases and factors $\norm{b(\tau)}$ \cite[Section~4.1]{an2023quantum}.
\end{itemize}
For cases when constructing oracles to initial states that approximately satisfy the constraint is significantly more efficient, or constructing a state that follows the constraint exactly is not possible, consider the following \cref{lem:ensuring-satisfaction}.
\begin{lemma}[Ensuring constraint-satisfying input states]\label{lem:ensuring-satisfaction}
Suppose we have an input oracle $\orac_{v,\varsigma}:\ket{0}\mapsto\ket{v_\varsigma(0)}$, where $\|v_\varsigma(0)\|=1$ and $\|\projc v_\varsigma(0)\|\le\varsigma<1$.
Furthermore, assume access to the reflection $\e^{-\imag\pi\projc}=\identity-2\projc$, implemented by the projection Hamiltonian simulation in \cref{eq:overall-hamsim-equation}.
Then we can produce $\ket{\proj v_\varsigma(0)}$ with success probability at least $1-\varsigma^2$ using one controlled reflection.
Amplitude amplification requires $O((1-\varsigma^2)^{-1/2})$ calls to the reflection and the input preparation and their inverses for constant success probability.
\end{lemma}
\begin{proof}
Applying the following circuit separates the two projected components:
\begin{equation}\label{eq:ensuring-nice-inputs}
\begin{quantikz}
    \lstick{$\ket{0}$} & \gate{{\rm Had}} & \ctrl{1} & \gate{{\rm Had}} & \qw \rstick{$0\text{ with probability }\|\proj v_\varsigma(0)\|^2$}\\
    \lstick{$\ket{v_\varsigma(0)}$} & \qwbundle{} & \gate{\e^{-\imag\pi\projc}} & \qw & \qw \rstick{$\ket{\proj v_\varsigma(0)}$}
\end{quantikz}
\end{equation}
Before measurement, the state is
\begin{equation}
    \ket{0}\otimes\proj v_\varsigma(0)+\ket{1}\otimes\projc v_\varsigma(0).
\end{equation}
Post-selecting the zero outcome prepares the constraint-satisfying state, with probability
\begin{equation}
    \|\proj v_\varsigma(0)\|^2\ge1-\varsigma^2.
\end{equation}
Finally, use fixed-point amplitude amplification~\cite[Theorem~27]{gilyen2019quantum}.
\end{proof}

\subsection{Complexity}
In this section, we summarize the overall complexity. We first provide the final result and subsequently walk through the analysis. \cref{tab:complexity-glossary} contains an overview of relevant variables that are introduced for the complexity analysis.
\begin{table}
\begin{center}
\caption{Glossary of terms used in complexity analysis}
\label{tab:complexity-glossary}
\small
\begin{tabular}{@{}p{0.22\linewidth}p{0.74\linewidth}@{}}
\toprule
\textbf{Symbol} & \textbf{Meaning} \\
\midrule
$T,\varepsilon$
& Final evolution time and target error in the normalized output state. \\
$\lambda$
& Strength of the constraint penalty $-\imag\lambda\projc$. \\
$a_T,\alpha_{A_0}$
& $a_T=\max_{t\in[0,T]}\|A_0(t)\|$;
  $\alpha_{A_0}\ge a_T$ is block-encoding normalization. \\
$B_{L^1},C_T$
& $B_{L^1}=\int_0^T\|b(t)\|\,\mathrm{d}t$ and
  $C_T=\|v_0\|+B_{L^1}$. \\
$v_{\min}$
& Lower bound on the norm of the exactly-constrained solution: $\|u(T)\|\ge v_{\min}>0$. \\
$E_T$
& Penalty-error constant used in \cref{result:error-new-kubo}; see \cref{tab:error-parameters-summary}. \\
$\Lambda_I,\Xi_I$
& Smoothness factors of the interaction-picture generator $A_I(t)$
  and forcing $b_I(t)$, respectively, defined \cref{thm:complexity-linear-constrained-ode}. \\
$\beta$
& LCHS kernel parameter, with $0<\beta<1$. \\
$M$
& Number of quadrature nodes for the LCHS integral over $k$. \\
$M'$
& Number of time-quadrature nodes for integral related to forcing term. \\
$M_D$
& Number of time-grid points within each short-time Hamiltonian simulation. \\
\bottomrule
\end{tabular}
\end{center}
\end{table}

\subsubsection{Final Result}
In \cref{tab:results-from-approximation-guarantees}, we summarize the necessary strengths of the penalty through lower bounds on $\lambda$. Then, as we show in the subsequent section after stating the main complexity result, this allows us to derive the smoothness parameters $\Lambda_I, \Xi_I$ as collected in \cref{tab:results-from-smoothness-factors}, which are part of the following theorem. 
\begin{table}[b]
\caption{Complexity-related parameters due to the constraint projection for different sets of assumptions}
\label{tab:complexity-parameters-table}
    \centering
    \begin{subtable}{\textwidth}
    \centering
    \begin{adjustbox}{max width=\linewidth}
    \begin{tabular}{lll}
        \toprule
        \textbf{Result} & \textbf{Error at most $\varepsilon$} & \textbf{Sufficient $\lambda$}\\
        \midrule
        \cref{result:error-new-kubo} & Error in feasible region $\|v(T)-u(T)\|$ & $(1+Ta_T)E_T/\varepsilon$\\
        \cref{result:error-new-kubo} & Constraint error $\|\projc v(T)\|$ & $E_T/\varepsilon$\\
        \cref{result:complexity} & Error in feasible region $\|\ket{v(T)}-\ket{u(T)}\|$ & $2(1+Ta_T)E_T/(\varepsilon v_{\min})$\\
        \cref{thm:complexity-linear-constrained-ode} & Total state error including solver error & $4(1+Ta_T)E_T/(\varepsilon v_{\min})$\\
        \bottomrule
    \end{tabular}
    \end{adjustbox}
    \caption{Sufficient penalties from \cref{result:error-new-kubo}, including time-dependent $A_0(t)$ and general forcing $b(t)$.
    For the normalized solution, $\|u(T)\|\ge v_{\min}>0$.
    The last row also allows a state error of $\varepsilon/2$ in the ODE solver.} 
    \label{tab:results-from-approximation-guarantees}
\end{subtable}
\begin{subtable}{\textwidth}
    \centering
    \begin{adjustbox}{max width=\linewidth}
        \begin{tabular}{ll|ll}
        \toprule
        \multicolumn{2}{c}{\textbf{Smoothness of dynamics} $\Lambda_I$} & \multicolumn{2}{c}{\textbf{Smoothness of forcing} $\Xi_I$} \\
        \midrule
        $A_0$ const.: & $\max\{\|A_0\|,2\lambda\}$ & $b$ const.: & $\max\{\|b\|,\lambda\}$\\
        $A_0(t)$: & $2\lambda+\sup_{p\ge0}\max_{t\le T}\|\partial_t^p A_0(t)\|^{1/(p+1)}$
        & $b(t)$: & $\lambda+\sup_{p\ge0}\max_{t\le T}\|\partial_t^p b(t)\|^{1/(p+1)}$\\
        \bottomrule
        \end{tabular}
        \end{adjustbox}
    \caption{Upper bounds for $\Lambda_I$ and $\Xi_I$ from \cref{lem:smoothness-factors}, when the displayed derivative bounds are finite.
    The penalty $\lambda$ can be chosen from \cref{tab:results-from-approximation-guarantees}.}
    \label{tab:results-from-smoothness-factors}
\end{subtable}
\end{table}
\begin{theorem}[Complexity of linear constrained ODE solution using LCHS, adapted from Theorem~15~in~\cite{an2023quantum}]\label{thm:complexity-linear-constrained-ode}
Consider the inhomogeneous ODE system in~\cref{prob:proj-ivp-bvp-defn}. 
    Suppose that $\tfrac12(A_{0}(\tau)+A_{0}(\tau)^\dagger)\preceq 0$ on $[0,T]$, $v(T)\ne0$, and we are given the oracles described in~\cref{subsec:oracles-for-algorithm}. 
    Let $\|A_0(\tau)\|\le\alpha_{A_0}$, $A_I(\tau)=\e^{\imag\lambda\tau\projc}A_0(\tau)\e^{-\imag\lambda\tau\projc}$, $b_I(\tau)=\e^{\imag\lambda\tau\projc}b(\tau)$ and $B_{L^1}=\int_0^T\D s\,\|b(s)\|$.
    Define the finite smoothness factors
    \begin{align*}
        \Lambda_I&=\sup_{p\ge0}\max_{\tau\in[0,T]}\|\partial_\tau^p A_I(\tau)\|^{1/(p+1)},\\
        \Xi_I&=\sup_{p\ge0}\max_{\tau\in[0,T]}\|\partial_\tau^p b_I(\tau)\|^{1/(p+1)},
    \end{align*}
    and fix $0<\beta<1$.
    Then we can prepare an $\varepsilon$-approximation, $0<\varepsilon<1$, of the normalized solution $\ket{v(T)}$ with probability $\Omega(1)$ and a flag indicating success, by choosing 
    \begin{equation}\label{eq:M-M-prime}
    \begin{aligned}
         M&\in O\left( (1+\alpha_{A_0}T)\left(\log\left(\e+\frac{\|v_0\|+B_{L^1}}{\|v(T)\|\varepsilon}\right)\right)^{1+1/\beta}\right),\\
         M'&\in \widetilde O\left((1+T(\Lambda_I+\Xi_I))\left(\log\left(\e+\frac{1+B_{L^1}}{\|v(T)\|\varepsilon}\right)\right)^{1+1/\beta}\right).
    \end{aligned}
    \end{equation}
    Different cases for $\Lambda_I$ and $\Xi_I$ are summarized in \cref{tab:results-from-smoothness-factors}.
    Further, this requires
        \begin{equation}\label{eq:final-query-comp}
            \widetilde{{O}}\left( \frac{\|v(0)\|+B_{L^1}}{\|v(T)\|} (1+\alpha_{A_0}T)\left(\log\left(\e+\frac{\|v_0\|+B_{L^1}}{\|v(T)\|\varepsilon}\right)\right)^{1+1/\beta}  \right)
        \end{equation}
        queries to the $\text{HAM-T}$ oracle and 
            \begin{equation}
            {O}\left( \frac{\|v(0)\|+B_{L^1}}{\|v(T)\|} \right)
        \end{equation}
        queries to the state preparation oracles $\orac_v$ and $\orac_b$, and the same order of calls to the preparations of the quadrature coefficients. 
For the normalized constrained solution $\ket{u(T)}$ in \cref{result:error-new-kubo}, suppose $\|u(T)\|\ge v_{\min}>0$.
Choosing $\lambda\ge4(1+T a_T)E_T/(\varepsilon v_{\min})$ and solving the penalized equation to state error $\varepsilon/2$ suffices for total state error $\varepsilon$.
The norm ratio in the complexity is then at most $4(\|v_0\|+B_{L^1})/(3v_{\min})$.
\end{theorem}
\begin{proof}
Unitary conjugation preserves $\tfrac12(A_0(\tau)+A_0(\tau)^\dagger)\preceq0$, the block-encoding normalization $\alpha_{A_0}$, and the vector norms $\|b_I(t)\|=\|b(t)\|$, $\|v_I(T)\|=\|v(T)\|$.
We apply \cite[Theorem~15]{an2023quantum} to $\ddt v_I=A_Iv_I+b_I$, choosing the quadrature and Hamiltonian simulation errors so that their sum in the unnormalized solution is at most $\varepsilon\|v(T)\|/2$.
The choices of $M$ and $M'$ in \cref{eq:M-M-prime} follow from \cite[Theorem~15]{an2023quantum}.
The LCHS coefficients have total absolute weight $O(\|v_0\|+B_{L^1})$, so that amplitude amplification incurs cost $O((\|v_0\|+B_{L^1})/\|v(T)\|)$.
Each repetition uses a constant number of state and coefficient preparations and the truncated Dyson series simulations in \cite[Corollary~4]{low2018hamiltonian}.
This proves the query bounds, including the final application of $\e^{-\imag\lambda T\projc}$.
For the constrained solution, \cref{result:error-new-kubo} implies
\begin{equation}
    \left\|\frac{v(T)}{\|v(T)\|}-\frac{u(T)}{\|u(T)\|}\right\|
    \le\frac{2\|v(T)-u(T)\|}{\|u(T)\|}
    \le\frac{2(1+Ta_T)E_T}{\lambda v_{\min}}\le\frac\varepsilon2.
\end{equation}
Moreover, $\|v(T)\|\ge\|u(T)\|-(1+Ta_T)E_T/\lambda\ge3v_{\min}/4$.
Combining this estimate with an ODE solver state error of $\varepsilon/2$ proves the last inequality.

For nonzero Dirichlet conditions, we also require preparation of $\ket{g_c(T)}$ and its norm to reconstruct $v'(T)+g_c(T)$.
We include this preparation as one further term in the LCHS linear combination before amplitude amplification.
The total coefficient weight is $O(\|v'_0\|+\int_0^T\D s\,\|b'(s)\|+\|g_c(T)\|)$, so the amplification factor becomes
\begin{equation}
    O\left(\frac{\|v'_0\|+\int_0^T\D s\,\|b'(s)\|+\|g_c(T)\|}{\|v'(T)+g_c(T)\|}\right).
\end{equation}
We include the cost of preparing the shift in each repetition.
Adding the same shift leaves the vector error unchanged, and we choose the penalty and numerical tolerances using a lower bound on the exact reconstructed solution norm.
The same construction applies to the function $u(t)$ used to shift Neumann data in \cref{subsubsec:neumann-construction}.
\end{proof}
The numbers $M$ and $M'$ count quadrature nodes, while $M_D$ counts time points within each short Hamiltonian simulation.
For $|k|\le K$, we have
\begin{equation}
    \left\|\ddt(kA_{I,\Re}(t)+A_{I,\Im}(t))\right\|
    \le(K+1)\left(\|\ddt A_0(t)\|+2\lambda\|A_0(t)\|\right).
\end{equation}
Thus the internal time discretization depends on $\lambda$ also when $b=0$.
For example, with $\Delta t\lesssim(K\alpha_{A_{0,\Re}}+\alpha_{A_{0,\Im}})^{-1}$, a sufficient choice is
\begin{align}
    M_D\in O\Bigg(1+\frac{T\Delta t(\|v_0\|+B_{L^1})}{\varepsilon\|v(T)\|}
    \Big[&(K+1)\max_{0\le t\le T}\big(\|\ddt A_0(t)\|+2\lambda\|A_0(t)\|\big)\nn
    &+(K\alpha_{A_{0,\Re}}+\alpha_{A_{0,\Im}})^2\Big]\Bigg),
\end{align}
by the time-discretization estimate for the truncated Dyson series~\cite[Lemma~5]{low2018hamiltonian}.
The corresponding time register uses $O(\log M_D)$ qubits.
Choosing the implementation error of each call to be $\varepsilon/4$ divided by the total number of calls bounds their combined error by $\varepsilon/4$. We reduce the ODE solver tolerance by the same constant factor.
An error in the projection phase changes its unitary by at most the absolute phase error.
For oracles whose cost is polynomial in the number of qubits used to encode the time-register and the number of precision bits, this dependence on $\lambda$ is polylogarithmic.

\begin{remark}[Non-normal or non-stable dynamics]
If $A_{0,\Re}(t)\preceq\gamma$ with $\gamma\ge0$, we apply the same algorithm to $\e^{-\gamma t}v(t)$.
In the LCHS query and quadrature bounds, we replace $A_0$ by $A_0-\gamma\identity$, $b(t)$ by $\e^{-\gamma t}b(t)$, $\alpha_{A_0}$ by $\alpha_{A_0}+\gamma$, and $\|v(T)\|$ by $\e^{-\gamma T}\|v(T)\|$.
The normalized final state is unchanged, and the state-preparation ratio becomes
\begin{equation}
    \frac{\e^{\gamma T}\|v_0\|+\int_0^T\D s\,\e^{\gamma(T-s)}\|b(s)\|}{\|v(T)\|}.
\end{equation}
\end{remark}

\subsubsection{Analysis of smoothness factors}
Intuitively, simulating $A_I(t)$ in the interaction picture with simulation parameter $\lambda t$ means that we simulate a highly oscillatory system, as $\lambda$ is large. Therefore, this will have impact on the size of the time steps to avoid aliasing. 
To that end, consider the smoothness parameters $\Lambda_I \ge \sup_{p\ge 0, t\in[0,T]} \lVert (\partial_t)^p A_I(t)\rVert^{\frac{1}{p+1}}, \Xi_I\ge \sup_{p\ge 0, t\in[0,T]} \lVert (\partial_t)^p b_I(t)\rVert^{\frac{1}{p+1}}$ from \cite{an2023quantum}, which need to be adjusted to the interaction picture simulation. 
We find the bounds summarized in the following lemma.
\begin{lemma}\label{lem:smoothness-factors}
Under the same assumptions as in \cref{thm:complexity-linear-constrained-ode}, the smoothness factors satisfy
\begin{align}
    \Lambda_I&\le2\lambda+\sup_{p\ge0}\max_{0\le t\le T}\|\partial_t^p A_0(t)\|^{1/(p+1)},\\
    \Xi_I&\le\lambda+\sup_{p\ge0}\max_{0\le t\le T}\|\partial_t^p b(t)\|^{1/(p+1)}.
\end{align}
For time-independent $A_0$ and $b$, these simplify to
\begin{equation}
    \Lambda_I\le\max\{\|A_0\|,2\lambda\},\quad
    \Xi_I\le\max\{\|b\|,\lambda\}.
\end{equation}
If $\projc b(t)=0$, then $b_I(t)=b(t)$ and $\Xi_I$ has no dependence on $\lambda$.
\end{lemma}
\noindent The derivation of these bounds can be found in \cref{app:analysis-of-smoothness}.

\subsubsection{Implications for the solution of discretized PDEs with boundary conditions}\label{paragraph:implications-for-disc-pde}
Typically, when quantifying the cost of  numerical methods, we are interested in the effect of the particular choice of discretization given a target error on the final cost ( with a `finer' discretization meaning more basis functions and thus more accurate representation). In our work, this `refinement' is expressed by a number of grid points $n^d$.
Upon choosing a suitable numerical method, there is also a relationship of approximation error with respect to the discretization.
The example below estimates the complexity through the number of discretization points $M'$ for the case of inhomogeneous solutions in \cref{thm:complexity-linear-constrained-ode}.

\begin{example}[Overhead estimate due to constraint for discrete heat equation]\label{example:overhead-heat}
As an example, we take the three-point finite difference stencil used in \cref{sec:numerical-experiments}.
On a fixed domain with $n$ points per coordinate, the error in the derivative approximation is $O(n^{-2})$ for smooth data~\cite{childs2021highprecision,kivlichan_bounding_2017}.
For this example, $A_0=D\mathbf L_h$ is time-independent, with $\|D\mathbf L_h\|\in O(Ddn^2)$, and we take a time-independent forcing term.
Then \cref{lem:smoothness-factors} implies
\begin{equation}
    \Lambda_I+\Xi_I\in O\big(\|D\mathbf L_h\|+\|b\|+\lambda\big).
\end{equation}
Suppose $\|u(T)\|\ge v_{\min}>0$ for the constrained heat equation.
Choosing $\lambda\ge4(1+Ta_T)E_T/(\varepsilon v_{\min})$ ensures the guarantee of \cref{thm:complexity-linear-constrained-ode}.
In particular,
\begin{equation}
    M'\in\widetilde O\left(\left(1+T\left(\|D\mathbf L_h\|+\|b\|+
    \frac{T\norm{D\mathbf L_h}E_T}{\varepsilon v_{\rm min}}
    \right)\right)
    \left(\log\left(\e+\frac{1+T\|b\|}{v_{\min}\varepsilon}\right)\right)^{1+1/\beta}\right).
\end{equation}
We fix $D, \beta$. 
Now, we want to give rough estimates when relating spatial discretization to the number of grid points, taking $n$ points per dimension $d$ when using a three-point finite difference Laplacian.
Then, $\norm{\mathbf{L}_h} \in O(dn^2)$, $\norm{v_0}\in O(n^{d/2})$ as well as $\norm{b}\in O(n^{d/2})$. 
Matching accuracies with the spatial discretization while assuming enough regularity on the solution, we have that $\varepsilon\in\Theta(d n^{-2})$.
With time-independent $\mathbf L_h, b$, and $\tfrac12(\mathbf L_h+\mathbf L_h^\dagger)\preceq 0$, 
$E_T = 2a_T (1+Ta_T)(\norm{v_0}+T\norm{b})+(2+Ta_T)\norm{\projc b}\in O\bigl(T^2d^2n^{4+d/2}\bigr)$.
Bounds on the final norm are provided in \cref{sec:lower-bound-on-norm}; we have that 
  $v_{\min}(T)\ge  \lVert v(0)\rVert - T\lVert D\mathbf L_h v(0) +b\rVert$. Thus consider $T$ so that $T\|D\mathbf L_h v_0+b\|\le\tfrac12\|v_0\|$. Then,  $\|v(T)\|,\|u(T)\|\ge \tfrac12\|v_0\|$ and therefore $v_{\min}\in\Omega(n^{d/2})$.

Putting everything together, we get 
    $M'
    \in\widetilde O
    \l( (Tdn^2 + T n^{d/2}
        +{T^4 d^2n^8})\log(1+T)
    \r)$.
Thus, the gate cost due to implementing the constraint in the case of the discretized heat equation grows as the logarithm of $M'$, namely, \begin{equation}
   \log M'\in  \widetilde O\l( d\log n + \log(1+T)    \r) .
\end{equation}
\end{example}

\section{Conclusion and Outlook}
In this work, we present a penalty projection method to enforce constraints such as boundary conditions in evolutionary discretized partial differential equations, using the LCHS algorithm as the underlying quantum ODE solver. Particularly, our approach enables arbitrary constraints and boundary conditions, up to representability via a fast-forwardable, Hermitian projector, for solving differential equations via the LCHS algorithm. 
The penalty projections have the advantage that they do not rely on adjustments of the block-encoding of the system matrix, which have the potential to make the algorithm more complicated when compiled to gates.
Moreover, our approach is able to tackle a very general class of constraints and boundary conditions, as well as interface conditions. 
Assuming that the constraint projections are orthogonal, our approach only requires fast-forwardable Hamiltonian simulation in addition to the usual ODE solution and interaction picture simulation of the constraint comes at a modest logarithmic overhead in the gate complexity when using the LCHS algorithm~\cite{an2023linear,an2023quantum} as ODE solver.
Whilst our analysis is tied to the LCHS algorithm, constraint projections could be equally employed with other quantum ODE solvers, in cases where those might be preferable. In this case, it would be necessary to investigate the details of consequences for the cost of the algorithm. 
The penalty strength $\lambda\ge (1+Ta_T)E_T/\varepsilon$ ensures error in the solution of at most $\varepsilon$ on $[0,T]$; see \cref{result:error-new-kubo}.
In a simple scenario of a uniform grid, directly enforcing constraints in the system matrix also attains a simple form for linear-systems based and time marching DE solvers. Yet, the algorithmic consequences, especially for LCHS, are not immediately obvious.
As our approach is a `plug-and-play' addition to an existing solver ODE solver, it does not address the typically very limiting aspects of initial state preparation and readout of results. Moreover, our construction is efficient so far only for the case where the discretization is carried out through a regular grid, which should be extended in a future work.
However, given these prerequisites and ODE approach where initial state preparation and final measurement are covered, there are no additional hurdles coming from adding constraints via our projection method.

Looking ahead, it would be interesting to work out specific costings involving also constant factors of the presented method as compared to other approaches to enforce constraints presented in the literature. 
Beyond this, extensions to other constraints that admit decomposition into orthogonal subspaces, such as more general symmetries, seem like a straightforward extension. Less obvious but even more interesting are symmetries with non-local support or situations that lead to non-orthogonal projections, where it is not clear if fast-forwarding in the interaction frame is possible. One could envision this within the framework of scattering theory and build on, e.g., \cite{riss1993calculation}.
Furthermore, evaluating the performance on  numerical examples of higher practical relevance would be an interesting subject of study. 
In particular, we found that our analytical error bounds were somewhat loose compared to the witnessed error in the simulation results of the heat and wave equations. Hence, one could look into exploring analytical methods that would enable tighter bounds, or perform a more thorough analysis of constant factors involved.

\section*{Code and Data Availability}
\addcontentsline{toc}{section}{Code and Data Availability}
For the simulations in \cref{sec:numerical-experiments}, we used Python with NumPy and SciPy as well as the \texttt{findiff}~\cite{findiff} package for finite difference discretization.  
The code used to generate the results below is available at~\url{https://github.com/xiang-yu/q-abs}.
\section*{Author Contributions Statement}
\addcontentsline{toc}{section}{Author contributions statement}
NW and PS conceived the project. PS wrote the manuscript and carried out the analysis, supported by XY, TK and NW. PS, TK and XY wrote the code and performed numerics.  All authors took part in discussing the project, and edited and revised the manuscript.
OpenAI’s GPT 6 Astra and 6.1 Sol were used during revisions to revise the proofs, in particular for the exact identity in Proposition 8, to simplify the presentation, and to extend numerics.

\section*{Acknowledgements}
\addcontentsline{toc}{section}{Acknowledgements}
PS thanks Lasse B.~Kristensen and Mohsen Bagherimehrab for helpful discussions in early stages of the project. PS further thanks Stephen P.~Jordan for useful comments through his thesis appraisal report.
PS was partially supported by the U.S. Department of Energy, Office of Science, Office of Advanced Scientific Computing Research under Contract No. DE-AC05- 00OR22725 through the Accelerated Research in Quantum Computing Program MACH-Q.
XYL and NW acknowledge the support from DOE, Office of Science, National Quantum Information Science Research Centers, Co-design Center for Quantum Advantage (C2QA) under Contract No. DE-SC0012704 (Basic Energy Sciences, PNNL FWP 76274) and Office of Fusion Energy Sciences, under Field Work Proposal 86966 and Contract No. DE-SCL0000299. This research was also supported by Pacific Northwest National Laboratory's Quantum Algorithms and Architecture for Domain Science (QuAADS) Laboratory Directed Research and Development (LDRD) Initiative. The Pacific Northwest National Laboratory is operated by Battelle for the U.S. Department of Energy under Contract DE-AC05-76RL01830.
AAG thanks Anders G. Frøseth for his generous support as well as the NSERC Industrial Research Chair and the Canada 150 Research Chairs programs.
This work is part of the University of Toronto's Acceleration Consortium, which receives funding from the CFREF-2022-00042 Canada First Research Excellence Fund.
This work is supported by the Novo Nordisk Foundation, Grant number NNF22SA0081175, NNF Quantum Computing Programme. Resources used in preparing this research were provided, in part, by the Province of Ontario, the Government of Canada through CIFAR, and companies sponsoring the Vector Institute \url{https://vectorinstitute.ai/partnerships/current-partners/}.
JPL acknowledges support from Tsinghua University and Beijing Institute of Mathematical Sciences and Applications.

\bibliographystyle{plainnat}
\bibliography{main_final}
\addcontentsline{toc}{section}{References}

\appendix
\section{Proof of \cref{lem:evolution-projection}}
\begin{lemma*}[Parametrized exponential of an orthogonal projection; \cref{lem:evolution-projection} in main text]
    Let $\sf P, Q$ be orthogonal projections on a vector space so that ${\sf P}+{\sf Q} = \identity$. Then, for any $\xi\in\complex$, 
    \begin{equation}
        \e^{ \xi {\sf P}} = {\sf Q} + \e^{ \xi }{\sf P}.
    \end{equation}
    For $\xi \in\imag\reals$, we retrieve Hamiltonian simulation.
    \begin{proof}
        We start by using the Taylor series of the operator exponential, 
        \begin{align}
            \e^{\xi {\sf P}} &= \sum_{k\ge 0} \frac{\xi^k}{k!} ({\sf P})^k \nn
            &= \identity + \sum_{k\ge 1} \frac{{\xi}^k}{k!} ({\sf P})^k \quad ; {\sf P}^k={\sf P} \, \forall k\ge 1 \nn
            &= \identity \pm {\sf P} + {\sf P} \sum_{k\ge 1} \frac{\xi^k}{k!} \nn
            &= {\sf Q} + \e^{\xi}{\sf P}.
        \end{align}
    \end{proof}
\end{lemma*}

\section{Derivatives for conjugated generators}\label{app:deriving-ipic-derivs}
Here we derive the identities in \cref{eq:interaction-pic-derivatives}, recalling that
\begin{align}
    {\rm ad}_{y}(x) &= [y,x]\nn
    {\rm ad}^m_y (x) &= \underbrace{[y,[y,\ldots, [y}_{m\text{ times}},x]]\cdots]    =\underbrace{{\rm ad}_y\circ\cdots\circ{\rm ad}_y}_{m\text{ times}}(x)  \nn
    {\rm ad}^0(x) &= x.
\end{align}
Let the conjugated operator be 
\begin{equation}
  A_C(t) = \e^{ B t} A(t) \e^{- B t}  ,
\end{equation}
with $A(t)$ smooth in time $t\in\reals$. In the main text, we use $B = \imag \projc$ and thus call this interaction picture $(\cdot)_I$.
Similarly, $b_C(t) = \e^{ B t}b(t)$ and smooth $b(t)$. We assume here that $B$ is constant in time. Therefore, the final expressions rely on a conjugation operation $T_t$ so that $ T_t A (T_t)^{-1}$  satisfies $\partial_t T_t = B T_t$ and $[B, T_t]=0$, which restricts us to $\e^{B t}$ with constant $B$.  
Then, we consider an arbitrary $p$th, $p\in\mathds{N}$, partial derivative with respect to time 
\begin{align*}
    (\partial_t)^p A_C(t) &=  (\partial_t)^{p-1}\left( (\dot{A}(t)_C + [ B, A_C(t)]  \right)  
    = (\partial_t)^{p-2} \left(
     [ B, [ B, A_C]] + 2 [ B, (\dot{A})_C] 
    + (\ddot{A})_C
    \right) = \ldots
\end{align*}
Using the $\rm ad$-notation introduced above, 
\begin{align*}
    (\partial_t)^p A_C(t) &=  (\partial_t)^{p-1}\left( (\dot{A}(t))_C + {\rm ad}_{ B}(A_C(t))  \right)  
    = (\partial_t)^{p-2} \left(
     {\rm ad}^2_{}(A_C(t)) + 2 {\rm ad}_{ B}((\dot{A})_C) 
    + (\ddot{A})_C
    \right) = \ldots 
\end{align*}
Then, as a direct consequence of the product rule of differentiation and the aforementioned condition that $\partial_t \e^{Bt} = B \e^{Bt} = \e^{Bt} B$, we can recognize a binomial structure of
\begin{equation}
  \partial_t^p ( A_C(t)   ) = \left((\partial_t + {\rm ad}_{ B})^p(A(t))\right)_C , \, p \in \mathds{N}.
\end{equation}
As $B$ is constant, $\partial_t$ and ${\rm ad}_B$ commute and we can apply the well-known identity $(x+y)^p = \sum_{0\le q\le p} \binom{p}{q} x^{q}y^{p-q}$. Hence 
\begin{equation}\label{eq:ad-formula-A}
    \left((\partial_t + {\rm ad}_B)^p(A(t))\right)_C 
    = \left( \sum_{0\le q\le p} \binom{p}{q} {\rm ad}_{ B}^q \, \partial_t^{(p-q)}A(t)\right)_C
    = \sum_{0\le q\le p} \binom{p}{q} {\rm ad}_{ B}^q \, \left(\partial_t^{(p-q)}A(t)\right)_C
\end{equation}
The same approach can be used for vectors $b(t)$, where the interaction picture rotation is only a left-multiplication by $\e^{ B t}$ rather than a conjugation. This means, simply replace $A(t)$ with $b(t)$ in \cref{eq:ad-formula-A} and replace ${\rm ad}_{y}^p(x)$ with left-multiplication instead of commutators, $y^p x$.

\section{Proof of \cref{prop:generalized-kubo}}\label{app:proof-of-gen-kubo}
\begin{proposition}[Restatement of \cref{prop:generalized-kubo}; Generalized Kubo formula]
Let $T>0$, let $H(t),V(t)$ be $n^d\times n^d$-dimensional time-dependent matrices, let $b(t)$ be a $n^d$-dimensional forcing, and let $P$ be a matrix.
Further, let $\zeta\in\reals$ be a perturbation parameter, independent of $H,V,b,P$ and the initial condition $v_0$.
We have the perturbed dynamics
\begin{equation}\label{eq:kubo-dynamics-appendix}
    \ddt{v_\zeta}(t) = \left(H(t) + \zeta V(t)\right)v_\zeta(t) + b(t)
\end{equation}
with initial condition $v_\zeta(0)=v_0$.
We write $v_H(t) $ for the unperturbed solution at $\zeta=0$.
We write $G_H(t,s)=\mathcal T\exp(\int_s^t\D r\,H(r))$ for $0\le s\le t\le T$.
Then, $v_H(t)=G_H(t,0)v_0+\int_0^t\D s\,G_H(t,s)b(s)$.

We have the density matrix $\sigma(t',t) = v_H(t')v_H(t)^\dagger$.
Further, $\<P(t)\>_\zeta\equiv\frac{\<v_\zeta(t),\,P v_\zeta(t)\>}{\<v_\zeta(t),\,v_\zeta(t)\>}$.
We assume $v_{\min}:=\min_{0\le t\le T}\|v_H(t)\|>0$ for this normalized expectation.
For sufficiently small $|\zeta|$, the normalized response is
\begin{equation}\label{eq:kubo-corrected-response}
 \< P(t)\>_\zeta-\< P(t)\>_0
 =\zeta\int_0^t \D s\, \chi(t,s)+R_P(t,\zeta).
\end{equation}
In particular, $\max_{0\le t\le T}|R_P(t,\zeta)|=O(\zeta^2)$.
The response function, for $0\le s\le t\le T$, is defined as
\begin{equation}\label{eq:kubo-corrected-kernel}
 \chi(t,s)
 =\frac{
 \tr[
 \left\{P-\< P(t)\>_0\identity,
 G_H(t,s)V(s)\sigma(s,t)\right\}_\sim
 ]}{\tr[\sigma(t,t)]},
\quad \chi(t,s)=0 \text{ for } s>t.
\end{equation}
We have introduced the modified anticommutator $\{X,Y\}_\sim:=XY+Y^\dagger X$.
We obtain the remainder by linearizing the exact response
\begin{equation}\label{eq:kubo-exact-response}
    \<P(t)\>_\zeta-\<P(t)\>_0
    =\int_0^\zeta\D\eta\,\int_0^t\D s\,\chi_\eta(t,s),
\end{equation}
where $\chi_\eta$ uses $G_\eta$, $v_\eta$, and the corresponding normalization.
\end{proposition}
\begin{proof}
Let $G_\eta(t,s)$ be the propagator with respect to $H(t)+\eta V(t)$, so that
\begin{equation}
    G_\eta(t,s)=\mathcal T\exp\!\left(\int_s^t\D r\,\bigl(H(r)+\eta V(r)\bigr)\right).
\end{equation}
Further, we have $\<P(t)\>_\eta$ as defined above.
Since $v_\eta(0)=v_0$ and $b(t)$ are independent of $\eta$, we differentiate \cref{eq:kubo-dynamics-appendix} and apply variation of constants to obtain
\begin{equation}\label{eq:d-eta-d-v}
  \partial_\eta v_\eta(t)=\int_0^t\D s\,G_\eta(t,s)V(s)v_\eta(s),
   \quad \partial_\eta v_\eta(0)=0.
\end{equation}
We write $(\mc G_\eta v_\eta)(t):=\int_0^t\D s\,G_\eta(t,s)V(s)v_\eta(s)$ for the integral in \cref{eq:d-eta-d-v}.

Next, we consider $\partial_\eta\<P(t)\>_\eta$ and apply the product and quotient rules of differentiation. 
\begin{align}
    \partial_\eta  \frac{\<v_\eta(t),\, P v_\eta(t)\>}{\<v_\eta(t), v_\eta(t)\>}
    &=
    \frac{1}{\<v_\eta(t),\,v_\eta(t)\>}
    \left(\begin{aligned}
    &\<\partial_\eta v_\eta(t),\,P v_\eta(t)\>
    +\<v_\eta(t),\,P\partial_\eta v_\eta(t)\>\\
    &-\<P(t)\>_\eta\,2\Re\<v_\eta(t),\,\partial_\eta v_\eta(t)\>
    \end{aligned}\right).
\end{align}
We define $\sigma_\eta(s,t)=v_\eta(s)v_\eta(t)^\dagger$, so that $\<v_\eta(t),\,v_\eta(s)\>=\tr[\sigma_\eta(s,t)]$, and substitute \cref{eq:d-eta-d-v}.
Using the modified anticommutator $\{X,Y\}_\sim=XY+Y^\dagger X$, we obtain
\begin{align}
    \partial_\eta\<P(t)\>_\eta
    &=\frac{\begin{aligned}
       &\<\partial_\eta v_\eta(t),\,[P-\<P(t)\>_\eta\identity]v_\eta(t)\>\\
       &+\<v_\eta(t),\,[P-\<P(t)\>_\eta\identity]\partial_\eta v_\eta(t)\>
       \end{aligned}}{\tr[\sigma_\eta(t,t)]}\\
    &=\int_0^t\D s\,
    \frac{\tr[\{P-\<P(t)\>_\eta\identity,\,G_\eta(t,s)V(s)\sigma_\eta(s,t)\}_\sim]}
         {\tr[\sigma_\eta(t,t)]}.
\end{align}

Then, with the fundamental theorem of calculus, we get that 
\begin{equation}
    \<P(t)\>_\zeta - \<P(t)\>_0 
    =
    \int_0^\zeta\D\eta\, 
    \int_0^t\D s\,\chi_\eta(t,s).
\end{equation}
Here, we defined the response function
\begin{equation}
    \chi_\eta(t,s):=
    \frac{\tr[\{P-\<P(t)\>_\eta\identity,\,G_\eta(t,s)V(s)\sigma_\eta(s,t)\}_\sim]}
         {\tr[\sigma_\eta(t,t)]},
\end{equation}
with $\chi_\eta(t,s)=0$ for $s>t$.
For the unnormalized quadratic form, the product rule also implies
\begin{equation}\label{eq:kubo-exact-unnormalized}
\begin{split}
 \<v_\zeta(t),\,P v_\zeta(t)\>-\<v_H(t),\,P v_H(t)\>
 ={}&\int_0^\zeta\D\eta\,\int_0^t\D s\,\\
 &\tr[\{P,\,G_\eta(t,s)V(s)\sigma_\eta(s,t)\}_\sim].
\end{split}
\end{equation}
This identity holds for every real $\zeta$, including when a solution vector vanishes.

Next, we linearize $\chi_\eta(t,s)$ around $\eta=0$.
With $\sigma(s,t)=v_H(s)v_H(t)^\dagger$, we obtain
\begin{equation}
    \chi_\eta(t,s)=\chi(t,s)+O(|\eta|),
    \quad
    \chi(t,s)=\frac{\tr[\{P-\<P(t)\>_0\identity,\,G_H(t,s)V(s)\sigma(s,t)\}_\sim]}
                   {\tr[\sigma(t,t)]}.
\end{equation}
Substitution in the exact response yields
\begin{equation}
    \<P(t)\>_\zeta-\<P(t)\>_0
    =\zeta\int_0^t\D s\,\chi(t,s)+R_P(t,\zeta).
\end{equation}
We now bound the Taylor remainder $R_P$ uniformly on $[0,T]$.

Let $(\mc G_H v)(t) = \int_0^t\D s\, G_H(t,s) V(s)v(s)$ and $v_H(t) = G_H(t,0)v_0+\int_0^t\D s\, G_H(t,s)b(s)$. Then, the exact dynamics can also be written implicitly as 
\begin{equation}
    v_\zeta(t)=v_H(t)+\zeta(\mc G_H v_\zeta)(t).
\end{equation}
We use this notion to write $v_\zeta$ as a Taylor series, $v_\zeta = \sum_{k\ge 0} \zeta^k \mc G_H^k v_H$.
Then, define $M\equiv \max_{0\le s \le t \le T}\norm{G_H(t,s)}$, $B\equiv \norm{v_0} + \int_0^T\D s\, \norm{b(s)}$ and $L\equiv \int_0^T\D s\, \norm{V(s)}$, which all are assumed to be finite.
We write $\|v\|_{\max}:=\max_{0\le t\le T}\|v(t)\|$.
Then $\|v_H\|_{\max}\le MB$, and the ordered integration region implies
\begin{equation}\label{eq:kubo-propagator-bound}
    \norm{\mc G^k_H v_H}_{\max}\le MB \frac{(ML)^k}{k!},
\end{equation}
so that the Taylor series above converges uniformly on $[0,T]$. 

We want to show that the residual term in a linearization of $v_\zeta$ is $O(\abs{\zeta}^2)$. To that end, consider
\begin{equation}
    v_\zeta(t) = v_H(t) + \zeta v_1(t) + r_\zeta(t),
\end{equation}
where $v_1(t)$ is the first-order term in $\zeta$, $v_1(t)=\int_0^t\D s\,G_H(t,s)V(s)v_H(s)=(\mc G_H v_H)(t)$.
Then, using \cref{eq:kubo-propagator-bound} and the definitions of $M,B,L$ above, we have that 
\begin{align}
    \norm{v_\zeta - v_H}_{\max} &\le MB(\e^{\abs{\zeta}ML}-1) \\
    \norm{r_\zeta}_{\max} &\le MB(\e^{\abs{\zeta}ML}-1 - \abs{\zeta}ML) \notag\\
    &\le \frac{\zeta^2}{2}M^3BL^2\e^{\abs{\zeta}ML}.
\end{align}
Further, set $C_0=MB$ and $a=ML$.
For $a>0$, we choose
\[
 \rho=\frac1a\log\left(1+\frac{v_{\min}}{2C_0}\right).
\]
Then $C_0(\e^{\rho a}-1)=v_{\min}/2$, so the preceding state estimate implies $\|v_\eta(t)\|\ge v_{\min}/2$ for $0\le t\le T$ and $|\eta|\le\rho$.
If $a=0$, the perturbation vanishes and no restriction on $\rho$ is needed.
Differentiating the convergent series also implies $\|\partial_\eta v_\eta\|_{\max}\le C_0a\e^{\rho a}$ and $\|\partial_\eta^2v_\eta\|_{\max}\le C_0a^2\e^{\rho a}$.
For the desired guarantee, consider $\abs{\partial^2_\eta \<P(t)\>_\eta}$, 
\begin{equation}
    \abs{\partial^2_\eta \<P(t)\>_\eta}
    \le20\norm{P}\frac{\norm{\partial_\eta v_\eta(t)}^2}{\norm{v_\eta(t)}^2}
    +4\norm{P}\frac{\norm{\partial_\eta^2v_\eta(t)}}{\norm{v_\eta(t)}}.
\end{equation}
Thus, by Taylor's theorem with $\lvert\zeta\rvert\le\rho$,
\begin{equation}
    \max_{0\le t\le T}|R_P(t,\zeta)|
    \le\|P\|\left(
       \frac{40C_0^2a^2\e^{2\rho a}}{v_{\min}^2}
       +\frac{4C_0a^2\e^{\rho a}}{v_{\min}}
    \right)\zeta^2.
\end{equation}

\end{proof}

\section{Proofs of Approximation Guarantees}
\subsection{Proof of \cref{lem:error-infeasible-dissipative}}\label{app:proof-error-infeasible-dissipative}
\begin{proof}
We have $v(t)=\e^{At}v(0)$ with $A=A_0-\imag\lambda\projc$ and $\projc v(0)=0$.
Thus,
\begin{equation}\label{eq:bound-on-initial-energy}
 |\<v(0),\,Av(0)\>|=|\<v(0),\,A_0v(0)\>|
 \le\|A_0\|\|v(0)\|^2=:E_0.
\end{equation}
Since $\Re(A)=\Re(A_0)\preceq0$, we have $\|\e^{At}\|\le1$ for $t\ge0$.
We also have $Av(0)=A_0v(0)$, so commutation of $A$ with its exponential implies
\begin{equation}
 |\<v(t),\,Av(t)\>|=|\<\e^{At}v(0),\,\e^{At}A_0v(0)\>|
 \le\|\e^{At}v(0)\|\,\|\e^{At}A_0v(0)\|\le E_0.
\end{equation}
We apply the reverse triangle inequality to
\begin{equation}\label{eq:lem-normalbc-reverse-triangle-2}
 \<v(t),\,Av(t)\>=\<v(t),\,A_0v(t)\>-\imag\lambda\<v(t),\,\projc v(t)\>.
\end{equation}
Using $\|v(t)\|\le\|v(0)\|$ and $\projc^2=\projc$, we obtain
\begin{equation}\label{eq:bound-on-first-error-lemma}
 \|\projc v(t)\|^2=\<v(t),\,\projc v(t)\>
 \le\frac{E_0+|\<v(t),\,A_0v(t)\>|}{\lambda}
 \le\frac{2\|A_0\|\|v(0)\|^2}{\lambda}.
\end{equation}
Taking the square root proves the claim.
\end{proof}

\subsection{Proof of \cref{lem:error-infeasible-dissipative-inhomo}}\label{app:proof-error-infeasible-dissipative-inhomo}
\begin{proof}
We apply the estimate proved in \cref{app:proof-error-infeasible-inhomo-dissipative-time} with constant $A_0$.
Here, $V_T=0$, $a_T=\|A_0\|$, and $K_T=2\|A_0\|+2T\|A_0\|^2$.
It follows that
\begin{align*}
 \max_{0\le t\le T}\|\projc v(t)\|
 &\le\frac1\lambda\Biggl[
 C_TK_T+2\max_{0\le t\le T}\|\projc b(t)\|\\
 &\quad+\int_0^T\D s\,\left(\left\|\projc\dd{s}{b(s)}\right\|+\|A_0\|\|\projc b(s)\|\right)
 \Biggr]
 =\frac{E_T}{\lambda}.
\end{align*}
This proves the stated choice of $\lambda$.
\end{proof}

\subsection{Proof of \cref{lem:error-feasible}}\label{app:proof-error-feasible}
\begin{proof}
We decompose the solution as $v_\perp(t)=\proj v(t)$ and $v_c(t)=\projc v(t)$.
The two components satisfy
\begin{align}
 \ddt{v_\perp(t)}&=\proj A_0\proj v_\perp(t)+\proj A_0\projc v_c(t),
 &v_\perp(0)&=v_0,\\
 \ddt{v_c(t)}&=(\projc A_0\projc-\imag\lambda\projc)v_c(t)+\projc A_0\proj v_\perp(t),
 &v_c(0)&=0.\label{eq:doubleUCdefinition}
\end{align}
Duhamel's principle implies
\begin{equation}
 v_\perp(t)-\e^{\proj A_0\proj t}v_0
 =\int_0^t\D s\,\e^{\proj A_0\proj(t-s)}\proj A_0\projc v_c(s).
\end{equation}
The propagator on the feasible space is contractive because $\Re(A_0)\preceq0$.
Consequently,
\begin{align*}
 \|v_\perp(t)-\e^{\proj A_0\proj t}v_0\|
 &\le\|\proj A_0\projc\|\int_0^t\D s\,\|v_c(s)\|\\
 &\le T\|A_0\|\|v_0\|\sqrt{\frac{2\|A_0\|}{\lambda}},
\end{align*}
where we used \cref{lem:error-infeasible-dissipative}.
\end{proof} 

\subsection{Proof of \cref{lem:error-infeasible-dissipative-time}}\label{app:proof-error-infeasible-dissipative-time}
\begin{proof}
We apply the exact response estimate proved in \cref{app:proof-error-infeasible-inhomo-dissipative-time} with $b=0$.
Then $C_T=\|v(0)\|$, the terms involving $\projc b$ vanish, and
\begin{equation}
 \max_{0\le t\le T}\|\projc v(t)\|
 \le\frac{\|v(0)\|}{\lambda}\left(2a_T+V_T+2Ta_T^2\right)
 =\frac{\|v(0)\|K_T}{\lambda}.
\end{equation}
The stated choice of $\lambda$ follows.
\end{proof}

\subsection{Proof of \cref{lem:error-infeasible-inhomo-dissipative-time}}\label{app:proof-error-infeasible-inhomo-dissipative-time}
\begin{proof}
We prove the estimate under the more general bound $\tfrac12(A_0(t)+A_0(t)^\dagger)\preceq\gamma$ for $\gamma\ge0$.
The dissipative case follows with $\gamma=0$.
We apply \cref{prop:generalized-kubo} with
\begin{align}
 H(t)&=\proj A_0(t)\proj+\projc A_0(t)\projc-\imag\lambda\projc,\\
 V(t)&=\proj A_0(t)\projc+\projc A_0(t)\proj.
\end{align}
For $0\le\eta\le1$, let $v_\eta$ solve $\ddt{v_\eta(t)}=(H(t)+\eta V(t))v_\eta(t)+b(t)$ with $v_\eta(0)=v_0$.
Here, $v_1=v$, while $\proj v_H$ solves the dynamics restricted to the feasible space.
The forcing $b(t)$ is the same for all $\eta$.
Using $P=\projc$ in \cref{eq:kubo-exact-unnormalized}, we obtain the exact response
\begin{equation}\label{eq:penalty-exact-kubo}
\begin{split}
 \|\projc v(t)\|^2-\|\projc v_H(t)\|^2
 ={}&\int_0^1\D\eta\,\int_0^t\D s\,\\
 &\tr[\{\projc,\,G_\eta(t,s)V(s)\sigma_\eta(s,t)\}_\sim].
\end{split}
\end{equation}
We estimate this response through the equations for the projected solutions.
For $0\le\eta\le1$, the Hermitian part of $H+\eta V$ is bounded by $\gamma$, being a convex combination of the Hermitian parts of $A_0$ and $\proj A_0\proj+\projc A_0\projc$.
Consequently,
\begin{equation}
 \|v_\eta(t)\|\le\e^{\gamma t}\left(\|v_0\|+\int_0^t\D s\,\|b(s)\|\right),
 \quad\|G_H(t,s)\|\le\e^{\gamma(t-s)}.
\end{equation}
The projected state equation has the exact form
\begin{equation}\label{eq:penalty-projected-response}
 \projc v_\eta(t)
 =\int_0^t\D s\,G_H(t,s)\projc b(s)
 +\eta\int_0^t\D s\,G_H(t,s)\projc A_0(s)\proj v_\eta(s).
\end{equation}
In particular, the first integral is $\projc v_H(t)$, which accounts for the forcing on the infeasible space.
We use
\[
 \partial_sG_H(t,s)\projc
 =G_H(t,s)\bigl(\imag\lambda\projc-\projc A_0(s)\projc\bigr)
\]
to integrate both terms in \cref{eq:penalty-projected-response} by parts.
This implies
\begin{align}
 \projc v_\eta(t)
 ={}&\frac1{\imag\lambda}
 \left[G_H(t,s)\bigl(\projc b(s)+\eta\projc A_0(s)\proj v_\eta(s)\bigr)\right]_{s=0}^{s=t}\nonumber\\
 &-\frac1{\imag\lambda}\int_0^t\D s\,G_H(t,s)
 \Biggl[\projc\dd{s}{b(s)}-\projc A_0(s)\projc b(s)\nonumber\\
 &\quad+\eta\projc\dd{s}{A_0(s)}\proj v_\eta(s)
 +\eta\projc A_0(s)\proj\dd{s}{v_\eta(s)}
 -\eta\projc A_0(s)\projc A_0(s)\proj v_\eta(s)\Biggr].
\end{align}
The projected derivative satisfies
\begin{equation}
 \left\|\proj\dd{s}{v_\eta(s)}\right\|
 \le a_T\|v_\eta(s)\|+\|\proj b(s)\|.
\end{equation}
Indeed, its matrix term is $\proj A_0(s)(\proj+\eta\projc)v_\eta(s)$ and $\|\proj+\eta\projc\|\le1$.
The endpoint terms involving $v_\eta$ contribute at most
$\eta\e^{\gamma t}a_T(\|v_0\|+C_T)/\lambda$.
Their derivative terms contribute at most
$\eta\e^{\gamma t}[(V_T+2Ta_T^2)C_T+a_T\int_0^T\D s\,\|b(s)\|]/\lambda$.
Using $C_T=\|v_0\|+\int_0^T\D s\,\|b(s)\|$, we obtain
\begin{align}
 \|\projc v_\eta(t)\|
 \le\frac{\e^{\gamma t}}{\lambda}\Biggl[
 &\eta C_TK_T+2\max_{0\le s\le T}\|\projc b(s)\|\nonumber\\
 &+\int_0^T\D s\,\left(\left\|\projc\dd{s}{b(s)}\right\|+a_T\|\projc b(s)\|\right)
 \Biggr].\label{eq:penalty-uniform-eta-bound}
\end{align}
Setting $\eta=1$ in \cref{eq:penalty-uniform-eta-bound} and using the definition of $E_T$ allows us to conclude that
\begin{equation}
 \max_{0\le t\le T}\|\projc v(t)\|
 \le\frac{E_T}{\lambda}.
\end{equation}
This also proves the constraint bound in \cref{result:error-new-kubo} for $\gamma\ge0$.
\end{proof}

\subsection{Proof of \cref{lem:error-feasible-time}}\label{app:proof-error-feasible-time}
\begin{proof}
We write $v_\perp(t)=\proj v(t)$ and $v_c(t)=\projc v(t)$.
Then,
\begin{align}
 \ddt{v_\perp(t)}&=\proj A_0(t)\proj v_\perp(t)+\proj A_0(t)\projc v_c(t)+\proj b(t),
 &v_\perp(0)&=v_0,\\
 \ddt{v_c(t)}&=(\projc A_0(t)\projc-\imag\lambda\projc)v_c(t)
 +\projc A_0(t)\proj v_\perp(t)+\projc b(t),
 &v_c(0)&=0.\label{eq:doubleUCdefinition-time}
\end{align}
Duhamel's principle implies
\begin{equation}
 v_\perp(t)-v_{\perp,\rm h}(t)-v_{\perp,\rm p}(t)
 =\int_0^t\D s\,\mathcal T\exp\!\left(\int_s^t\D r\,\proj A_0(r)\proj\right)\proj A_0(s)\projc v_c(s).
\end{equation}
For $\tfrac12(A_0(t)+A_0(t)^\dagger)\preceq\gamma$, the propagator on the feasible space has norm at most $\e^{\gamma(t-s)}$.
Using the pointwise bound \cref{eq:penalty-uniform-eta-bound} at $\eta=1$, we have $\norm{v_c(s)}\le \e^{\gamma(s-T)}E_T/\lambda$ and obtain
\begin{align*}
 \|v_\perp(t)-v_{\perp,\rm h}(t)-v_{\perp,\rm p}(t)\|
 &\le a_T\int_0^t\D s\,\e^{\gamma(t-s)}\|v_c(s)\|\\
 & \le \frac{ta_T\e^{\gamma(t-T)}E_T}{\lambda}
 &\le\frac{Ta_TE_T}{\lambda}.
\end{align*}
This proves the lemma with $\gamma=0$.
Moreover, $u=v_{\perp,\rm h}+v_{\perp,\rm p}$, so the triangle inequality and \cref{eq:penalty-uniform-eta-bound} imply
\begin{equation}
 \max_{0\le t\le T}\|v(t)-u(t)\|
 \le\frac{(1+Ta_T)E_T}{\lambda}.
\end{equation}
This proves the solution bound in \cref{result:error-new-kubo}.
\end{proof}

\clearpage
\section{Analysis of smoothness factors in LCHS discretization}\label{app:analysis-of-smoothness}
We aim to derive the bounds for $\Lambda_I$ and $\Xi_I$ in \cref{lem:smoothness-factors}.
Recall that
\begin{equation}
    A_I(t)=\e^{\imag\lambda\projc t}A_0(t)\e^{-\imag\lambda\projc t},\quad
    b_I(t)=\e^{\imag\lambda\projc t}b(t).
\end{equation}
By the product rule, as derived in \cref{app:deriving-ipic-derivs},
\begin{align}
    \partial_t^p A_I(t)
    &=\sum_{q=0}^p\binom pq {\rm ad}_{\imag\lambda\projc}^q
      \big((\partial_t^{p-q}A_0(t))_I\big),\label{eq:interaction-pic-matrix-derivatives}\\
    \partial_t^p b_I(t)
    &=\sum_{q=0}^p\binom pq(\imag\lambda\projc)^q
      (\partial_t^{p-q}b(t))_I.\label{eq:interaction-pic-derivatives}
\end{align}
Here ${\rm ad}_y(x)=[y,x]$, ${\rm ad}_y^q$ denotes $q$ nested commutators, and the zeroth powers in both formulas are the identity.
Since unitary conjugation preserves the norm and $\|[\projc,X]\|\le2\|X\|$,
\begin{equation}\label{eq:intermediate-getting-Lambda-I-for-timedependent}
    \|\partial_t^p A_I(t)\|
    \le\sum_{q=0}^p\binom pq(2\lambda)^q\|\partial_t^{p-q}A_0(t)\|.
\end{equation}
The binomial theorem now implies
\begin{equation}
    \|\partial_t^p A_I(t)\|^{1/(p+1)}
    \le2\lambda+\sup_{r\ge0}\max_{0\le s\le T}\|\partial_s^r A_0(s)\|^{1/(r+1)}.
\end{equation}
To see this, each derivative of order $p-q$ in \cref{eq:intermediate-getting-Lambda-I-for-timedependent} is bounded by the $(p-q+1)$st power of the displayed supremum.
The sum is therefore bounded by that supremum times its sum with $2\lambda$, raised to the $p$th power.
Taking the $(p+1)$st root proves the claim.
Taking the supremum over $p$ proves the bound for $\Lambda_I$.
For constant $A_0$, only the $q=p$ term remains, so
\begin{equation}
    \Lambda_I\le\sup_{p\ge0}(2\lambda)^{p/(p+1)}\|A_0\|^{1/(p+1)}
    \le\max\{2\lambda,\|A_0\|\}.
\end{equation}

For the source, \cref{eq:interaction-pic-derivatives} similarly implies
\begin{equation}\label{eq:above-other-smoothness-term}
    \|\partial_t^p b_I(t)\|
    \le\sum_{q=0}^p\binom pq\lambda^q\|\partial_t^{p-q}b(t)\|.
\end{equation}
The same binomial argument proves
\begin{equation}
    \Xi_I\le\lambda+\sup_{p\ge0}\max_{0\le t\le T}\|\partial_t^p b(t)\|^{1/(p+1)}.
\end{equation}
For constant $b$, we obtain
\begin{equation}\label{eq:intermediate-Xi-I}
    \Xi_I\le\max\{\lambda,\|b\|\}.
\end{equation}
If $\projc b(t)=0$, the source is unchanged by the interaction picture transformation.
These bounds hold for the chosen $\lambda$; inserting a sufficient penalty from \cref{tab:results-from-approximation-guarantees} then determines the discretization parameters.

\clearpage

\section{A lower bound on the final solution norm for the discrete heat equation}\label{sec:lower-bound-on-norm}
\begin{lemma}\label{lem:lower-bound-on-norm}
Let $v(t)$ satisfy $\ddt v(t)= (A_0-\imag\lambda\projc)v(t)+b$ with $A_0=D\mathbf L_h$ be Hermitian and negative semidefinite and $D>0$.
Further, $\projc v(0)=0$ and define $\alpha:=-\lambda_{\min}(\mathbf{L}_h)$ as well as  $v_{\min}=\min_{0\le t \le T} \norm{v(t)}$.
Then, we have that
\begin{equation}
    v_{\min}(T)\ge \max\l\{0, \, \lVert v(0)\rVert - T\lVert D\mathbf L_h v(0) +b\rVert, \,  \e^{-D\alpha T } \norm{v(0)} - \frac{1-\e^{-D\alpha T}}{D\alpha}\norm{b}  \r\}.
\end{equation}
\end{lemma}

\begin{proof}
Consider 
\begin{equation}
    \frac12 \ddt \norm{v(t)}^2 = \< v(t), \mathbf DL_h v(t) \> + \Re \<v(t), b\>
    \ge -D\alpha \norm{v(t)}^2 - \norm{b}\norm{v(t)}.
\end{equation}
Then, we obtain after integration that 
\begin{equation}
    v_{\rm min} \ge \e^{-D\alpha T } \norm{v(0)} - \frac{1-\e^{-D\alpha T}}{D\alpha}\norm{b}.
\end{equation}
Additionally, consider 
\begin{equation}
    v(t)-v(0)=\int_0^t\D s\, \e^{(D\mathbf L_h-\imag\lambda\projc) s} \bigl( D\mathbf L_h v(0)+b\bigr),
\end{equation}
which gives us the lower bound 
\begin{equation}
    \norm{v(t)}\ge \norm{v(0)} - \lVert v(t)-v(0)\rVert \ge \lVert v(0)\rVert -t\lVert D\mathbf L_hv(0)+b\rVert.
\end{equation}
\end{proof}

\section{Numerical Experiments}\label{sec:numerical-experiments}

In the following, we are presenting numerical experiments as proof-of-concept  validation of the penalty projections method via interaction picture based on classical simulation of the dynamics. In addition, we are interested in tightness of our bounds on $\lambda$ derived in \cref{subsec:constraint-error-bounds}.
To that end, we consider the heat equation  and the wave equation as two of the canonical examples for `simple' PDEs.
While for the heat equation, a quantum speedup beyond a quadratic one coming from amplitude amplification cannot be expected in general when evaluating an expectation value with respect to the final state (see~\cite{linden2022quantum}), it still serves as an illustrative example. 
However, as shown in \cite{costa2019quantum} and recovered as the isotropic case in \cite{babbush2023exponential}, the wave equation, which in contrast to the heat equation is not dissipative and obeys conservation of energy, allows for an at least quartic speedup.

\subsection{Setup}
In the computations below, we restricted the possible range of $\lambda$ to `small' values (up to $10^6$) to avoid aliasing in the numerical solution of the highly oscillatory system in interaction picture. Note that this is not as much of a problem in the quantum implementations, as we will outline below, as the necessary time discretization only grows with $\log(\lambda)$. 
For spatial discretization, we use a simple central three-point finite difference formula in periodic boundary conditions form to represent the Laplacian.
We use the Runge-Kutta scheme RK23~\cite{bogacki19893} to simulate the time evolution. Even though this is not high-accuracy, it proves sufficient for the regimes we are looking at; in order to capture high-frequency effects of the interaction picture simulation of $A_I(t)$, a relatively small time-step is already required, hence the accuracy is sufficient here to ensure that the penalty error is the dominant source of error.
We are solving the following discretized PDEs by time-stepping the interaction picture equations from \cref{eq:ipic-equation-two} with respect to the discretized differential operators. 

\subsection{Simulation results}
In what follows, we present numerical experiments for dimensionless heat and wave equations, discretized by finite differences.
\subsubsection{Heat equation with Dirichlet and Neumann boundary conditions}
We consider the following 2D-problem for the heat equation:
\begin{align}
    \partial_t {u}(x,y;t) &=  D  \Delta u(x,y;t)   + f(x,y;t) \\
    u(x,y;t)|_{x,y\in\Gamma_D} &= g(x,y;t)\\
    \partial_n\cdot u(x,y;t)|_{x,y\in\Gamma_N} &= h(x,y;t)\\
    u(x,y;t=0) &= u_0(x,y).
\end{align}
with the temperature distribution $u(x,y;t)$ and diffusion coefficient $D>0$; for the sake of simplicity, we choose  isotropic diffusion here. The initial conditions are described by $u_0$, $f$ is a forcing term (heat source) and $g$ describes the temperature along the boundary $\Gamma_D$, where $h$ describes the temperature flux across the boundary $\Gamma_N$.
Then, we introduce an equidistant uniform grid for space to obtain $\mathbf{u}_h(t)\in \reals^{N^2}$ using $N$ points to represent $x$ and $y$, and a fourth-order finite difference stencil to approximate $\Delta$ via $\mathbf{L}_h\in\reals^{N^2\times N^2}$ with truncation error $O(N^{-4})$.
Then, we have the discretized ODE system
\begin{equation}
    \ddt \mathbf{u}_h(t)  = D \mathbf{L}_h \mathbf{u}_h(t) + \mathbf{f}(t),
\end{equation}
and use the constraint projections as defined earlier in \cref{eq:def-dirichlet-projection,eq:projector-for-neumann} and use the homogenization strategies outlined in \cref{subsec:discretized-pdes-discussion} for non-zero boundary conditions.

\begin{figure}
    \centering
    \begin{subfigure}[c]{.38\textwidth}
    \centering
    \includegraphics[width=\textwidth]{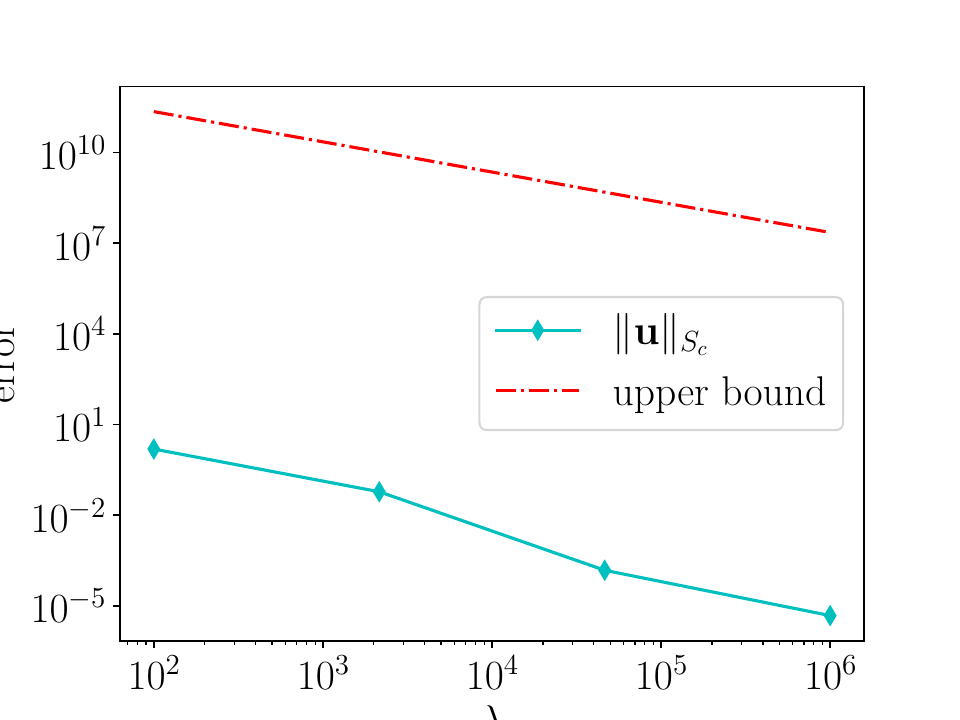}
    \vfill
    \caption{Boundary error $\norm{\mathbf{u}_h}_{\domcomp}$ vs penalty parameter; upper bound follows \cref{lem:error-infeasible-dissipative-inhomo}}
    \end{subfigure}
    \hfill
    \begin{subfigure}[c]{.49\textwidth}
    \centering
    \includegraphics[width=\textwidth]{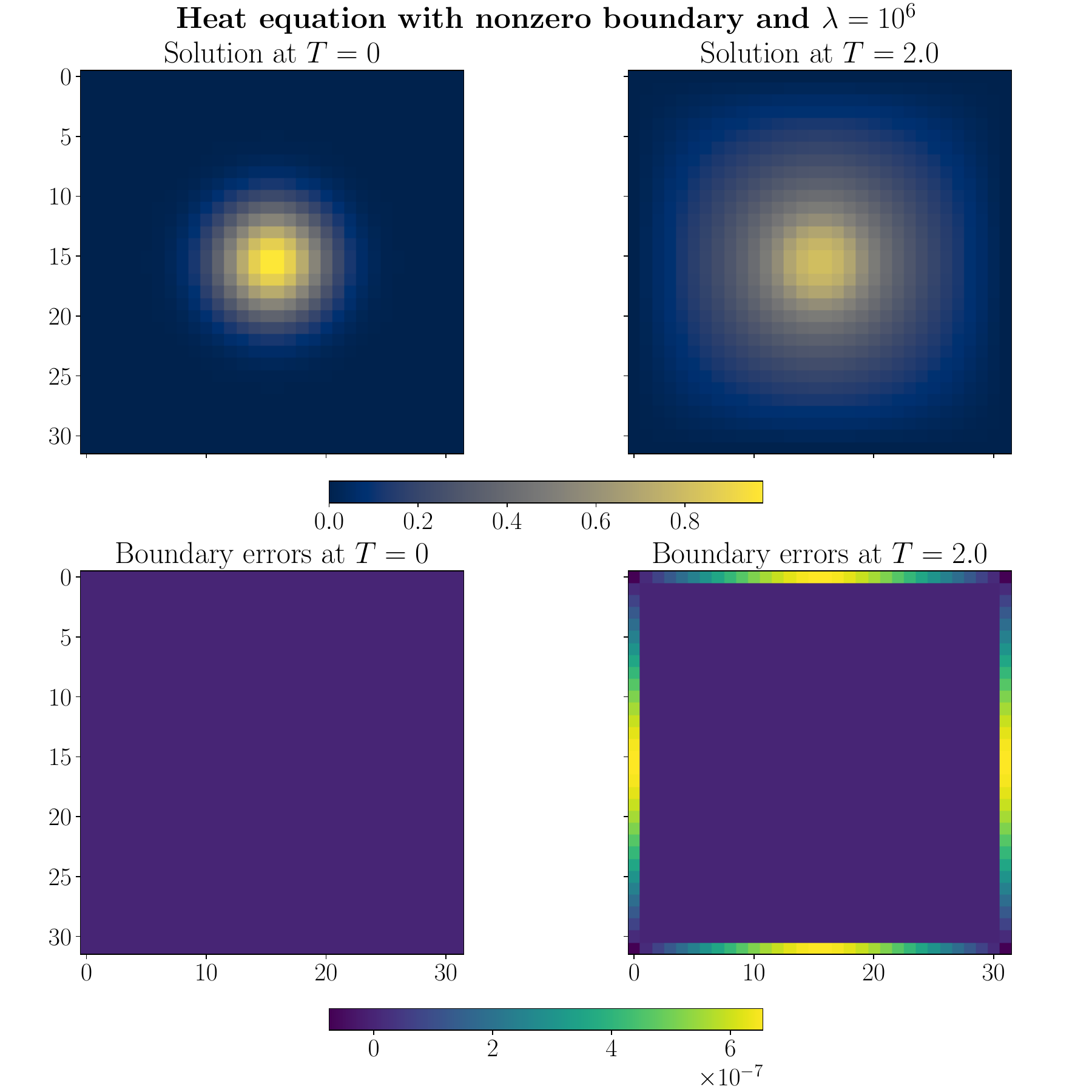}
    \caption{Solution and boundary values at initial vs final time.}
    \end{subfigure}
    \caption{Numerical simulations of heat equation with $\projcD \mathbf{u}_h = 0$, $D=4$, $N=2^5$, $t=1$, and $\Delta t = 10^{-5}$. The boundary is defined as the outer grid points (`wall').}
    \label{fig:numerics-zero-bc-errors-heat}
\end{figure}

\begin{figure}
    \centering
     \begin{subfigure}[c]{.38\textwidth}
    \centering
    \includegraphics[width=\linewidth]{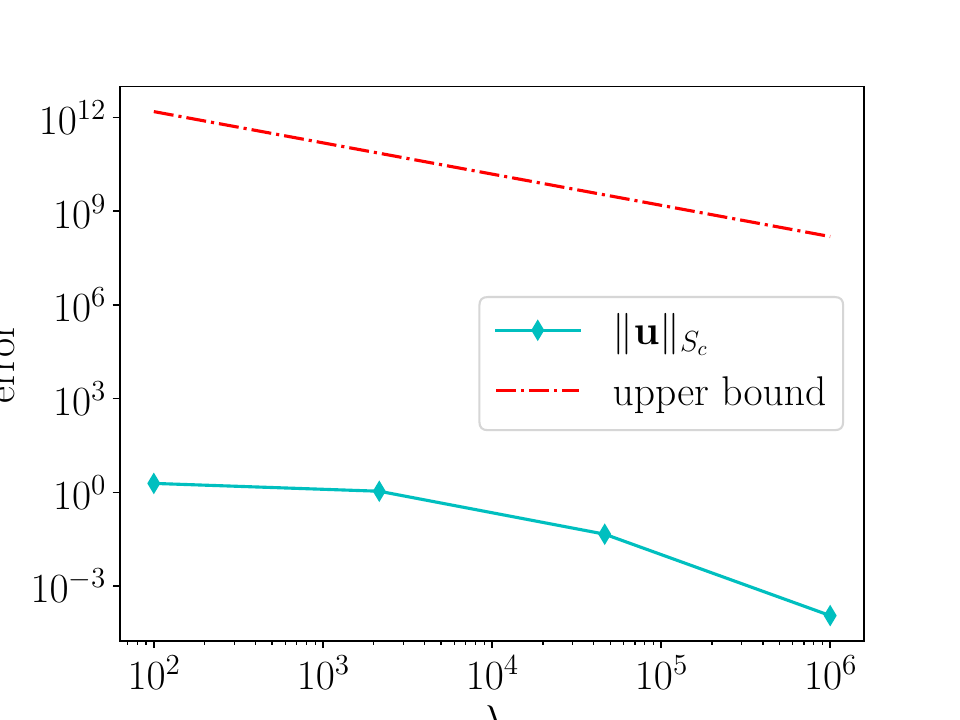}
    \caption{Boundary error $\norm{\mathbf{u}_h}_{\domcomp}$ for $N=2^5, \Delta t = 10^{-5}$, $D=4$ and boundary along the four walls upper bound follows \cref{lem:error-infeasible-dissipative-inhomo}.}
    \end{subfigure}
    \hfill
    \begin{subfigure}[c]{.49\textwidth}
    \centering
    \includegraphics[width=\linewidth]{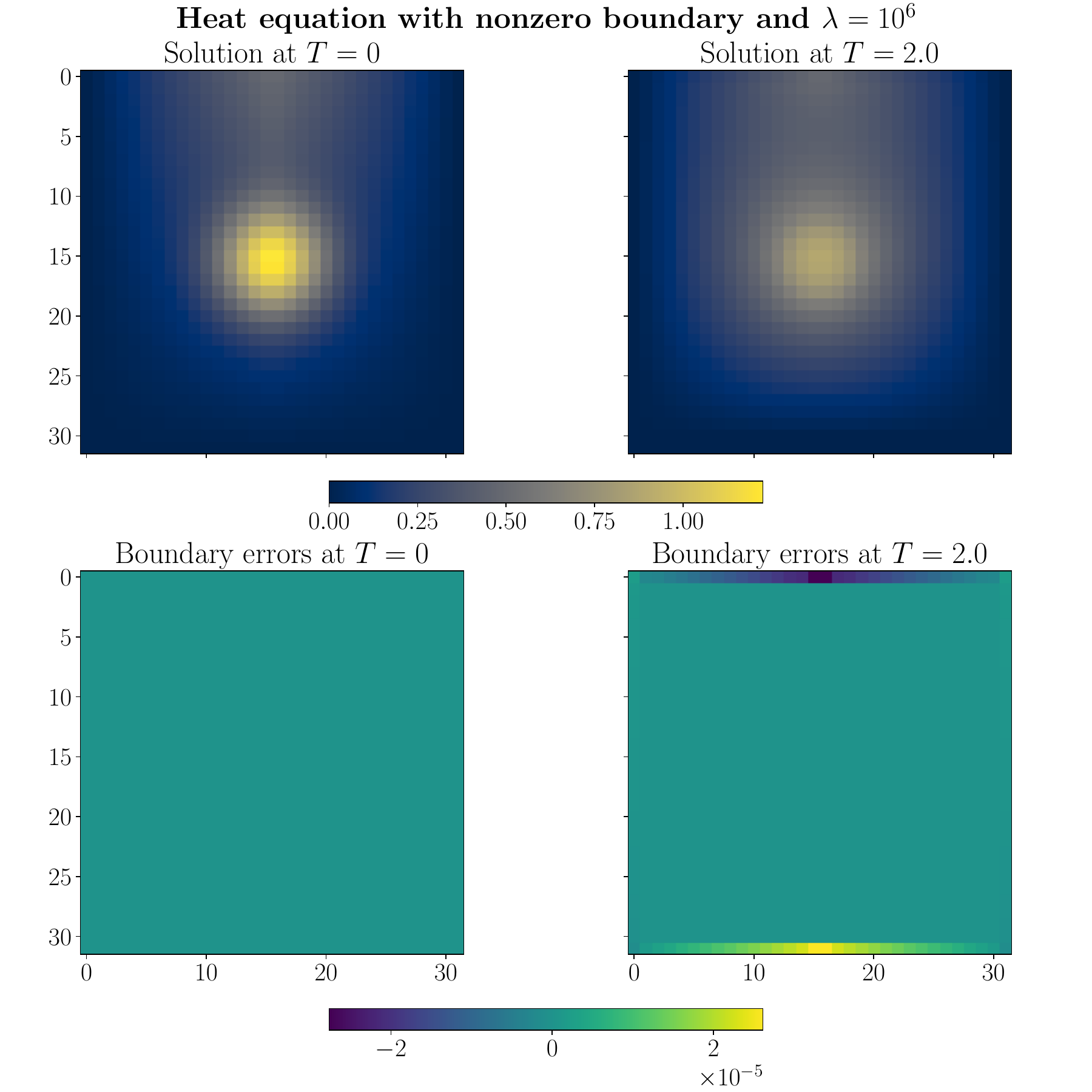}
    \caption{Solution and boundary values at initial vs final time.}
    \end{subfigure}
    \caption{Numerical simulations of heat equation with $\projcD \mathbf{u}_h = \mathbf{g}$ from \cref{eq:g-numerics-defn}, $D=4$, $N=2^5$, $t=1$, and $\Delta t = 10^{-5}$. The boundary is defined to be the outer grid points (`wall').}
    \label{fig:numerics-nonzero-bc-errors-heat}
\end{figure}

\begin{figure}
    \centering
     \begin{subfigure}[c]{.38\textwidth}
    \centering
    \includegraphics[width=\linewidth]{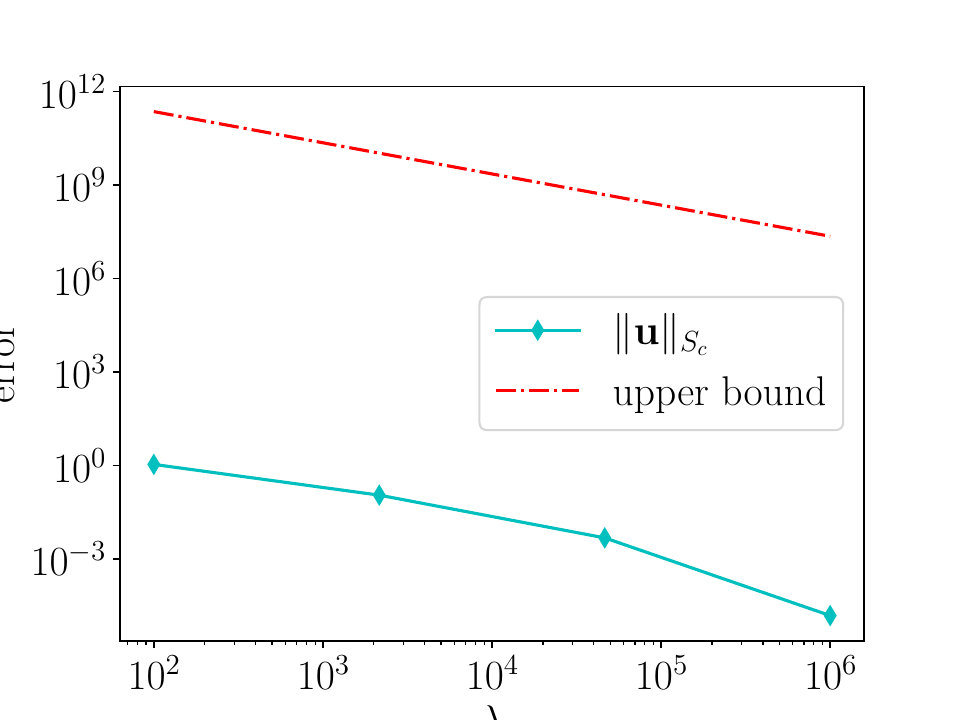}
    \caption{Boundary error $\norm{\mathbf{u}_h}_{\domcomp}$ for $N=2^5, \Delta t = 10^{-5}$, $D=4$ and boundary along a circle boundary \cref{lem:error-infeasible-dissipative}}
    \end{subfigure}
    \hfill
     \begin{subfigure}[c]{.49\textwidth}
    \centering
    \includegraphics[width=\linewidth]{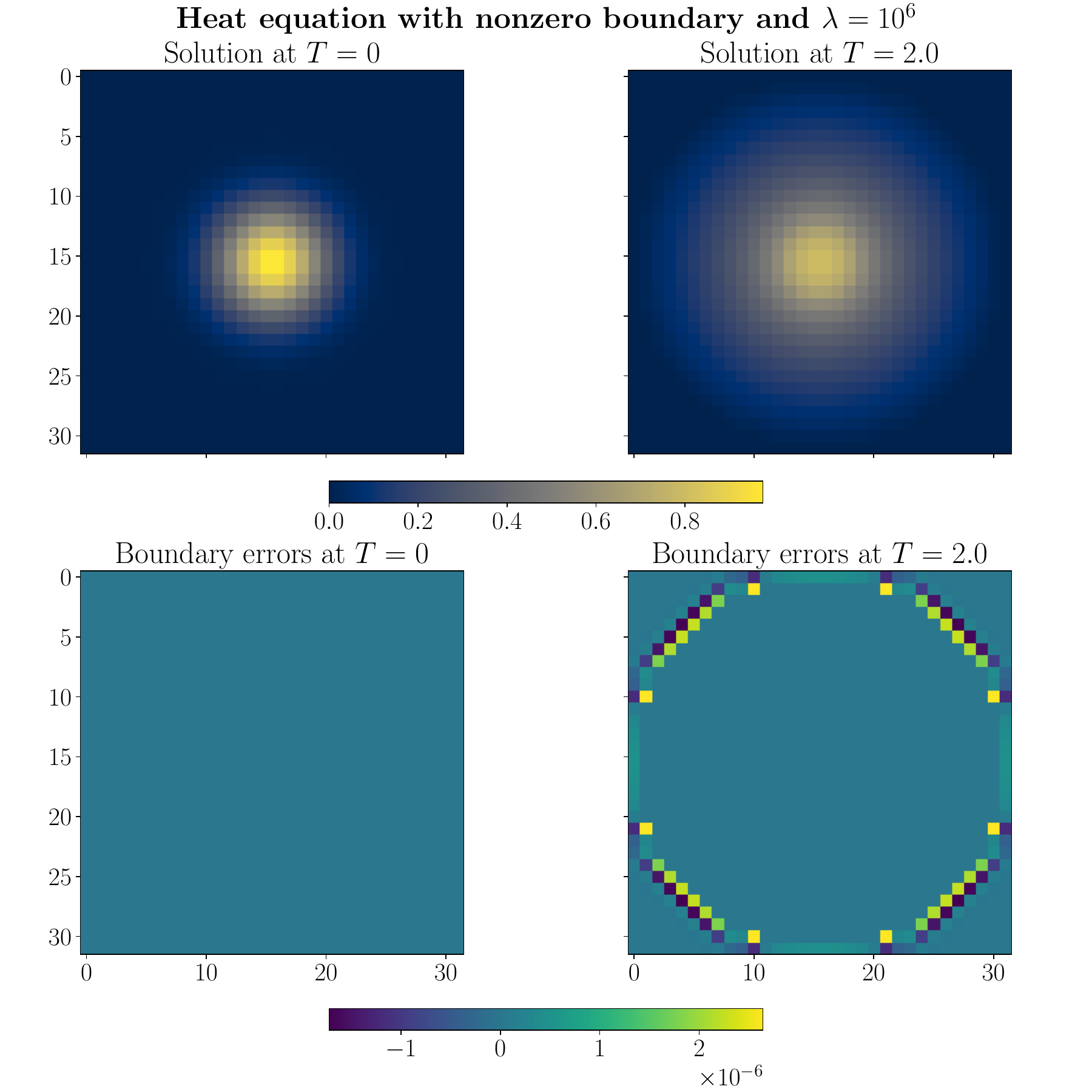}
    \caption{Solution and boundary values at initial vs final time.}
    \end{subfigure}
    \caption{Numerical simulation of heat equation with $\projcD \mathbf u = \mathbf{g}$ from \cref{eq:g-numerics-defn}, $D=4$, $N=2^5$, $t=1$, and $\Delta t = 10^{-5}$. The boundary is defined to be on a circle of radius $\frac{1}{2}$.}
    \label{fig:numerics-circle-bc-errors-heat}
\end{figure}

\begin{figure}
    \centering
     \begin{subfigure}[c]{.38\textwidth}
    \centering
    \includegraphics[width=\linewidth]{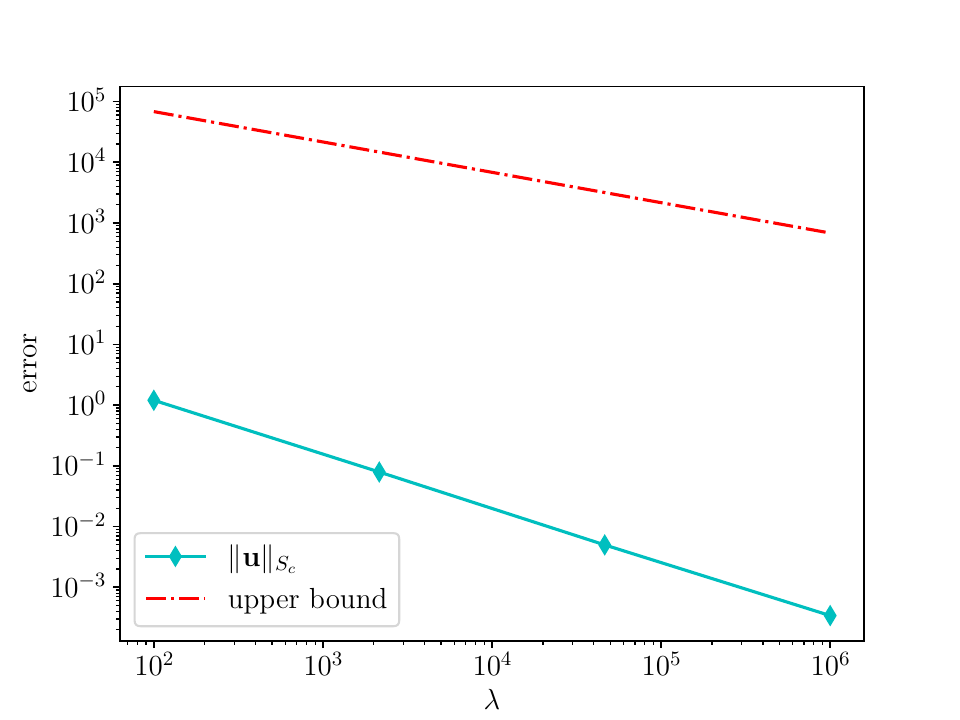}
    \caption{Neumann boundary error $\norm{\mathbf{u_h}}_{\domcomp}$ on wall boundary with respect to `derivative constraint' $\projcN$}
    \end{subfigure}
    \hfill
     \begin{subfigure}[c]{.49\textwidth}
    \centering
    \includegraphics[width=\linewidth]{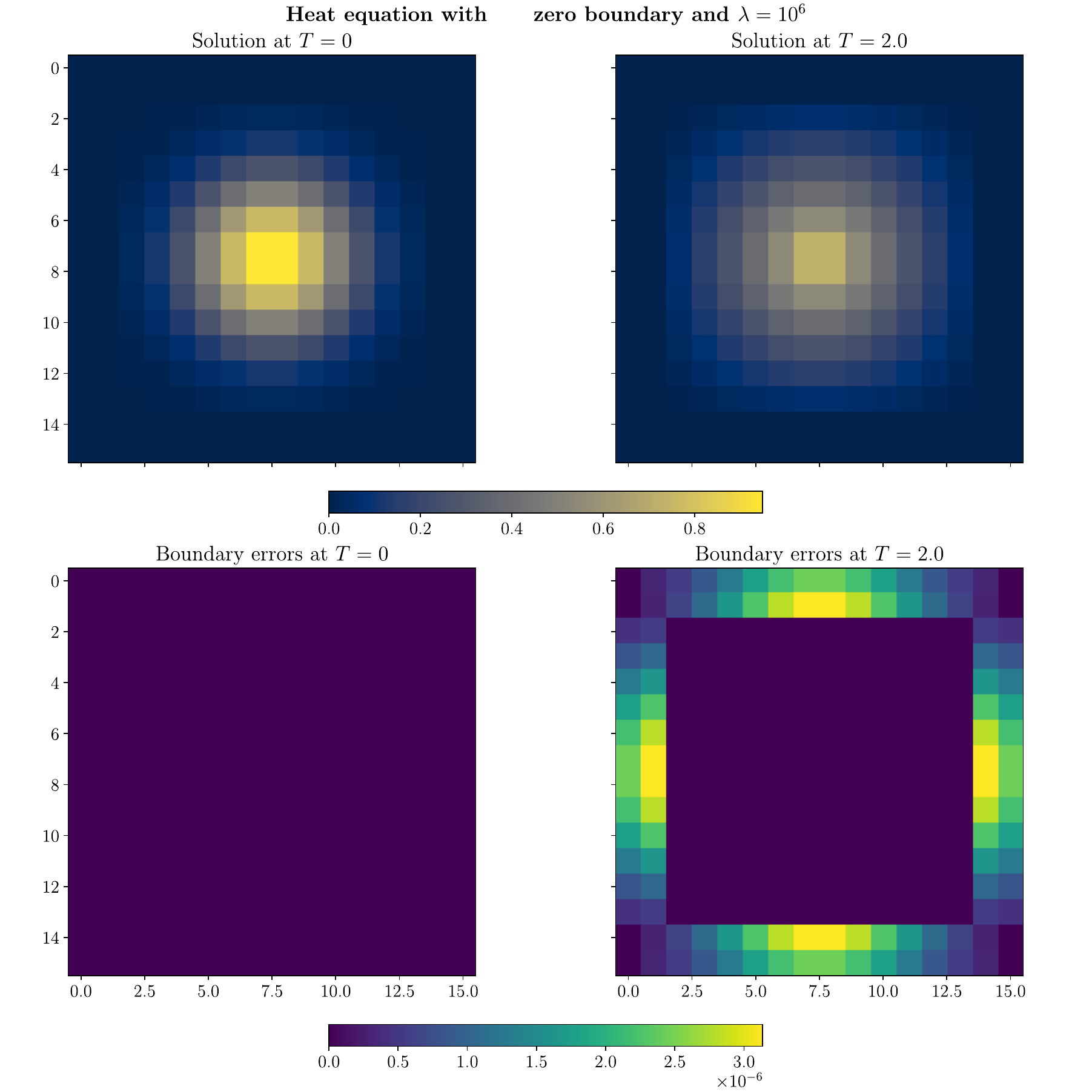}
    \caption{Solution and boundary values at initial vs final time.}
    \end{subfigure}
    \caption{Numerical simulations of heat equation with $\projcN\mathbf{u}_h=0$ from \cref{eq:g-numerics-defn}, $D=4$, $N=2^4$, $t=2$, and $\Delta t = 10^{-3}$. The constraint region is formed by the differences of the outer points (`wall') with their closest neighbours inwards. The corner points as well as one of the points next to the corner are skipped so that every inner point is only neighbour to at most one corner point.}
    \label{fig:numerics-neumann-bc-errors-heat}
\end{figure}

We examine the following choices of boundary conditions:
\begin{itemize}
    \item Vanishing boundary conditions, $\projcD \mathbf{u}_h(t) = 0$. See \cref{fig:numerics-zero-bc-errors-heat,fig:numerics-circle-bc-errors-heat}.
    \item Non-zero boundary conditions, $\projcD\mathbf{u}_h(t) = g(\mathbf{x},\mathbf{y}) $
    \begin{equation}\label{eq:g-numerics-defn}
        \mathbf{g} = 
        (1-\mathbf y)\odot 
    \begin{cases}
       \mathbf x, & \mathbf{x} \le \frac{1}{2} \\  
       (1-\mathbf x), & \mathbf{x} > \frac{1}{2}, 
    \end{cases}
    \end{equation}
    where $\mathbf{x} = [0, \Delta x, 2\Delta x, \ldots,]^T$ and $\mathbf{y}= [0, \Delta y, 2\Delta y, \ldots,]^T$ are the grid representation of the $x,y$ domain and $\odot$ denotes element-wise multiplication. This is implemented via the approach outlined in \cref{subsec:nonzero-dirichlet}.
    Results are shown in \cref{fig:numerics-nonzero-bc-errors-heat}.
    \item No heat in/out-flux: $\projcN\mathbf{u}_h(t) = 0$. The boundary values are zero in compliance to the initial condition. 
    The numerical results are depicted in \cref{fig:numerics-neumann-bc-errors-heat}.
\end{itemize}
All simulations are dimensionless, and as initial state we use a centred, isotropic Gaussian with height 1 and define a cut-off so that the initial condition respects the boundary constraints.
Additionally, we use a point-source $\mathbf{f}(t) = 298$ at the centre element. 


\subsubsection{Wave equation with Dirichlet and Neumann boundary conditions}
\begin{figure}[H]
    \centering
    \begin{subfigure}[c]{.38\textwidth}
    \centering
    \includegraphics[width=\textwidth]{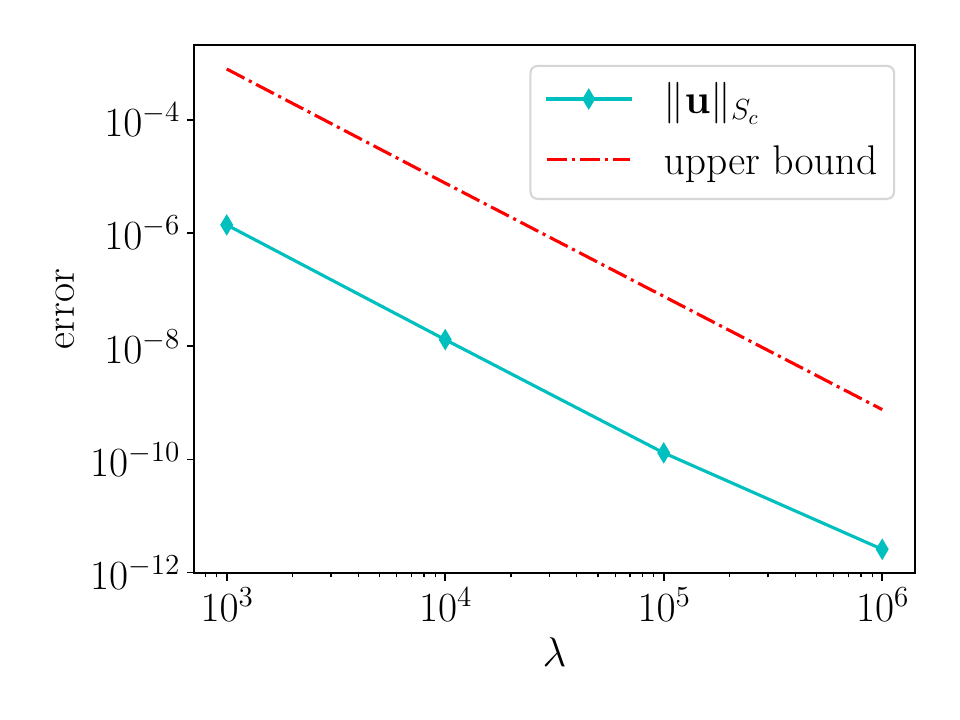}
    \caption{Boundary error $\|\mathbf{u}\|_{\domcomp}$, upper bound follows \cref{lem:state-dep-wave-bound}.}
    \end{subfigure}
    \hfill
    \begin{subfigure}[c]{.49\textwidth}
    \includegraphics[width=\textwidth]{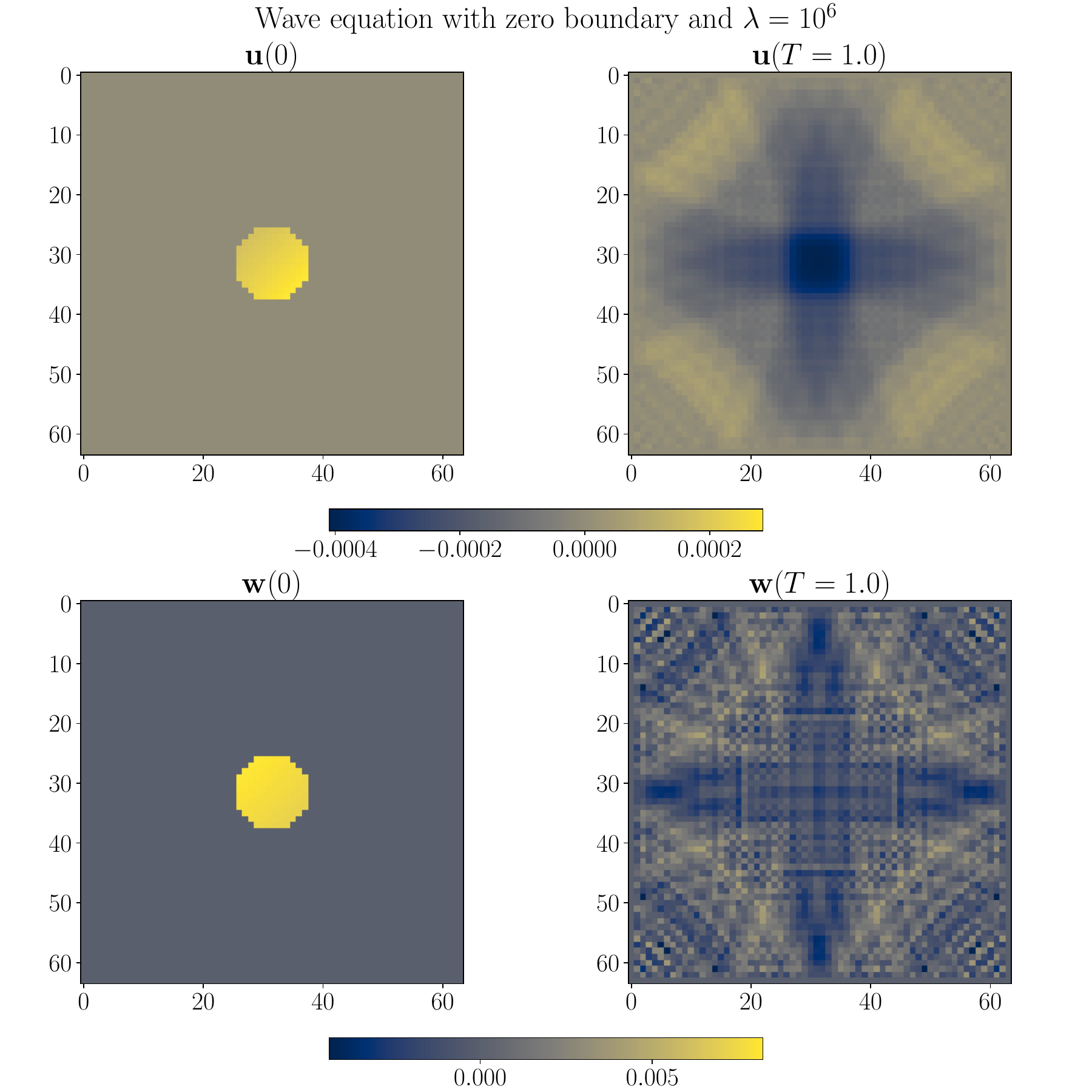}
    \caption{$\mathbf{u}_h$ and $\mathbf{w}_h$ at $t=0$ and final time $t=1$.}
    \end{subfigure}
    \caption{Numerical experiments for wave equation with zero Dirichlet boundary conditions at radius $\frac{1}{2}$. $N=2^6$, $\Delta t = 10^{-5}$, $t=1$, $c^2=1$.}
    \label{fig:numerics-wave}
\end{figure}

The 2-D wave equation is defined as
\begin{equation}\label{eq:wave-eq}
    \partial_{tt} u(x,y;t) = c^2 \Delta u(x,y;t),
\end{equation}
with a speed-of-sound parameter $c^2>0$. 
In order to solve the wave equation, let $w(x,y;t) = \partial_t u(x,y;t)$. 
Then, \cref{eq:wave-eq} is equivalent to the system
\begin{equation}\label{eq:systemized-wave-eq}
    \partial_t \mathbf{v}(t) = \partial_t \begin{bmatrix}
       u(t) \\ w(t) 
    \end{bmatrix} 
    = 
    \begin{bmatrix}
     0 & \identity \\     
     c^2\Delta & 0
    \end{bmatrix} 
    \begin{bmatrix}
       u(t) \\ w(t) 
    \end{bmatrix} = \mathbf{A}  \mathbf{v}(t),
\end{equation}
with initial data $\mathbf{v}(t) = \l[ \begin{smallmatrix}
    u(x,y;t=0)\\ \partial_t u(x,y;t=0)\end{smallmatrix}\r]$.
Upon discretizing space on a uniform, regular grid with $N$ points per spatial dimension, we obtain a matrix $\mathbf{L}_h\in\reals^{N^2\times N^2}$ and overall system matrix $\mathbf{A}_h\in\reals^{2N^2\times 2N^2}$, $\mathbf{v}_h(t) \in\reals^{2N^2}$.
For the constrained system,  
\begin{equation*}
    \ddt \mathbf{v}_h(t) = \l(\mathbf{A}_h - \imag\lambda \mathbf{P}_c \r)\mathbf{v}_h(t)
\end{equation*}
we apply a penalty projection on both the original variable $\mathbf u$ and $\mathbf w$,
$\mathbf{P}_c = \left[\begin{smallmatrix}
        \projc & 0 \\ 0 & \projc
    \end{smallmatrix}\right]$. 
As the value of interest is only $\mathbf{u}(t)$, we measure the error via $\| \mathbf u \|_{\ell_2, \domcomp}^2 = \< \mathbf{v}, \widetilde{\mathbf{P}}_c \mathbf{v}\>$ so that  
$  \widetilde{\mathbf{P}}_c = \left[ \begin{smallmatrix}
        \projc & 0 \\ 0 & 0
    \end{smallmatrix}\right]$.
Note that the generator for the wave equation is not dissipative. With 
\begin{equation}\label{eq:wave-generator-systemized}
    \mathbf{A}_0 = \begin{bmatrix}
            0 & \identity \\ c^2 \mathbf{L}_h & 0 
        \end{bmatrix}
\end{equation}
we get 
\begin{equation}
    \frac12 \l(  \mathbf{A}_0 + \mathbf{A}_0^\dagger  \r)
    = 
    \frac12
 \begin{bmatrix}
            0 & \identity + c^2 \mathbf{L}_h \\   \identity + c^2 \mathbf{L}_h  & 0 
        \end{bmatrix},
\end{equation}
which implies an amplification of $\gamma = \tfrac12\norm{\identity+ c^2\mathbf{L}_h}$ when applying \cref{result:error-new-kubo}. 
The initial condition is given by 
\begin{align}
    \begin{bmatrix}
    \mathbf{v}_0 \\  \mathbf{w}_0
    \end{bmatrix}
     = 
    \begin{bmatrix}
        0.001 \, \sin(\mathbf{x})\sin(\mathbf{y})
        \\
        0.01 \, \cos(\mathbf{x})\cos(\mathbf{y})
    \end{bmatrix} 
    \text{  if  }
    (\mathbf{x}-\tfrac12)^2 + (\mathbf{y}-\tfrac12)^2 \le 0.1, \text{  else  } 0.
\end{align}
The simulation results are shown in \cref{fig:numerics-wave}.

\begin{figure}[H]
    \centering
    \begin{subfigure}[c]{.38\textwidth}
    \centering
    \includegraphics[width=\textwidth]{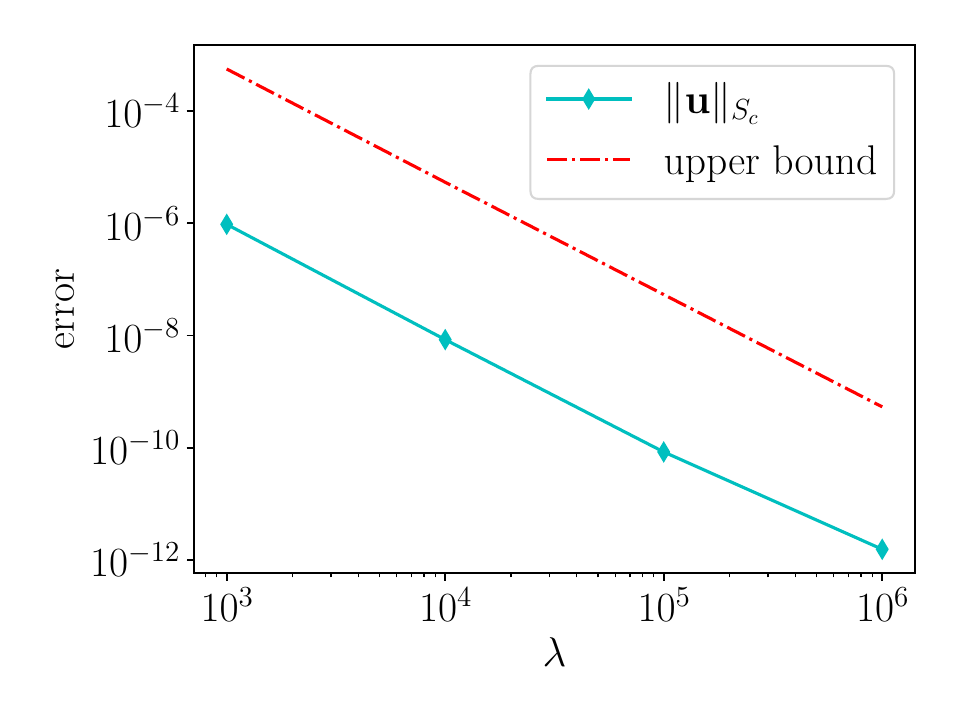}
    \caption{Boundary error $\|\mathbf{u}\|_{\domcomp}$, upper bound follows \cref{lem:state-dep-wave-bound}}
    \end{subfigure}
    \hfill
    \begin{subfigure}[c]{.49\textwidth}
    \includegraphics[width=\textwidth]{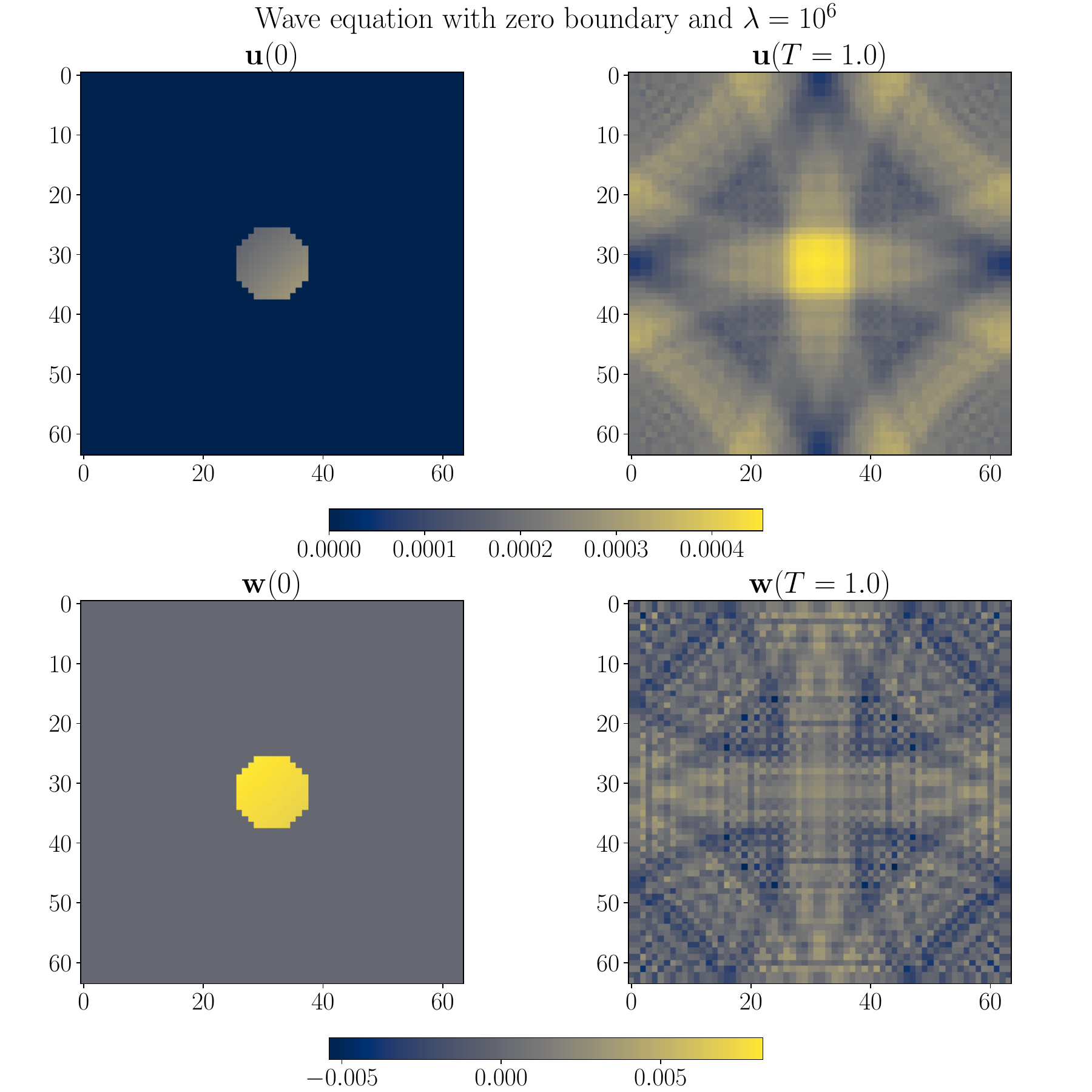}
    \caption{$\mathbf{u}_h$ and $\mathbf{w}_h$ at $t=0$ and final time $1$.}
    \end{subfigure}
    \caption{Numerical experiments for wave equation with zero Neumann boundary conditions. $N=2^6$, $\Delta t = 10^{-5}$, $t=1$, $c^2=1$.}
    \label{fig:numerics-wave-neumann}
\end{figure}

As the wave equation is non-dissipative, the incurred factor $\e^{T\gamma}$ in the upper bound makes the resulting bound rather loose. For that reason, we provide a state-dependent bound in this case to validate our numerical results. 
\begin{lemma}[State-dependent bound on the constraint error for the wave equation]\label{lem:state-dep-wave-bound}
    Consider the wave equation in \cref{eq:systemized-wave-eq}, using the finite difference discretization as in \cref{eq:wave-generator-systemized}. 
We have that $\mathbf K=-c^2\mathbf L_h$ is Hermitian and positive semidefinite.
The orthogonal projection $\projc$ with complement $\proj$ represents the constraint space, with compatible initial conditions $\projc\mathbf{u}(0)=\projc\mathbf{w}(0)=0$.
Further, let $\mathbf B=(\projc \mathbf K \projc)\mid_{\mathrm{ran}(\projc)}$ have minimal eigenvalue $\mu_->0$ and largest eigenvalue $\mu_+$, and choose $\lambda>\sqrt{\mu_+}$.

We get a coupling term from the feasible into the infeasible space $f(t)=-\projc \mathbf K \proj \mathbf u(t)$, together with $f'(t)=-\projc \mathbf K \proj \mathbf w(t)$ and $f''(t)= \projc \mathbf K \proj \mathbf K \mathbf u(t)$.
For $j=0,1,2$, define $F_j = \max_{0\le t\le T}\norm{f^{(j)}(t)}$.

Then, 
\begin{equation}
    \max_{0\le t\le T} \norm{\projc \mathbf u(t)}_{\ell_2} \le \frac{1}{\lambda^2-\mu_{+}}
    \l(
    (1+\frac{\lambda}{\sqrt {\mu_-}})\norm{f(0)}
    + \frac{\norm{f'(0)}}{\sqrt{\mu_-}} + F_0 + 2TF_1 + \frac{TF_2}{\sqrt{\mu_-}}
    \r)
\end{equation}
If $f(0)=f'(0)=0$ and $F_j$, $j=0,1,2$ are bounded independently of $\lambda$, we get overall $O(\tfrac1{\lambda^2})$.
\end{lemma}
\begin{proof}
   Let $\mathbf y=\projc\mathbf u$ and $\mathbf z = \projc \mathbf w$, with dynamics $ \mathbf y' = \mathbf z -\imag\lambda\mathbf y$ and $ z' = -\mathbf B \mathbf y + f - \imag\lambda \mathbf z$. 

   Then, eliminating $\mathbf z$ gives
   \begin{equation}
        \mathbf y'' + 2\imag\lambda  \mathbf y '+ (\mathbf{B}-\lambda^2\identity)\mathbf y = \mathbf f,
   \end{equation}
   with initial conditions $\mathbf{ y} (0)= \mathbf y'(0) = 0$.

   Using $S(t)=\mathbf{B}^{-\tfrac12}\sin(t\sqrt{\mathbf B})$, $C(t)=\cos(t\sqrt{\mathbf B})$ and $H(t)=\e^{-\imag\lambda t}S(t)$.
   Then, we get the solution 
   \begin{equation}\label{eq:wave-bound-sol-formula}
       \mathbf y(t) = \int_0^t\D s\, H(t-s)f(s), 
   \end{equation}
   where $H$ solves the homogeneous DE with $H(0)=0$ and $H'(0)=\identity$.
   We continue by integration by parts. To that end, introduce $J(t)=\e^{-\imag\lambda t}C(t)$ and $R(t)=J(t)+\imag\lambda H(t)$.

   With $S'=C$ and $C'=-BS$, $H'=J-\imag\lambda H$ as well as $R'=(\lambda^2\identity-\mathbf B)H$ with $R(0)=\identity$. 
   Now, integrating \cref{eq:wave-bound-sol-formula} once using that $R'=(\lambda^2\identity-\mathbf B)H$, 
   \begin{equation}
       (\lambda^2 \identity - \mathbf{B})\mathbf{y}(t) = R(t)f(0) - f(t)+\int_0^t\D s\, R(t-s)f'(s).
   \end{equation}
   Upon substituting $R=2J-H'$, integrating \cref{eq:wave-bound-sol-formula} a second time leads to 
   \begin{equation}
       \int_0^t\D s\, H'(t-s)f'(s) = H(t)f'(0)+\int_0^t\D s\, H(t-s)f''(s),
   \end{equation}
   keeping in mind that $H(0)=0$. 
   We can then conclude that 
   \begin{equation}
       (\lambda^2\identity-\mathbf B)\mathbf y(t) = 
       R(t)f(0) -  H(t)f'(0) - f(t) + 2\int_0^t\D s\, J(t-s)f'(s) - \int_0^t\D s\, H(t-s)f''(s).
   \end{equation}
   Finally, given $\mu_-\preceq \mathbf B \preceq \mu_+$, we have that $\norm{J(t)}\le 1$,
   $\norm{H(t)}\le {\mu_-}^{-1/2}$ and thus
   $\norm{R(t)}\le 1+\lambda {\mu_-}^{-1/2}$.
   Thus, 
   \begin{equation}
       \norm{(\lambda^2\identity-\mathbf B)^{-1}}\le \frac{1}{\lambda^2-\mu_{+}}.
   \end{equation}
   Using this bound in the solution for $y(t)$ above proves the claim.
\end{proof}

\subsubsection{Discussion}
The results to our simulations are shown in \cref{fig:numerics-zero-bc-errors-heat,fig:numerics-circle-bc-errors-heat,fig:numerics-nonzero-bc-errors-heat,fig:numerics-neumann-bc-errors-heat} for the heat equation and \cref{fig:numerics-wave,fig:numerics-wave-neumann} for the wave equation. 
For all simulations, we observe that the error decay rate aligns very well with the upper bounds derived previously, i.e., roughly linearly, $\norm{\mathbf u}_{\domcomp}\sim\tfrac 1 \lambda$. 
The measured errors are several orders of magnitude smaller than the upper bounds that we derived. Even though the asymptotic behaviour seems to be well-captured in the bounds, they appear to be quite loose.  
Given non-dissipativity of the wave equation, we derived additional state-dependent upper bounds in this case which are less loose but still roughly one order of magnitude larger than the perturbative bounds.

The results in \cref{fig:numerics-nonzero-bc-errors-heat} validates the efficacy of the procedure of homogenizing/shifting the solution in order to solve for non-zero constraints. 
Moreover, we tested a circle boundary in \cref{fig:numerics-circle-bc-errors-heat} as a simple example for a non-trivial geometry. In particular, we notice that the solution does not penetrate further than two grid points into the constraint region within the chosen discretization.
Additionally note that the constraint error decays faster in the case of the wave equation in \cref{fig:numerics-wave,fig:numerics-wave-neumann}; roughly linear if not including the first point at $\lambda=100$ in the Dirichlet case. 
In the case of the heat equation, we observe that error performance is more reliable and more favourable in the case of derivative/Neumann constraints (\cref{fig:numerics-neumann-bc-errors-heat}) -- this holds also for the wave equation in \cref{fig:numerics-wave-neumann}. 
Therefore, using this setup in combination with `ghost points' that store a desired value and are out of the range of the generator $A_0$ could be an interesting alternative to directly enforcing the value constraints.  

\end{document}